\documentclass[final]{LMCS}

\def\doi{8 (3:04) 2012}
\lmcsheading%
{\doi}
{1--49}
{}
{}
{Oct.~27, 2011}
{Aug.~10, 2012}
{}

\usepackage{enumerate}
\usepackage{hyperref}
\usepackage{amssymb}

\def\eg{{\em e.g.}}
\def\cf{{\em cf.}}
\def\ie{{\em i.e.}}

\newenvironment{againtheorem}[1][Theorem]{\smallskip\noindent{\bf #1.}\it }{}

\def\Serbanuta{{\c S}erb{\u a}nu{\c t}{\u a}}
\def\Rosu{Ro{\c s}u}

\def\Cc{{\mathcal{C}}}
\def\cD{{\mathcal{D}}}
\def\cF{{\mathcal{F}}}
\def\cG{{\mathcal{G}}}
\def\cT{{\mathcal{T}}}

\def\cV{{\mathcal{V}}}
\def\tto{\twoheadrightarrow}
\def\pto{\rightrightarrows}
\def\Condition#1#2{#1 \approx #2}
\def\Arity{{\tt arity}}

\def\Pos{{\mathcal{P}os}}

\def\evbto{{%
	\vtop{\ialign{##\crcr%
		\noalign{\nointerlineskip}%
		\rightarrowfill\crcr%
		\noalign{\nointerlineskip}%
		$\hfil\scriptstyle{\ \sf evs \ }\hfil$\crcr%
	}}%
        {}}%
}
\def\evbpto{{%
	\vtop{\ialign{##\crcr%
		\noalign{\nointerlineskip}%
		$\rightrightarrows$\crcr%
		\noalign{\nointerlineskip}%
		$\hfil\scriptstyle{\sf evs \,}\hfil$\crcr%
	}}%
        {}}%
}
\def\Root{{\tt root}}
\def\Var{{\mathcal{V}ar}}
\def\EVar{{\mathcal{EV}ar}}
\def\Subst{{\mathcal{S}ub}}
\def\Dom{{\mathcal{D}om}}
\def\Range{{\mathcal{R}an}}

\def\Blank{\cdot}

\def\Tuple{\mathsf{tp}}
\def\U{{\mathbb{U}}}
\def\Uj{\mathbb{U}^r_{\rm J}}
\def\Ujl{\mathbb{U}_{\rm J}}
\def\Un{\mathbb{U}^r_{\rm N}}
\def\Unl{\mathbb{U}_{\rm N}}

\def\Uopt{{{\U}_{\rm opt}}}
\def\Usym{U}
\def\LaterUsedVar{Y}
\def\AppearedVar{X}
\def\AppearedLaterUsedVar{Z}
\def\SRTd{\mathbb{SR}}

\def\Tjn{{\mathcal{N}\!orm}}
\def\Tjd{\mathcal{D}et}

\def\TRSmult{R_1}
\def\TRSinvmult{R_2}
\def\Uadd{\mathsf{U}_1}
\def\Umulti{\mathsf{U}_2}
\def\Umultii{\mathsf{U}_3}
\def\TRSmarchiori{R_3}
\def\TRSmarchioriconstants{R_0}
\def\Umarchiori{\mathsf{U}_4}
\def\TRScs{R_4}
\def\Ucsfgx{\mathsf{U}_5}
\def\Ucsfgy{\mathsf{U}_6}
\def\Ucsg{\mathsf{U}_7}
\def\TRSmem{R_5}
\def\Umemfx{\mathsf{U}_8}
\def\Umemfb{\mathsf{U}_9}
\def\TRSmarchiorill{R_6}
\def\Umarchiorillf{\mathsf{U}_{10}}
\def\Umarchiorillg{\mathsf{U}_{11}}
\def\Umarchiorillh{\mathsf{U}_{12}}
\def\TRSqsort{R_7}
\def\Usplita{\mathsf{U}_{13}}
\def\Usplitb{\mathsf{U}_{14}}
\def\Usplitc{\mathsf{U}_{15}}
\def\Usplitd{\mathsf{U}_{16}}
\def\TRSne{R_8}
\def\Uneg{\mathsf{U}_{17}}
\def\Uney{\mathsf{U}_{18}}
\def\TRSquad{R_9}
\def\TRSinvquad{R_{10}}
\def\TRSfeh{R_{10}}
\def\Ufeh{\mathsf{U}_{19}}
\def\TRSsnoc{R_{11}}
\def\TRSinvsnoc{R_{20}}
\def\Uinvsnoc{\mathsf{U}_{21}}
\def\TRSoddeven{R_{12}}
\def\Uodda{\mathsf{U}_{22}}
\def\Uoddb{\mathsf{U}_{23}}
\def\Uevena{\mathsf{U}_{24}}
\def\Uevenb{\mathsf{U}_{25}}

\begin{document}

\title[Soundness of Unravelings for CTRSs via Ultra-Properties]%
{Soundness of Unravelings for Conditional Term Rewriting Systems via
Ultra-Properties \\Related to Linearity\rsuper*} 

\author[N.~Nishida]{Naoki Nishida}
\address{Graduate School of Information Science, Nagoya University
}
\email{\{nishida, sakai, sakabe\}@is.nagoya-u.ac.jp}

\author[M.\ Sakai]{Masahiko Sakai}

\author[T.~Sakabe]{Toshiki Sakabe}

\titlecomment{{\lsuper*}This research has been partially supported by 
{\it MEXT KAKENHI}\/ \#17700009, \#20300010 and \#21700011,
and 
{\it Kayamori Foundation of Informational Science Advancement}.
} 

\begin{abstract}
 \ 
Unravelings are transformations from conditional term rewriting systems
(CTRSs) into unconditional term rewriting systems (TRSs) over extended
signatures. They are complete, but in general, not sound w.r.t.\
reduction. Here, soundness w.r.t.\ reduction for a CTRS means that for
every term over the original signature of the CTRS, if the
corresponding unraveled TRS reduces the term to a term over the
original signature, then so does the original CTRS. In this paper, we
show that an optimized variant of Ohlebusch's unraveling for
deterministic CTRSs is sound w.r.t.\ reduction if the corresponding
unraveled TRSs are left-linear, or both right-linear and
non-erasing. Then, we show that soundness of the variant implies
soundness of Ohlebusch's unraveling, and show that soundness of
Marchiori's unravelings for join and normal CTRSs also implies
soundness of Ohlebusch's unraveling. Finally, we show that soundness of
a transformation proposed by {\Serbanuta} and {\Rosu} for deterministic
CTRSs implies soundness of Ohlebusch's unraveling.
\end{abstract}

\maketitle

\section{Introduction}
\label{sec:intro}

{\em Unravelings}\/ are transformations from conditional term rewriting
systems (CTRSs) into unconditional term rewriting
systems (TRSs) over extended signatures of the original signatures for
the CTRSs.
They are complete w.r.t.\ 
reduction sequences of the original CTRSs~\cite{Mar96},
\ie, for every derivation of the CTRSs, there exists a corresponding
derivation of the unraveled TRSs.
In this respect, the unraveled TRSs are over-approximations of the
original CTRSs w.r.t.\ reduction, and the unraveled TRSs are useful
for analyzing the properties of the original CTRSs, such as syntactic
properties, {\em modularity}, and {\em operational termination}\/ since
TRSs are in general much easier to handle than CTRSs.  

Marchiori proposed unravelings for {\em join}\/ and {\em normal}\/ CTRSs
in order to analyze {\em ultra-properties}\/ and modularity of the
CTRSs~\cite{Mar96}, and he also proposed an unraveling for {\em
deterministic}\/ CTRSs (DCTRSs)~\cite{Mar97}.
The transformation technique used in his unravelings originates
from~\cite{BK86,GM87}.  
Afterwards, Ohlebusch presented an improved variant of Marchiori's unraveling
for DCTRSs in order to analyze termination of logic
programs~\cite{Ohl01}|Marchiori's and Ohlebusch's unravelings are
called {\em sequential unravelings}~\cite{GGS12}.
Termination of the unraveled TRSs is a sufficient condition for proving
{\em operational termination}\/ of the original CTRSs~\cite{LMM05}.  
Later, a variant of Ohlebusch's unraveling was proposed in
both~\cite{Nishida04phd} and~\cite{DLMMU04}. 
This variant is sometimes called {\em optimized}, in the sense
that the variable-carrying arguments of {\em U symbols}\/ introduced by
means of the application of the unraveling are optimized, \ie, 
U symbols propagate only values received by variables that are referred
later.  

Although the mechanism of unconditional rewriting is much simpler than
that of conditional rewriting, the reduction of the unraveled TRS has
never been used instead of the original CTRS in order to reduce terms
over the original signature, until being used in program inversion
methods~\cite{Nishida04phd,NSS05rta,NSS05ieice} described later. 
This is because unravelings are not {\em sound w.r.t.\ reduction}\/ in
general~\cite{Mar96,Ohl01} while they are complete.   
Here, soundness w.r.t.\ reduction (simply, {\em soundness}) for a
CTRS means that, 
for every term over the original signature of the CTRS, if the unraveled
TRS reduces the term to a term over the original signature, then so does
the original CTRS~\cite{Mar96}.  
Several studies have been made on soundness conditions of
unravelings|some syntactic properties and particular reduction
strategies for the unraveled TRSs. 
Marchiori showed that his unraveling for normal CTRSs is sound for
left-linear ones~\cite{Mar96}, and he also showed that his unraveling
for DCTRSs is sound for DCTRSs that are semi-linear or
confluent~\cite{Mar97}.  
Nishida et al.\ showed that the combined
reduction restriction of the {\em membership condition}~\cite{Toy87} and {\em
context-sensitive condition}~\cite{Luc98} determined by means of the
application of the optimized unraveling is sufficient for
soundness~\cite{NSS05rta}.  
Later, Schernhammer and Gramlich showed that the same
context-sensitive condition without the membership condition is
sufficient for soundness of Ohlebusch's unraveling~\cite{SG07,SG10}
and Gmeiner et al.\ showed that Marchiori's unraveling
for normal CTRSs is sound for confluent, non-erasing, or weakly
left-linear ones, and they presented some properties that are not
sufficient for soundness~\cite{GGS10}.

As another kind of transformation from CTRSs to TRSs, 
{\Serbanuta} and {\Rosu} proposed a complete transformation ({\em SR
transformation}) from strongly or syntactically DCTRSs into
TRSs~\cite{SR06,SR06b}. 
The SR transformation is sound if the DCTRSs are semi-linear or
confluent, where function symbols in the original signatures are
completely extended by increasing the arities of some function
symbols.
The SR transformation is based on Viry's approach~\cite{Vir99} that is
another direction of developing transformations from CTRSs into TRSs,
and that has been further studied in~\cite{ABH03,Ros04}. 
The SR transformation provides 
{\em computationally equivalent}\/ TRSs to the original DCTRSs if the
original DCTRSs are operationally terminating and either
semi-linear or (ground) confluent. 
On the other hand, the theoretical relationship between the SR
transformation and the existing unravelings has never been discussed. 

In this paper, we show two sufficient conditions of DCTRSs for soundness
of the optimized unraveling:
one condition is {\em ultra-left-linearity}, \ie, that the unraveled TRSs are
left-linear, and the second condition is the combination of {\em
ultra-right-linearity}\/ and {\em ultra-non-erasingness}, \ie, that the
unraveled TRSs are right-linear and non-erasing. 
We also provide necessary and sufficient conditions of DCTRSs under
which the corresponding unraveled TRSs are left-linear, right-linear,
and non-erasing, respectively. 
All the conditions are syntactic and it is decidable whether a
DCTRS satisfies the conditions.
Moreover, we show that soundness of the optimized unraveling
implies soundness of Ohlebusch's unraveling, \ie, if the 
optimized unraveling is sound for a DCTRS, then so is Ohlebusch's
unraveling.
Finally, we show that 
soundness of the existing unravelings and the SR transformation
respectively imply soundness of Ohlebusch's unraveling. 
This paper is different from the preliminary version~\cite{NSS11rta} in
that we present
\begin{enumerate}[$\bullet$]
 \item abstract comparison methods for soundness of two
       transformations from CTRSs into TRSs
       (Lemmas~\ref{lem:soundness_of_two-unravelings},~\ref{lem:ex-soundness_of_two-unravelings}
       and Theorem~\ref{th:CTRS-transformations}),
 \item a comparison with other unravelings for join and
       normal DCTRSs (Subsection~\ref{subsec:soundness_of_Uj-Un}) in
       terms of soundness, and    
 \item a comparison with the SR transformation
       (Section~\ref{sec:comparison}) in terms of soundness.
\end{enumerate}

The optimized unraveling has been employed in the ({\em full}\/ or {\em
partial}\/) {\em program inversion} methods for constructor
TRSs~\cite{Nishida04phd,NSS05rta,NSS05ieice}. 
The methods first transform a constructor TRS into a DCTRS that defines
(full or partial) inverses of functions defined in the constructor TRS,
and then {\em unravel}\/ the DCTRS into a TRS (see Example~\ref{ex:U}).  
The resulting TRS may have extra variables since the intermediate
DCTRS may have extra variables that occur on the right-hand side, but not
in the conditional part. 
For this reason, this paper allows TRSs to have extra variables.
In applying a rewrite rule, extra variables of the rule are allowed to
be instantiated with arbitrary terms.
Since many instantiated terms of extra variables are meaningless and
sometimes cause non-termination, we limit reduction sequences to
meaningful ones by giving a restriction to reduction sequences of the
resulting TRS.  
The restriction is {\em
EV-safeness}~\cite{NSS03entcs,Nishida04phd,NSS04ss} that is a relaxed 
variant of the {\em basicness}\/ property~\cite{Hul80,MH94} of reduction
sequences: 
when a TRS has extra variables, any redex introduced by means of
extra variables is not reduced anywhere in the reduction sequences.
In this paper, we discuss soundness of unravelings w.r.t.\ EV-safe
derivations of the unraveled TRSs.

It has been shown that the optimized unraveling is sound for the
intermediate DCTRSs of the inversion methods
in~\cite{Nishida04phd,NSS05rta,NSS05ieice}, where conditional rules of
the intermediate DCTRSs are of the restricted form:
$l\to r \Leftarrow s_1 \tto t_1;\ldots; s_k \tto t_k$ where 
$r,t_1,\ldots,t_k$ are non-variable constructor terms and
$s_1,\ldots,s_k$ are rooted by defined symbols. 
Although the optimized unraveling is known to be sound for the
intermediate DCTRSs, studies on soundness conditions of the (optimized)
unraveling would be useful when the intermediate DCTRSs are further
transformed into more relaxed forms, \eg, DCTRSs obtained by removing a
unary {\em tuple}\/ symbol $\mathsf{tp}_1$ (see Example~\ref{ex:quad}).  
Roughly speaking, in applying the inversion method, the resulting TRS
is often right-linear if the input constructor TRS is left-linear.
Moreover, the resulting TRS is non-erasing if the input constructor TRS
is fully inverted, and, in addition, the resulting TRS has no extra
variable if the input is non-erasing.
Note that injective functions are often defined by non-erasing TRSs and 
the class of injective functions is the most interesting as an
object of program inversion.
For the reasons mentioned above, the sufficient conditions shown in this
paper can 
 be used to guarantee that the resulting TRSs of the inversion method
 for left-linear constructor TRSs are definitely inverses of the
 constructor TRSs (see Example~\ref{ex:inversion}).

As mentioned previously, Ohlebusch's unraveling is sound for any
DCTRS if we introduce the particular context-sensitive
restriction to the reduction of the corresponding unraveled TRSs.
Since recently context-sensitive reduction has been well investigated
(\eg, techniques to prove context-sensitive termination) and its
interpreter can be easily implemented, the unraveled TRSs with the 
particular context-sensitivity can be used instead of the original CTRSs
to completely reduce terms over the original signature to terms over the
original signature. 
However, sufficient (syntactic) properties for
soundness without the restriction to the reduction are
useful for the use of the unraveled TRSs instead of the original CTRSs
since context-sensitivity makes the reduction more complicated than
ordinary reduction. 
Moreover, if the unraveling used
in~\cite{Nishida04phd,NSS05rta,NSS05ieice} is sound for the resulting
TRS obtained by the inversion method without context-sensitivity, then
we can apply the restricted version of {\em completion}~\cite{NS09entcs}
to the resulting TRS to make the resulting TRS convergent or to
provide useful information for transforming the intermediate DCTRS into
an equivalent 
functional program.   
For these reasons, soundness of unravelings without any
restriction to 
the reduction is meaningful in order to employ the reduction of the
unraveled TRSs instead of the original CTRSs.

In summary, the main contribution of this paper is to show the
following: 
\begin{enumerate}[$\bullet$]
 \item the optimized unraveling is sound for a DCTRS that is
       ultra-left-linear, or both ultra-right-linear and ultra-non-erasing
       (Theorems~\ref{th:LL-soundness},~\ref{th:RLNE-soundness}), 
 \item soundness of the existing unravelings and the SR transformation
       respectively implies soundness of Ohlebusch's 
       unraveling (Corollary~\ref{cor:U-soundness_from_Uopt} and
       Theorems~\ref{th:soundness_by_Un-full},~\ref{th:SRTd-soundness}), and 
 \item abstract comparison methods for soundness of two
       transformations from CTRSs into TRSs
       (Lemmas~\ref{lem:soundness_of_two-unravelings},~\ref{lem:ex-soundness_of_two-unravelings}
       and Theorem~\ref{th:CTRS-transformations}).
\end{enumerate}
All the soundness conditions are summarized at the end of this paper
(Table~\ref{tbl:all-results} in Subsection~\ref{subsec:comparison}).

This paper is organized as follows.
In Section~\ref{sec:preliminaries}, we recall basic notions and
notations of term rewriting.
In Section~\ref{sec:unraveling}, we review the existing unravelings for
DCTRSs, and present syntactic properties of DCTRSs for some ultra-properties.
In Section~\ref{sec:soundness}, we show that the optimized unraveling is
sound for a DCTRS if the corresponding unraveled TRS is left-linear, or
both right-linear and non-erasing. 
In Section~\ref{sec:soundness_of_unravelings}, we show that 
soundness of the existing unravelings for join, normal, and deterministic
CTRSs respectively 
implies soundness of Ohlebusch's unraveling. 
In Section~\ref{sec:comparison}, we compare soundness of Ohlebusch's unraveling
with soundness of the SR transformation. 
In Section~\ref{sec:summary_and_related-work}, we briefly describe
related work and summarize soundness conditions of unravelings and the
SR transformation.
In Section~\ref{sec:conclusion}, we conclude this paper and briefly
describe future work on unravelings. 
Proofs of some technical results are included in the appendix.

\section{Preliminaries}
\label{sec:preliminaries}

In this section, we recall basic notions and notations of term
rewriting~\cite{BN98,Ohl02}.  

Let $\to_L$ be a binary relation (over a set of $A$) with a label $L$.
The reflexive closure of $\to_L$ is denoted by $\to^=_L$,
the transitive closure of $\to_L$ by $\to^+_L$, 
and the reflexive and transitive closure of $\to_L$ by $\to^*_L$.
The {\em joinability}\/ relation w.r.t.\ $\to_L$ is denoted by $\downarrow_L$:
$\downarrow_L$ $=$ $\to_L^* \cdot \gets_L^*$.
An element $a$ $\in$ $A$\/ is called a {\em normal form w.r.t.\ $\to_L$}\/
(or {\em w.r.t.\ $L$}\/) if there exists no element $b$ $\in$ $A$\/ such
that $a$ $\to_L$ $b$.

Throughout the paper, we use $\cV$\/ as a countably infinite set of {\em
variables}. 
Let $\cF$\/ be a {\em signature}, a finite set of {\em function symbols}\/
each of which has its own fixed arity, and $\Arity(f)$ be the arity of
function symbol $f$. 
The set of {\em terms}\/ over $\cF$\/ and $\cV$\/ is denoted by
$\cT(\cF,\cV)$, and the set of variables appearing in any of terms
$t_1,\ldots,t_n$ is denoted by $\Var(t_1,\ldots,t_n)$. 
A term $t$\/ is called {\em ground}\/ if $\Var(t)$ $=$ $\emptyset$.
A term is called {\em linear}\/ if any variable occurs in the term at most
once, and called {\em linear w.r.t.\ a variable}\/ if the variable appears
at most once in $t$. 
The set of {\em positions}\/ of term $t$\/ is denoted by $\Pos(t)$.
The set of positions for function symbols in $t$\/ is denoted by
$\Pos_\cF(t)$, and the set of positions for variables in $t$\/ is denoted
by $\Pos_\cV(t)$. 
For term $t$\/ and position $p$\/ of $t$, the notation $t|_p$
represents the {\em subterm}\/ of $t$\/ at position $p$.
The function symbol at the {\em root}\/ position $\varepsilon$\/ of term
$t$\/ is denoted by $\Root(t)$. 
Given an $n$-hole {\em context}\/ $C[~]$ with parallel positions
$p_1,\ldots,p_n$, the notation $C[t_1,\ldots,t_n]_{p_1,\ldots,p_n}$
represents the term obtained by replacing hole $\Box$
at position $p_i$ with term $t_i$ for all $1$ $\leq$ $i$ $\leq$ $n$.
We may omit the subscription ${}_{p_1,\ldots,p_n}$ from
$C[\ldots]_{p_1,\ldots,p_n}$. 
For positions $p$\/ and $p'$ of a term, we write $p'$ $\geq$ $p$\/ if
$p$\/ is a prefix of $p'$ (\ie, there exists a sequence $q$\/ such that $pq$ $=$
$p'$\/). 
Moreover, we write $p'$ $>$ $p$\/ if $p$\/ is a proper prefix of $p'$.

The {\em domain}\/ and {\em range}\/ of a {\em substitution}\/ $\sigma$ are
denoted by $\Dom(\sigma)$ and $\Range(\sigma)$, respectively.
We may denote $\sigma$ by $\{ x_1 \mapsto t_1, \ldots, x_n \mapsto t_n
\}$ if $\Dom(\sigma)$ $=$ $\{x_1,\ldots,x_n\}$ and $\sigma(x_i)$
$=$ $t_i$ for all $1$ $\leq$ $i$ $\leq$ $n$.
For a signature $\cF$, the set of {\em substitutions}\/ whose ranges are
over $\cF$\/ and $\cV$\/ is denoted by $\Subst(\cF,\cV)$:
$\Subst(\cF,\cV)$ $=$ $\{ \sigma \mid \Range(\sigma) \subseteq
\cT(\cF,\cV) \}$.  
For a substitution $\sigma$\/ and a term $t$, the application $\sigma(t)$ of
$\sigma$\/ to $t$\/ is abbreviated to $t\sigma$, and $t\sigma$\/ is called an
{\em instance}\/ of $t$.
Given a set $X$\/ of variables, $\sigma|_X$ denotes the {\em restricted}\/
substitution of $\sigma$ w.r.t.\ $X$\/:
$\sigma|_X$ $=$ $\{ x \mapsto x\sigma \mid x \in \Dom(\sigma) \cap X\}$.
The {\em composition}\/ $\sigma\theta$\/ of substitutions $\sigma$\/ and
$\theta$\/ is defined as $x(\sigma\theta)$ $=$ $(x\sigma)\theta$.

A {\em conditional rewrite rule}\/ over a signature $\cF$\/ is a
triple $(l,r,c)$, denoted by $l \to r \Leftarrow c$, such that the {\em
left-hand side}\/ $l$\/ is a non-variable term in $\cT(\cF,\cV)$, the {\em
right-hand side}\/ $r$\/ is a term in $\cT(\cF,\cV)$, and the {\em
conditional part}\/ $c$\/ is a sequence $\Condition{s_1}{t_1}; \ldots;
\Condition{s_k}{t_k}$\/ of 
term pairs ($k$ $\geq$ $0$) where all of $s_1,t_1,\ldots,s_k,t_k$\/ are
terms in $\cT(\cF,\cV)$.  
In particular, a conditional rewrite rule is called {\em unconditional}\/ if the
conditional part is the empty sequence (\ie, $k$ $=$ $0$), 
and we may abbreviate it to $l \to r$.
The conditional rewrite rule is called {\em extended}\/ if the condition
``\,$l$ $\not\in$ $\cV$\,'' is not imposed.
We sometimes attach a unique label $\rho$\/ to the conditional rewrite
rule $l \to r \Leftarrow c$\/ by denoting $\rho: l \to r \Leftarrow c$,
and we use the label to refer to the rewrite rule.
The sets of variables in $c$\/ and $\rho$\/ are denoted by $\Var(c)$ and
$\Var(\rho)$, respectively: 
$\Var( \Condition{s_1}{t_1}; \ldots; \Condition{s_k}{t_k} )$ $=$
$\Var(s_1,t_1,\ldots,s_k,t_k)$ and $\Var(\rho)$ $=$ $\Var(l,r) \cup \Var(c)$.
A variable occurring in $r$\/ or $c$\/ is called an {\em extra
variables of $\rho$}\/ if it does not occur in $l$.
The set of extra variables of $\rho$\/ is denoted by $\EVar(\rho)$:
$\EVar(\rho)$ $=$ $(\Var(r) \cup \Var(c)) \setminus \Var(l)$. 

A {\em conditional term rewriting system}\/ (CTRS) over a
signature $\cF$\/ is a set of conditional rules over $\cF$.
In particular, a CTRS is called an {\em EV-TRS}\/ if all of its rules are
unconditional, and called an {\em extended CTRS}\/ (eCTRS) if the
condition ``\,$l$ $\not\in$ $\cV$\,'' of conditional rewrite rules $l \to r
\Leftarrow c$\/ is not imposed. 
Moreover, a CTRS is called an (unconditional) {\em term rewriting 
system}\/ (TRS) if every rule $l \to r \Leftarrow c$\/ in
the CTRS is unconditional and satisfies $\Var(l)$ $\supseteq$ $\Var(r)$. 
Note that an eCTRS is called an {\em eTRS}\/ if all of its rules are
unconditional. 
The {\em underlying unconditional system}\/ of a CTRS $R$\/ is denoted by
$R_u$\/: $R_u$ $=$ $\{ l \to r \mid l \to r \Leftarrow c \in R \}$.

A CTRS $R$\/ is called {\em oriented}\/ if the symbol $\Condition{}{}$ in
the conditions of its rewrite rules is interpreted as {\em
reachability}\/: 
the {\em reduction relation}\/ of $R$\/ is defined as $\to_R$ $=$
$\bigcup_{n \geq 0} \to_{(n),R}$ where
\begin{enumerate}[$\bullet$]
 \item $\to_{(0),R}$ $=$ $\emptyset$, and
 \item $\to_{(i+1),R}$ $=$ $\to_{(i),R}$ $\cup$ $\{
       (C[l\sigma]_p,C[l\sigma]_p) \mid \rho: l \to r \Leftarrow
       \Condition{s_1}{t_1}; \ldots; \Condition{s_k}{t_k} \in R, ~
       s_1\sigma \to^*_{(i),R} t_1\sigma, ~\ldots,~ s_k\sigma
       \to^*_{(i),R} t_k\sigma \}$ for $i$ $\geq$ $0$. 
\end{enumerate}
Rewrite rules $l \to r \Leftarrow \Condition{s_1}{t_1}; \ldots;
\Condition{s_k}{t_k}$\/ of oriented CTRSs are written as $l \to r
\Leftarrow s_1 \tto t_1; \ldots; s_k \tto t_k$. 
To specify the applied rule $\rho$\/ and the position $p$\/ where $\rho$\/ is
applied, we may write $\to_{p,\rho}$ or $\to_{p,R}$ instead of $\to_R$. 
Moreover, we may write $\to_{>\varepsilon,R}$ instead of $\to_{p,R}$ if
$p$ $>$ $\varepsilon$. 
The {\em parallel reduction}\/ $\pto_R$ is defined as
$\pto_R$ $=$ $\{
(C[s_1,\ldots,s_n]_{p_1,\ldots,p_n},C[t_1,\ldots,t_n]_{p_1,\ldots,p_n})
\mid s_1 \to_R t_1, ~\ldots, ~ s_n \to_R t_n \}$.
To specify the positions $p_1,\ldots,p_n$ in the definition, we may
write $\pto_{\{p_1,\ldots,p_n\},R}$ instead of
$\pto_R$, and we may
write $\pto_{>\varepsilon,R}$ instead of
$\pto_R$ if
$p_i$ $>$ $\varepsilon$ for all $1$ $\leq$ $i$ $\leq$ $n$.
Moreover, for a set $P$\/ of parallel positions, we may write $\pto_{\geq
P,R}$ instead of $\pto_R$ if, for each position $p_i$ $\in$
$\{p_1,\ldots,p_n\}$, there exists a position $p$ $\in$ $P$\/ such that
$p$ $\leq$ $p_i$.  

For an eCTRS $R$, a substitution $\sigma$\/ is called {\em normalized
w.r.t.\ $R$}\/ if $x\sigma$\/ is a normal form w.r.t.\ $R$\/ for every
variable $x$ $\in$ $\Dom(\sigma)$. 

An (extended) conditional rewrite rule $l \to r \Leftarrow c$\/ is
called
\begin{enumerate}[$\bullet$]
 \item {\em left-linear}\/ (LL) if $l$\/ is linear, 
 \item {\em right-linear}\/ (RL) if $r$\/ is linear,
 \item {\em non-erasing}\/ (NE) if $\Var(l)$ $\subseteq$ $\Var(r)$,  
 \item {\em non-collapsing}\/ or {\em non-right-variable}\/ (non-RV) if
       the right-hand side $r$\/ is not a variable, and   
 \item {\em non-left-variable}\/ (non-LV) if $l$\/ is not a
       variable.
\end{enumerate}
For a syntactic property {\it P}\/ of conditional rewrite rules, we say that
an eCTRS has the property {\it P}\/ if all of its rules have the
property {\it P}, \eg, an eCTRS is called {\em left-linear}\/ (LL) if
all of its rules are LL. 
Note that a non-LV eCTRS is a CTRS. 

An (extended) conditional rewrite rule $\rho: l \to r \Leftarrow s_1
\tto t_1; \ldots; s_k \tto t_k$\/ is called {\em deterministic}\/ if
$\Var(s_i)$ $\subseteq$ $\Var(l,t_1,\ldots,t_{i-1})$ for all $1$ $\leq$ $i$
$\leq$ $k$. 
An (e)CTRS is called {\em deterministic}, an {\em (e)DCTRS}\/ for short, if
all of its rules are deterministic. 
Conditional rule $\rho$\/ is classified according to the distribution of
variables in the rule as follows:
\begin{enumerate}[$\bullet$]
 \item {\em Type 1}\/ if $\Var(r,s_1,t_1,\ldots,s_k,t_k)$ $\subseteq$ $\Var(l)$,
 \item {\em Type 2}\/ if $\Var(r)$ $\subseteq$ $\Var(l)$, 
 \item {\em Type 3}\/ if $\Var(r)$ $\subseteq$
       $\Var(l,s_1,t_1,\ldots,s_k,t_k)$, and
 \item {\em Type 4}\/ otherwise. 
\end{enumerate}
An (eD)CTRS is called an {\em {\it i}-(eD)CTRS}\/ if all of its rules are
of Type {\it i}.
An eDCTRS $R$ is called {\em normal}\/ (or a {\em normal CTRS}\/) if, for
every rule $l \to r \Leftarrow s_1 \tto t_1;\ldots; s_k \tto t_k$ $\in$
$R$, all of $t_1,\ldots,t_k$ are ground normal forms w.r.t.\ $R_u$.

Let $R$\/ be a CTRS over a signature $\cF$.
The sets of {\em defined symbols}\/ and {\em constructors}\/ of $R$\/ are
denoted by $\cD_R$ and $\Cc_R$, respectively:
$\cD_R$ $=$ $\{ \Root(l) \mid l \to r \Leftarrow c \in R\}$ and $\Cc_R$
$=$ $\cF \setminus \cD_R$.
Terms in $\cT(\Cc_R,\cV)$ are {\em constructor terms of $R$}.
$R$\/ is called a {\em constructor system}\/ if all proper subterms of the
left-hand sides in $R$\/ are constructor terms of $R$.

Let $R$\/ be a CTRS.
Two conditional rewrite rules $l_1 \to r_1 \Leftarrow c_1$ and $l_2 \to
r_2 \Leftarrow c_2$ in $R$\/ are called {\em overlapping}\/ if there
exists a context $C[~]$ and a non-variable term $t$\/ such that $l_2$ $=$
$C[t]$ and 
$l_1$ and $t$\/ are unifiable, 
where we assume w.l.o.g.\ that these rules share no variable.
Then, a conditional pair of terms $((C[r_1])\theta,r_2\theta)
\Leftarrow c_1\theta;c_2\theta$ is called a {\em critical pair}\/ of $R$\/
where $\theta$\/ is a {\em most general unifier}\/ of $l_1$ and $t$.
A critical pair $(s,t) \Leftarrow c$\/ is called {\em trivial}\/ if $s$
$=$ $t$, and called {\em infeasible}\/ if for any
substitution $\sigma$, $c$\/ contains a condition $u \tto v$\/ such that
$u\sigma$ $\not\to^*_R$ $v\sigma$~\cite{Kap87} (\cf,~\cite{Ohl02}).

Let $\cF_1,\cF_2$ be signatures, $\cG$ $\subseteq$ $\cF_1\cap\cF_2$, and
$\to_1,\to_2$ be binary relations on terms in $\cT(\cF_1,\cV)$ and
$\cT(\cF_2,\cV)$, 
respectively. 
We say that {\em $\to_1$ $\subseteq$ $\to_2$ on terms in $\cT(\cG,\cV)$}\/
if, for all terms $s,t$ $\in$ $\cT(\cG,\cV)$, $s$ $\to_2$ $t$\/ whenever $s$
$\to_1$ $t$.

\section{Unravelings for Deterministic CTRSs}
\label{sec:unraveling}

In this section, we first recall 
unravelings for DCTRSs, 
and then show some syntactic properties of DCTRSs, that are related to
the syntactic properties of the unraveled TRSs. 
The unravelings and some results in this section are straightforwardly
extended to eDCTRSs. 

We first recall the notion of unravelings. 
A computable transformation $U$\/ from eCTRSs into eTRSs is called an {\em
unraveling}\/ if for every eCTRS $R$, we have $\to_R$ $\subseteq$
$\to^*_{U(R)}$ and $U(R' \cup R)$ $=$ $R' \cup U(R)$ whenever $R'$ is an
eTRS~\cite{Mar96,Mar97}.%
\footnote{\label{fnt:unraveling-definition} In the original
definition~\cite{Mar96}, not the property $\to_R$ $\subseteq$ 
$\to^*_{U(R)}$ but the property $\downarrow_R$ $\subseteq$
$\downarrow_{U(R)}$ is imposed.
Under this property, unravelings are not complete in general.
For example, if $\to_R$ $\subseteq$ $\gets^*_{U(R)}$, then $U$\/ is
an unraveling.
However, all the existing unravelings are designed so as to satisfy
$\to_R$ $\subseteq$ $\to^*_{U(R)}$, that is implicitly required of
unravelings. 
For this reason, this paper imposes the more restrictive property
$\to_R$ $\subseteq$ $\to^*_{U(R)}$.}
Unraveling $U$\/ is called {\em tidy}\/ if it has {\em compositionality}\/
($U(R_1 \cup R_2)$ $=$ $U(R_1) \cup U(R_2)$), {\em finiteness}\/
(if $R$\/ is finite, then so is $U(R)$), and {\em emptiness}\/ (if $R$\/ is
empty, then so is $U(R)$)~\cite{Mar96}.
Let $R$\/ be an eCTRS over a signature $\cF$, and $\Rightarrow_{U(R)}$ be a
subrelation of $\to_{U(R)}$. 
$U$\/ is called {\em sound w.r.t.\ reduction for $R$ w.r.t.\
$\Rightarrow_{U(R)}$}\/ ({\em simulation-sound}~\cite{NSS04ss,NSS05rta}, or
simply {\em sound for $R$ w.r.t.\ $\Rightarrow_{U(R)}$}\/) if
$\Rightarrow^*_{U(R)}$ $\subseteq$ $\to^*_R$ on terms in $\cT(\cF,\cV)$
(\ie, for all terms $s,t$\/ in $\cT(\cF,\cV)$,  if $s$
$\Rightarrow^*_{U(R)}$ $t$, then $s$ $\to^*_R$ $t$\/).  
$U$\/ is called {\em complete w.r.t.\ reduction for $R$ w.r.t.\
$\Rightarrow_{U(R)}$}\/ (or simply {\em complete for $R$ w.r.t.\
$\Rightarrow_{U(R)}$}\/) if $\to^*_R$ $\subseteq$ $\Rightarrow^*_{U(R)}$
on terms in $\cT(\cF,\cV)$.
We omit ``w.r.t.\ $\to_{U(R)}$'' if $\Rightarrow_{\U(R)}$ $=$ $\to_{U(R)}$.  

Next, we recall an unraveling for DCTRSs, proposed by
Ohlebusch~\cite{Ohl01} that is a natural improvement of Marchiori's
unraveling~\cite{Mar97}.
For a finite set $T$ $=$ $\{t_1,\ldots,t_n\}$, given some fixed ordering
$\prec$ such that $t_1 \prec \cdots \prec t_n$, $\overrightarrow{T}$ denotes 
the unique sequence $t_1,\ldots,t_n$ of elements in $T$.
\begin{defi}[unraveling $\U$~\cite{Ohl01}]\label{def:U}
Let $R$\/ be an eDCTRS over a signature $\cF$.
For every conditional rule $\rho: l \to r \Leftarrow s_1 \tto t_1;
 \ldots; s_k \tto t_k$\/ in $R$, we prepare $k$\/ fresh function symbols
 $\Usym^\rho_1,\ldots,\Usym^\rho_k$, called {\em U symbols}, 
 that do not appear in $\cF$. 
 We transform $\rho: l \to r \Leftarrow s_1 \tto t_1; \ldots; s_k \tto t_k$
 into a set $\U(\rho)$ of $k+1$ unconditional rewrite rules as follows:
\[
 \U(\rho) =
 \left\{
 \begin{array}{r@{\>}c@{\>}l}
  l & \to & \Usym^\rho_1(s_1,\overrightarrow{\AppearedVar_1}) \\
  \Usym^\rho_1(t_1,\overrightarrow{\AppearedVar_1}) & \to &
   \Usym^\rho_2(s_2,\overrightarrow{\AppearedVar_2}) \\
  & \vdots \\
  \Usym^\rho_k(t_k,\overrightarrow{\AppearedVar_k}) & \to & r \\
 \end{array}
 \right\}
\]
where $\AppearedVar_i$ $=$ $\Var(l,t_1,\ldots,t_{i-1})$ for all $1$
 $\leq$ $i$ $\leq$ $k$. 
Note that $\U(l' \to r')$ $=$ $\{ l' \to r' \}$.
$\U$ is extended to eDCTRSs (\ie, $\U(R)$ $=$ $\bigcup_{\rho \in R}
 \U(\rho)$) and $\U(R)$ is an eTRS over the 
 extended signature $\cF_{\U(R)}$ $=$ $\cF \cup \{
 \Usym^\rho_i \mid \rho: l \to r \Leftarrow s_1 \tto t_1; \ldots; s_k
 \tto t_k \in R,~ 1 \leq i \leq k \}$.
\end{defi}
\noindent
It is clear that $\to_R$ $\subseteq$ $\to^*_{\U(R)}$, and $\U(R' \uplus
R)$ $=$ $R' \cup \U(R)$ if $R'$ is unconditional. 
Moreover, by definition, $\U$ has compositionality, finiteness, and
emptiness. 
Thus, $\U$ is a tidy unraveling for eDCTRSs.

The variant $\Uopt$ of Ohlebusch's unraveling $\U$ is proposed
in both~\cite{Nishida04phd} and~\cite{DLMMU04}. 
For a conditional rewrite rule $\rho: l \to r \Leftarrow s_1 \tto
t_1;\ldots; s_k \tto t_k$, 
the set $\Uopt(\rho)$ of unconditional rewrite rules is defined by
replacing $\overrightarrow{\AppearedVar_i}$ in $\U(\rho)$ by
$\overrightarrow{\AppearedVar_i \cap
\Var(r,t_i,s_{i+1},t_{i+1},\ldots,s_k,t_k)}$ for all $1$ $\leq$ $i$
$\leq$ $k$:  
\[
 \Uopt(\rho) =
 \left\{
  \begin{array}{r@{\>}c@{\>}l}
   l & \to & \Usym^\rho_1(s_1,\overrightarrow{\AppearedVar_1 \cap \LaterUsedVar_1}) \\
   \Usym^\rho_1(t_1,\overrightarrow{\AppearedVar_1 \cap \LaterUsedVar_1}) & \to &
    \Usym^\rho_2(s_2,\overrightarrow{\AppearedVar_2 \cap \LaterUsedVar_2}) \\
   & \vdots \\
   \Usym^\rho_k(t_k,\overrightarrow{\AppearedVar_k \cap \LaterUsedVar_k}) & \to & r \\
  \end{array}
 \right\}
\]
where $\LaterUsedVar_i$ $=$ $\Var(r,t_i,s_{i+1},t_{i+1},\ldots,s_k,t_k)$
for all $1$ $\leq$ $i$ $\leq$ $k$. 
Note that $\Uopt(R)$ is an eTRS over the extended signature $\cF_{\Uopt(R)}$
 where $\cF_{\Uopt(R)}$ $=$ $\cF \cup \{ \Usym^\rho_i \mid \rho: l \to r
 \Leftarrow s_1 \tto t_1; \ldots; s_k \tto t_k \in R,~ 1 \leq i \leq k
 \}$.%
\footnote{ The extended signatures $\cF_{\U(R)}$ and $\cF_{\Uopt(R)}$ are
not equivalent in terms of the arities of U symbols (see, \eg,
Example~\ref{ex:U}).
We distinguish between these extended signatures since we deal with
mappings from $\cT(\cF_{\U(R)},\cV)$ to $\cT(\cF_{\Uopt(R)},\cV)$ in
Subsection~\ref{subsec:soundness_of_U}.} 
Note also that $\Uopt$ is a tidy unraveling for eDCTRSs.
$\LaterUsedVar_i$ above, 
the set of variables appearing in any of
$r,t_i,s_{i+1},t_{i+1},\ldots,s_k,t_k$, 
is the set of variables that are referred after $s_i$ is considered. 
Thus, $\AppearedVar_i \cap \LaterUsedVar_i$ is the set of variables that
appear in any of $l,t_1,\ldots,t_{i-1}$ and also appear after $s_i$ is
considered, and hence $\overrightarrow{\AppearedVar_i \cap
\LaterUsedVar_i}$ is used for propagating only the variables that are
referred later. 
On the other hand, $\overrightarrow{\AppearedVar_i}$ in
Definition~\ref{def:U} is used for propagating {\em all}\/ the appeared variables.
This is the only difference between $\U$ and $\Uopt$, and the
reason why $\Uopt$ is sometimes called an {\em optimized}\/ variant of
$\U$.
Note that all of the following are equivalent:
\begin{enumerate}[$\bullet$]
 \item $R$\/ is of Type~3, 
 \item $\U(R)$ has no extra variables, and 
 \item $\Uopt(R)$ has no extra variables.
\end{enumerate}

In the rest of the paper, unless noted otherwise, we use the label
$\rho$\/ for presenting a conditional rewrite rule $l \to r \Leftarrow
s_1 \tto t_1;\ldots; s_k \tto t_k$, and we denote the sets
$\Var(l,t_1,\ldots,t_{i-1})$,
$\Var(r,t_i,s_{i+1},t_{i+1},\ldots,s_k,t_k)$, and $\AppearedVar_i \cap
\LaterUsedVar_i$ by $\AppearedVar_i$, $\LaterUsedVar_i$, and
$\AppearedLaterUsedVar_i$, respectively.
\begin{exa}\label{ex:U}
Consider the following TRS defining addition and multiplication of
 natural numbers encoded as
 $\mathsf{0},\mathsf{s}(\mathsf{0}),\mathsf{s}(\mathsf{s}(\mathsf{0})),\ldots$:
\[
 \TRSmult = 
 \left\{
 \begin{array}{r@{\>}c@{\>}l}
  \mathsf{add}(\mathsf{0},y) & \to & y \\
  \mathsf{add}(\mathsf{s}(x),y) & \to & \mathsf{s}( \mathsf{add}(x,y) ) \\
  \mathsf{mult}(\mathsf{0},y) & \to & \mathsf{0} \\
  \mathsf{mult}(x,\mathsf{0}) & \to & \mathsf{0} \\
  \mathsf{mult}(\mathsf{s}(x),\mathsf{s}(y)) & \to & \mathsf{s}( \mathsf{add}(\mathsf{mult}(x,\mathsf{s}(y)),y) ) \\
 \end{array}
 \right\}
\]
The inversion method in~\cite{Nishida04phd} 
inverts this TRS to the following DCTRS $\TRSinvmult$ where
 $\mathsf{add}^{-1}$ and 
 $\mathsf{mult}^{-1}$ are function symbols that define the 
 inverse relation of $\mathsf{add}$ and $\mathsf{mult}$, respectively,%
 \footnote{ As inverse computation of
 $\mathsf{add}(\mathsf{s}^m(\mathsf{0}),\mathsf{s}^n(\mathsf{0}))$
 $\to^*_{\TRSmult}$ $\mathsf{s}^{m+n}(\mathsf{0})$ and
 $\mathsf{mult}(\mathsf{s}^m(\mathsf{0}),\mathsf{s}^n(\mathsf{0}))$
 $\to^*_{\TRSmult}$ $\mathsf{s}^{m \times n}(\mathsf{0})$, we have the
 derivations $\mathsf{add}^{-1}(\mathsf{s}^{m+n}(\mathsf{0}))$
 $\to^*_{\TRSinvmult}$
 $\Tuple_2(\mathsf{s}^m(\mathsf{0}),\mathsf{s}^n(\mathsf{0}))$ and
 $\mathsf{mult}^{-1}(\mathsf{s}^{m \times n}(\mathsf{0}))$
 $\to^*_{\TRSinvmult}$
 $\Tuple_2(\mathsf{s}^m(\mathsf{0}),\mathsf{s}^n(\mathsf{0}))$.}
 and $\Tuple_2$ is a binary constructor for representing tuples of two
 terms: 
 \[
 \TRSinvmult = 
 \left\{
  \begin{array}{@{\>}r@{\>}c@{\>}l@{}}
   \mathsf{add}^{-1}(y) & \to & \Tuple_2(\mathsf{0},y) \\
   \mathsf{add}^{-1}(\mathsf{s}(z)) & \to & \Tuple_2(\mathsf{s}(x),y)
    \Leftarrow \mathsf{add}^{-1}(z) \tto \Tuple_2(x,y) \\
   \mathsf{mult}^{-1}(\mathsf{0}) & \to & \Tuple_2(\mathsf{0},y)  \\
   \mathsf{mult}^{-1}(\mathsf{0}) & \to & \Tuple_2(x,\mathsf{0}) \\
   \mathsf{mult}^{-1}(\mathsf{s}(z)) & \to & \Tuple_2(\mathsf{s}(x),\mathsf{s}(y))
    \Leftarrow \mathsf{add}^{-1}(z) \tto \Tuple_2(w,y);~
    \mathsf{mult}^{-1}(w) \tto \Tuple_2(x,\mathsf{s}(y)) \\
  \end{array}
 \right\}
 \]
This DCTRS is unraveled by $\U$ and $\Uopt$ as follows:
\[
 \U(\TRSinvmult) =
 \left\{
  \begin{array}{r@{\>}c@{\>}l}
   & \vdots \\
   \mathsf{add}^{-1}(\mathsf{s}(z)) & \to & \Uadd(\mathsf{add}^{-1}(z),z) \\
   \Uadd(\Tuple_2(x,y),z) & \to & \Tuple_2(\mathsf{s}(x),y) \\
   & \vdots \\   
   \mathsf{mult}^{-1}(\mathsf{s}(z)) & \to & \Umulti(\mathsf{add}^{-1}(z),z) \\
   \Umulti(\Tuple_2(w,y),z) & \to & \Umultii(\mathsf{mult}^{-1}(w),z,w,y) \\
   \Umultii(\Tuple_2(x,\mathsf{s}(y)),z,w,y) & \to & \Tuple_2(\mathsf{s}(x),\mathsf{s}(y)) \\
  \end{array}
 \right\}
\]
\[
  \Uopt(\TRSinvmult) = 
 \left\{
  \begin{array}{r@{\>}c@{\>}l}
   & \vdots \\
   \mathsf{add}^{-1}(\mathsf{s}(z)) & \to & \Uadd(\mathsf{add}^{-1}(z)) \\
   \Uadd(\Tuple_2(x,y)) & \to & \Tuple_2(\mathsf{s}(x),y) \\
   & \vdots \\
   \mathsf{mult}^{-1}(\mathsf{s}(z)) & \to & \Umulti(\mathsf{add}^{-1}(z)) \\
   \Umulti(\Tuple_2(w,y)) & \to & \Umultii(\mathsf{mult}^{-1}(w),y) \\
   \Umultii(\Tuple_2(x,\mathsf{s}(y)),y) & \to & \Tuple_2(\mathsf{s}(x),\mathsf{s}(y)) \\
  \end{array}
 \right\}
\]
\end{exa}

Unravelings are not sound in general.
The CTRS shown in the following example is a counterexample against
soundness of both $\U$ and $\Uopt$.
\begin{exa}[\cite{Mar96,Ohl02}]\label{ex:marchiori}
Consider the following 3-DCTRS $\TRSmarchiori$ and its unraveled TRSs:
\[
 \TRSmarchiori = 
 \left\{
 \begin{array}{r@{\>}c@{\>}l}
  \mathsf{f}(x) & \to & x \Leftarrow x \tto \mathsf{e} \\
  \mathsf{g}(\mathsf{d},x,x) & \to & \mathsf{A} \\
  \mathsf{h}(x,x) & \to & \mathsf{g}(x,x,\mathsf{f}(\mathsf{k})) \\
 \end{array}
 \right\}
 \cup \TRSmarchioriconstants
\]
\[
 \U(\TRSmarchiori) = \Uopt(\TRSmarchiori) = 
 \left\{
 \begin{array}{r@{\>}c@{\>}l}
  \mathsf{f}(x) & \to & \Umarchiori(x,x) \\
  \Umarchiori(\mathsf{e},x) & \to & x \\
  & \vdots \\
 \end{array}
 \right\} 
 \cup \TRSmarchioriconstants
\]
where
\[
 \TRSmarchioriconstants = 
 \left\{
 \begin{array}{r@{\>}c@{\>}l@{\quad\quad}r@{\>}c@{\>}l@{\quad\quad}r@{\>}c@{\>}l@{\quad\quad}r@{\>}c@{\>}l@{\quad\quad}r@{\>}c@{\>}l}
  \mathsf{a} & \to & \mathsf{c} &
  \mathsf{b} & \to & \mathsf{c} &
  \mathsf{c} & \to & \mathsf{e} &
  \mathsf{k} & \to & \mathsf{l} &
  \mathsf{d} & \to & \mathsf{m} \\
  \mathsf{a} & \to & \mathsf{d} &
  \mathsf{b} & \to & \mathsf{d} &
  \mathsf{c} & \to & \mathsf{l} &
  \mathsf{k} & \to & \mathsf{m} \\
 \end{array}
\right\}
\]
We have a reduction sequence of $\Uopt(\TRSmarchiori)$ from
 $\mathsf{h}(\mathsf{f}(\mathsf{a}),\mathsf{f}(\mathsf{b}))$ to
 $\mathsf{A}$:
\[
 \begin{array}{l@{\>}c@{\>}l}
  \mathsf{h}(\mathsf{f}(\mathsf{a}),\mathsf{f}(\mathsf{b})) 
   & \to^*_{\Uopt(\TRSmarchiori)} & 
   \mathsf{h}(\Umarchiori(\mathsf{c},\mathsf{d}),\Umarchiori(\mathsf{c},\mathsf{d})) 
   \to_{\Uopt(\TRSmarchiori)} 
    \mathsf{g}(\Umarchiori(\mathsf{c},\mathsf{d}),\Umarchiori(\mathsf{c},\mathsf{d}),\mathsf{f}(\mathsf{k}))
    \\ 
   & \to^*_{\Uopt(\TRSmarchiori)} & 
    \mathsf{g}(\mathsf{d},\Umarchiori(\mathsf{l},\mathsf{m}),\Umarchiori(\mathsf{l},\mathsf{m})) 
    \to_{\Uopt(\TRSmarchiori)} 
    \mathsf{A} \\
 \end{array}
\]
 However, we have no similar reduction sequence of $\TRSmarchiori$, 
 \ie, $\mathsf{h}(\mathsf{f}(\mathsf{a}),\mathsf{f}(\mathsf{b}))$
 $\not\to^*_{\TRSmarchiori}$ $\mathsf{A}$.  
Thus, neither $\U$ nor $\Uopt$
is sound for $\TRSmarchiori$.
Note that being \mbox{(ultra-)}overlapping-systems is not sufficient for
 soundness of $\Uopt$ and $\U$ since $\Uopt(\TRSmarchiori)$ ($=$
 $\U(\TRSmarchiori)$) is an overlapping system. 
\end{exa}

Soundness of $\U$ can be recovered by restricting
the reduction of the unraveled TRSs to the {\em context-sensitive}\/
reduction~\cite{Luc98} with the {\em replacement mapping}\/ determined by means
of the application of $\U$~\cite{SG07,SG10}:
$\U$ is sound for a 3-DCTRS $R$\/ if the reduction
of $\U(R)$ is restricted to context-sensitive rewriting with the
replacement mapping $\mu$
such that $\mu(\Usym^\rho_i)$ $=$ $\{1\}$ for any U symbol
$\Usym^\rho_i$|the replacement mapping forbids reducing any redex inside
the second or later arguments of U symbols. 
This holds for $\Uopt$ by restricting
the context-sensitive reduction to the reduction with the {\em membership
condition}~\cite{Toy87},
a very complicated restriction that soundness of $\U$ does not require.  
In this respect, $\Uopt$ does not look like an ``optimized'' variant of
$\U$.  
The following examples show that neither the context-sensitive nor
membership conditions above is sufficient on its own for soundness
of $\Uopt$.
\begin{exa}
\label{ex:mem-cs-unsound}
Consider the following DCTRS and its unraveled TRSs:
\[
 \TRScs =
 \left\{
 \begin{array}{r@{\>}c@{\>}l@{\quad\quad}r@{\>}c@{\>}l}
  \mathsf{f}(x,y) & \to & x \Leftarrow \mathsf{g}(x) \tto z;~
   \mathsf{g}(y) \tto z 
   &
  \mathsf{g}(x) & \to & \mathsf{c} \Leftarrow \mathsf{d} \tto \mathsf{c}\\
 \end{array}
 \right\}
\]
\[
 \U(\TRScs) =
 \left\{
 \begin{array}{r@{\>}c@{\>}l@{\quad}r@{\>}c@{\>}l@{\quad}r@{\>}c@{\>}l}
  \mathsf{f}(x,y) & \to & \Ucsfgx(\mathsf{g}(x),x,y) &
  \Ucsfgx(z,x,y) & \to & \Ucsfgy(\mathsf{g}(y),x,y,z) &
  \Ucsfgy(z,x,y,z) & \to &  x \\
  \mathsf{g}(x) & \to & \Ucsg(\mathsf{d},x) &
   \Ucsg(\mathsf{c},x) & \to & \mathsf{c} \\
 \end{array}
 \right\}
\]
\[
 \Uopt(\TRScs) =
 \left\{
 \begin{array}{r@{\>}c@{\>}l@{\quad\quad}r@{\>}c@{\>}l@{\quad\quad}r@{\>}c@{\>}l}
  \mathsf{f}(x,y) & \to & \Ucsfgx(\mathsf{g}(x),x,y) &
  \Ucsfgx(z,x,y) & \to & \Ucsfgy(\mathsf{g}(y),x,z) &
  \Ucsfgy(z,x,z) & \to &  x \\
  \mathsf{g}(x) & \to & \Ucsg(\mathsf{d}) &
   \Ucsg(\mathsf{c}) & \to & \mathsf{c} \\
 \end{array}
 \right\}
\]
For the context-sensitive condition mentioned above, we forbid reducing
 any redex inside the second or third arguments of $\Ucsfgx$ and $\Ucsfgy$.   
We have the derivation $\mathsf{f}(\mathsf{a},\mathsf{b})$
 $\to^*_{\Uopt(\TRScs)}$
 $\Ucsfgy(\Ucsg(\mathsf{d}),\mathsf{a},\Ucsg(\mathsf{d}))$
 $\to_{\Uopt(\TRScs)}$ $\mathsf{a}$ under the context-sensitive
 condition, but this derivation is not possible in $\TRScs$.
 Therefore, the context-sensitive condition is not sufficient on its own
 for soundness of $\Uopt$. 
 Note that the derivation $\mathsf{f}(\mathsf{a},\mathsf{b})$
 $\to^*$ $\mathsf{a}$ does not hold in $\U(\TRScs)$
 under the context-sensitive condition, either, since
 $\mathsf{f}(\mathsf{a},\mathsf{b})$ can be reduced to both
 $\Ucsfgy(\Ucsg(\mathsf{d},\mathsf{b}),\mathsf{a},\mathsf{g}(\mathsf{a}))$
 and
 $\Ucsfgy(\Ucsg(\mathsf{d},\mathsf{a}),\mathsf{a},\Ucsg(\mathsf{d},\mathsf{b}))$, 
 but they are not reduced any more. 
\end{exa}
\begin{exa}
Consider the following DCTRS $\TRSmem$ and its unraveled TRSs:
\[
 \TRSmem = 
 \left\{
 \begin{array}{r@{\>}c@{\>}l@{\quad\quad}r@{\>}c@{\>}l}
  \mathsf{f}(x) & \to & x \Leftarrow 
   x \tto \mathsf{a};~\mathsf{b} \tto x 
   &
  \mathsf{a} & \to & \mathsf{b} \\
 \end{array}
 \right\}
\]
\[
 \U(\TRSmem) = \Uopt(\TRSmem) =
 \left\{
 \begin{array}{r@{\>}c@{\>}l@{\quad\quad}r@{\>}c@{\>}l@{\quad\quad}r@{\>}c@{\>}l@{\quad\quad}c}
  \mathsf{f}(x) & \to & 
   \Umemfx(x,x) 
   & \Umemfx(\mathsf{a},x) & \to & \Umemfb(\mathsf{b},x) &
  \Umemfb(x,x) & \to & x 
  & \ldots \\
 \end{array}
 \right\}
\]
For the membership condition, we forbid reducing any redex that has a
 proper subterm containing U symbols.
We have the derivation $\mathsf{f}(\mathsf{a})$ $\to^*_{\Uopt(\TRSmem)}$
$\Umemfb(\mathsf{b},\mathsf{a})$ $\to_{\Uopt(\TRSmem)}$
 $\Umemfb(\mathsf{b},\mathsf{b})$ $\to_{\Uopt(\TRSmem)}$ $\mathsf{b}$
 under the membership condition, but this derivation is not possible in $\TRSmem$. 
Therefore, the membership condition is not sufficient on its own for
soundness of either $\Uopt$ or $\U$.
\end{exa}

To analyze syntactic relationships between eDCTRS and the corresponding
unraveled eTRSs, we recall {\em ultra-properties}\/ of
DCTRSs~\cite{Mar96,Mar97}, extending them to eDCTRSs. 
\begin{defi}[ultra-property~\cite{Mar96,Mar97}]
\label{def:ultra-property}
Let {\it P}\/ be a property on (extended) conditional rewrite rules, and
 $U$ be an unraveling. 
An (extended) conditional rewrite rule $\rho$\/ is said to be {\em
 ultra-{\it P}\/ w.r.t.\ $U$}\/ ($U$-{\it P}\/) if all the
 rules in $U(\rho)$ satisfy the property {\it P}. An eDCTRS $R$\/ is said
 to be {\em ultra-{\it P} w.r.t.\ $U$}\/ ($U$-{\it P}\/) if all the
 rules in $R$\/ are $U$-{\it P}.
\end{defi}
\begin{exa}
 \label{ex:ultra-P}
 The DCTRS $\TRSinvmult$ in Example~\ref{ex:U}
 is non-LV and non-RV w.r.t.\ both $\U$
 and $\Uopt$,
 but $\TRSinvmult$ is not $\U$-LL, $\U$-RL, or $\U$-NE either, 
 while 
 $\TRSinvmult$ is $\Uopt$-RL and $\Uopt$-NE, but not $\Uopt$-LL.
\end{exa}
\noindent
Note that the $\Uopt$-LL property is the same as {\em semi-linearity}\/
in~\cite{Mar97}.  
Roughly speaking, the conditional parts of $\Uopt$-LL conditional rules
correspond to the \texttt{let} structures of functional programs. 

The $\Uopt$-LL, $\Uopt$-RL, and $\Uopt$-NE properties of conditional
rewrite rules are identical with the following syntactic properties of
DCTRSs, respectively.  
\begin{thm}
\label{th:Uopt-P}
 Let $\rho: l \to r \Leftarrow s_1 \tto t_1; \ldots; s_k \tto t_k$ be an
 extended deterministic conditional rewrite rule.
 Then, all of the following hold:
\begin{enumerate}[\em(1)]
 \item $\rho$ is $\Uopt$-LL iff all of $l,t_1,\ldots,t_k$ are linear and
       $\Var(t_i) \cap \AppearedVar_i$ $=$ $\emptyset$ for
       all\/ $1$ $\leq$ $i$ $\leq$ $k$, 
 \item $\rho$ is $\Uopt$-RL iff all of $r,s_1,\ldots,s_k$ are linear and
       $\Var(s_i) \cap \LaterUsedVar_i$ $=$
       $\emptyset$ for all\/ $1$ $\leq$ $i$ $\leq$ $k$, and   
 \item $\rho$ is $\Uopt$-NE iff\/
       $\Var(l)$ $\subseteq$ $\Var(r,s_1,\ldots,s_k)$ and
       $\Var(t_i)$ $\subseteq$ $\Var(r,s_{i+1},\ldots,s_k)$
       for all\/ $1$ $\leq$ $i$ $\leq$ $k$.
\end{enumerate}
\end{thm}
\proof
The proof can be seen in Appendix~\ref{subsec:proof:th:Uopt-P}. 
\qed
\noindent
The sufficient and necessary condition for the $\Uopt$-NE property in
Theorem~\ref{th:Uopt-P} is equivalent to the one shown
in~\cite{Nishida04phd,NSS05ieice} since the following are equivalent: 
\begin{enumerate}[$\bullet$]
 \item $\Var(l)$ $\subseteq$ $\Var(r,s_1,t_1,\ldots,s_k,t_k)$ and
       $\Var(t_i)$ $\subseteq$ $\Var(r,s_{i+1},t_{i+1},\ldots,s_k,$
       $t_k)$ for all $1$ $\leq$ $i$ $\leq$ $k$, and
 \item $\Var(l)$ $\subseteq$ $\Var(r,s_1,\ldots,s_k)$ and $\Var(t_i)$
       $\subseteq$ $\Var(r,s_{i+1},\ldots,s_k)$ for all $1$ $\leq$ $i$
       $\leq$ $k$.
\end{enumerate}
Neither the second nor third claims in Theorem~\ref{th:Uopt-P} holds
for $\U$ (\cf, Examples~\ref{ex:U},~\ref{ex:ultra-P}), while the
first one holds for $\U$.
Quite restricted variants of the second and third claims hold for $\U$.
\begin{thm}
\label{th:U-P}
Let $\rho: l \to r \Leftarrow s_1 \tto t_1; \ldots; s_k \tto t_k$ be an
 extended deterministic conditional rewrite rule.
Then, all of the following hold:
\begin{enumerate}[\em(1)]
 \item $\rho$ is $\U$-LL iff all of $l,t_1,\ldots,t_k$ are linear and
       $\Var(t_i) \cap \AppearedVar_i$ $=$ $\emptyset$ for all\/
       $1$ $\leq$ $i$ $\leq$ $k$, 
 \item $\rho$ is $\U$-RL iff\/ $r$ is linear and all of
       $s_1,\ldots,s_k$ are ground, and
 \item $\rho$ is $\U$-NE iff\/
       $\Var(l,t_1,\ldots,t_k)$ $\subseteq$ $\Var(r)$.
\end{enumerate}
\end{thm}
\proof
The proof can be seen in Appendix~\ref{subsec:proof:th:U-P}. 
\qed
\noindent
Note that the $\U$-LL and $\Uopt$-LL properties are equivalent.
Theorems~\ref{th:Uopt-P},~\ref{th:U-P} lead to the following
relationship between the ultra-RL and ultra-NE 
properties w.r.t.\ $\U$ and $\Uopt$.
\begin{cor}
 \label{col:U-P_implies_Uopt-P}
The $\U$-RL and\/ $\U$-NE properties imply the\/ $\Uopt$-RL
 and\/ $\Uopt$-NE properties, respectively.
\end{cor}

As for the non-LV and non-RV properties, we have the
following relationships between eDCTRSs and the corresponding unraveled
eTRSs. 
\begin{prop}
 \label{prop:NC-LV}
 Let $U$ be either\/ $\U$ or\/ $\Uopt$, $R$ be an eDCTRS, and\/ $\rho$
 be an (extended) conditional rewrite rule.
 Then, all of the following hold:
 \begin{enumerate}[$\bullet$]
  \item $\rho$ is non-LV iff so is $U(\rho)$,
  \item $\rho$ is non-RV iff so is $U(\rho)$, 
  \item $R$ is non-LV iff so is $U(R)$, and
  \item $R$ is non-RV iff so is $U(R)$.
 \end{enumerate}
\end{prop}
\proof
Trivial by definition.
\qed
\noindent
We recognize from Proposition~\ref{prop:NC-LV} that, for both $\U$
and $\Uopt$, the 
non-LV and non-RV properties are equivalent to the ultra-non-LV
and ultra-non-RV properties, respectively.

\section{Soundness of the Optimized Unraveling}
\label{sec:soundness}

In this section, we first show that the optimized unraveling $\Uopt$ is
sound for $\Uopt$-LL 3-DCTRSs. 
Then, we show that $\Uopt$ is sound for DCTRSs that are both $\Uopt$-RL
and $\Uopt$-NE. 
Finally, we extend the result on soundness for $\Uopt$-LL 3-DCTRSs
to $\Uopt$-LL DCTRSs, \ie, $\Uopt$ is sound for a $\Uopt$-LL DCTRS
if the reduction of the corresponding unraveled EV-TRS is restricted to
{\em EV-safe}\/ ones (see Definition~\ref{def:EV-safe}). 
In the rest of this paper, we write the terminology ``RLNE''
for ``RL and NE''.

\subsection{Soundness on Ultra-Left-Linearity}
\label{subsec:LL-soundness}

In this subsection, we first show that the LL property is not a
soundness condition of either $\Uopt$ or $\U$, and then we show that
$\Uopt$ is sound for $\Uopt$-LL 3-DCTRSs. 
This result also holds for arbitrary DCTRSs under some restriction to
reduction.
Although we first show the case of 3-DCTRSs to make the essential scheme
of the proof clear, we will extend the result in this subsection to
DCTRSs in Subsection~\ref{subsec:soundness_of_LL-evb}. 

As described in Section~\ref{sec:intro}, the LL property is a soundness
condition for unravelings for normal CTRSs.
In contrast, the LL property is not a soundness condition for either
$\Uopt$ or $\U$.
\begin{exa}
\label{ex:marchiorill}
Consider the following DCTRS obtained from $\TRSmarchiori$ by
 left-linearizing:
\[
 \TRSmarchiorill = 
 \left\{
 \begin{array}{r@{\>}c@{\>}l}
  \mathsf{f}(x) & \to & x \Leftarrow x \tto \mathsf{e} \\
  \mathsf{g}(\mathsf{d},x,y) & \to & \mathsf{A} \Leftarrow y \tto x \\
  \mathsf{h}(x,y) & \to & \mathsf{g}(x,y,\mathsf{f}(\mathsf{k})) \Leftarrow y \tto x \\
 \end{array}
 \right\}
 \cup \TRSmarchioriconstants
\]
$\TRSmarchiorill$ is unraveled by $\Uopt$ and $\U$ to the following
 TRSs:
\[
 \Uopt(\TRSmarchiorill) = 
 \left\{
 \begin{array}{r@{\>}c@{\>}l@{\quad\quad}r@{\>}c@{\>}l}
  \mathsf{f}(x) & \to & \Umarchiorillf(x,x) 
   & \Umarchiorillf(\mathsf{e},x) & \to & x \\
  \mathsf{g}(\mathsf{d},x,y) & \to & \Umarchiorillg(y,x) 
   & \Umarchiorillg(x,x) & \to & \mathsf{A} \\
  \mathsf{h}(x,y) & \to & \Umarchiorillh(y,x,y) 
   & \Umarchiorillh(x,x,y) & \to & \mathsf{g}(x,y,\mathsf{f}(\mathsf{k})) \\
 \end{array}
 \right\}
 \cup \TRSmarchioriconstants
\]
\[
 \U(\TRSmarchiorill) = 
 \left\{
 \begin{array}{r@{\>}c@{\>}l@{\quad\quad}r@{\>}c@{\>}l}
  \mathsf{f}(x) & \to & \Umarchiorillf(x,x) 
   & \Umarchiorillf(\mathsf{e},x) & \to & x \\
  \mathsf{g}(\mathsf{d},x,y) & \to & \Umarchiorillg(y,x,y) 
   & \Umarchiorillg(x,x,y) & \to & \mathsf{A} \\
  \mathsf{h}(x,y) & \to & \Umarchiorillh(y,x,y) 
   & \Umarchiorillh(x,x,y) & \to & \mathsf{g}(x,y,\mathsf{f}(\mathsf{k})) \\
 \end{array}
 \right\}
 \cup \TRSmarchioriconstants
\]
As in Example~\ref{ex:marchiori}, we have the
 derivations $\mathsf{h}(\mathsf{f}(\mathsf{a}),\mathsf{f}(\mathsf{b}))$
 $\to^*_{\Uopt(\TRSmarchiorill)}$ $\mathsf{A}$ and
 $\mathsf{h}(\mathsf{f}(\mathsf{a}),\mathsf{f}(\mathsf{b}))$ 
 $\to^*_{\U(\TRSmarchiorill)}$ $\mathsf{A}$, but
 $\mathsf{h}(\mathsf{f}(\mathsf{a}),\mathsf{f}(\mathsf{b}))$ cannot be
 reduced by $\TRSmarchiorill$ to $\mathsf{A}$. 
Therefore, neither $\Uopt$ nor $\U$ is sound for $\TRSmarchiorill$.
\end{exa}

The LL property of normal CTRSs is equivalent to the $\Uopt$-LL property
since the right-hand sides $n_i$ of conditions $s_i \tto n_i$ are
ground.
In contrast, the LL property of DCTRSs is not equivalent to the ultra-LL
property in general (see $\Uopt(\TRSmarchiorill)$ and
$\U(\TRSmarchiorill)$ in Example~\ref{ex:marchiorill}). 
Moreover, the LL property of the unraveled TRSs plays an important role
in the existing proof of soundness.
Thus, the ultra-LL property seems a soundness condition for $\Uopt$
(and also for $\U$). 

The soundness result of this subsection is a consequence of
the following key lemma:
given a derivation $s$ $\to^*_{\Uopt(R)}$ $t\sigma$ with $s,t$ $\in$
$\cT(\cF,\cV)$, the lemma guarantees the existence of an intermediate
term $t\theta$ $\in$ $\cT(\cF,\cV)$ such that $s$ $\to^*_R$ $t\theta$
$\to^*_{\Uopt(R)}$ $t\sigma$ and, moreover, $t\theta$ $=$ $t\sigma$
whenever $t\sigma$ $\in$ $\cT(\cF,\cV)$. 
\begin{lem}
\label{lem:LL-soundness}
Let $R$ be a\/ $\Uopt$-LL 3-eDCTRS over a signature $\cF$, $s$ be a term in
 $\cT(\cF,\cV)$, $t$ be a linear term in $\cT(\cF,\cV)$, and
 $\sigma$ be a substitution in $\Subst(\cF_{\Uopt(R)},\cV)$. 
Suppose that $R$ is non-LV or non-RV.
 If $s$ $\pto^n_{\Uopt(R)}$ $t\sigma$ for some $n$ $\geq$
 $0$, then there exists a substitution $\theta$ in $\Subst(\cF,\cV)$
 such that 
 \begin{enumerate}[$\bullet$]
  \item $s$ $\to^*_R$ $t\theta$ $\pto^{n'}_{\geq \Pos_\cV(t),\Uopt(R)}$
	$t\sigma$ for some $n'$ $\leq$ $n$, and  
  \item if $t\sigma$ $\in$ $\cT(\cF,\cV)$, then $t\theta$ $=$ $t\sigma$.
 \end{enumerate}
\end{lem}
\proof
The proof can be seen in Appendix~\ref{subsec:proof:lem:LL-soundness}.
\qed

As a consequence of Lemma~\ref{lem:LL-soundness}, we show the main
theorem of this subsection. 
\begin{thm}
\label{th:LL-soundness}
$\Uopt$ is sound for\/ $\Uopt$-LL 3-eDCTRSs that are non-LV or
 non-RV. 
\end{thm}
\proof
Let $R$\/ be a 3-eDCTRS over a signature $\cF$, that is non-LV or
non-RV. 
Suppose that $s$ $\to^*_{\Uopt(R)}$ $t$ and $s,t$ $\in$
 $\cT(\cF,\cV)$.
 Since a single step of $\to_{\Uopt(R)}$ can be considered a single
 step of the parallel reduction, we have the derivation $s$
 $\pto^*_{\Uopt(R)}$ $t$.
 Let $x$\/ be a variable and $\sigma$\/ be a substitution such that
 $x\sigma$ $=$ $t$.
 Then, it follows from Lemma~\ref{lem:LL-soundness} that $s$ $\to^*_R$
 $x\sigma$ $=$~$t$.
\qed
\begin{exa}
 \label{ex:qsort}
Consider the following $\Uopt$-LL and non-LV DCTRS to define 
a splitting function $\mathsf{split}$ for lists of non-negative integers
 encoded as
 $\mathsf{0},\mathsf{s}(\mathsf{0}),
 \ldots$,
 \eg, $\mathsf{split}(3,[2,5,1,4,3])$ $=$ $([2,1],[5,4,3])$:
\[
 \begin{array}{@{}l@{}}
  \TRSqsort = \\
 \left\{
 \begin{array}{@{}r@{\>}c@{\>}l@{}}
  \mathsf{split}(x,\mathsf{nil}) & \to &
   \mathsf{tp}_2(\mathsf{nil},\mathsf{nil}) \\
  \mathsf{split}(x,\mathsf{cons}(y,ys)) & \to &
   \mathsf{tp}_2(zs_1,\mathsf{cons}(y,zs_2)) 
   \!\Leftarrow \mathsf{split}(x,ys) \!\tto\! \mathsf{tp}_2(zs_1,zs_2);
    ~\mathsf{le}(x,y) \!\tto\! \mathsf{true} \\
  \mathsf{split}(x,\mathsf{cons}(y,ys)) & \to &
   \mathsf{tp}_2(\mathsf{cons}(y,zs_1),zs_2) 
   \!\Leftarrow \mathsf{split}(x,ys) \!\tto\! \mathsf{tp}_2(zs_1,zs_2);
    ~\mathsf{le}(x,y) \!\tto\! \mathsf{false} \\
  \mathsf{le}(\mathsf{0},y) & \to & \mathsf{true} \\
  \mathsf{le}(\mathsf{s}(x),\mathsf{0}) & \to & \mathsf{false} \\
  \mathsf{le}(\mathsf{s}(x),\mathsf{s}(y)) & \to & \mathsf{le}(x,y) \\
 \end{array} 
 \right\} \\
 \end{array}
\]
$\TRSqsort$ is unraveled by $\Uopt$ and $\U$ into the following TRSs: 
\[
 \Uopt(\TRSqsort) =
 \left\{
 \begin{array}{r@{\>}c@{\>}l}
  & \vdots \\  
  \mathsf{split}(x,\mathsf{cons}(y,ys)) & \to &
   \Usplita(\mathsf{split}(x,ys),x,y) \\
  \Usplita(\mathsf{tp}_2(zs_1,zs_2),x,y) & \to & 
   \Usplitb(\mathsf{le}(x,y),y,zs_1,zs_2) \\
  \Usplitb(\mathsf{true},y,zs_1,zs_2) & \to &
   \mathsf{tp}_2(zs_1,\mathsf{cons}(y,zs_2)) \\ 
  \mathsf{split}(x,\mathsf{cons}(y,ys)) & \to &
   \Usplitc(\mathsf{split}(x,ys),x,y) \\
  \Usplitc(\mathsf{tp}_2(zs_1,zs_2),x,y) & \to & 
   \Usplitd(\mathsf{le}(x,y),y,zs_1,zs_2) \\
  \Usplitd(\mathsf{false},y,zs_1,zs_2) & \to & 
   \mathsf{tp}_2(\mathsf{cons}(y,zs_1),zs_2) \\
  & \vdots \\ 
 \end{array}
 \right\}
\]
We recognize from Theorem~\ref{th:LL-soundness} that $\Uopt$ is
 sound for $\TRSqsort$. 
\end{exa}

Due to the technical proof of Lemma~\ref{lem:LL-soundness}, we assumed
that eDCTRSs are non-LV or non-RV. 
It is not known yet whether this assumption can be relaxed (removed) or not.
However, this assumption is not so restrictive since
every DCTRS is non-LV.
Theorem~\ref{th:LL-soundness} is not a direct consequence of the
result in~\cite{Mar97} on soundness for $\Uopt$-LL 3-DCTRSs
since U symbols introduced by $\Uopt$ have less arguments than those
introduced by the unraveling in~\cite{Mar97}. 

\subsection{Observing Unsoundness of Marchiori's Counterexample to Soundness}
\label{subsec:observation}

In the previous subsection, we conjectured and proved that the ultra-LL
property is a soundness condition of $\Uopt$ since the property is
already known to be a soundness condition of Marchiori's unraveling for
normal CTRSs.  
To find other soundness conditions, in this subsection, we take a close
look at the derivation
$\mathsf{h}(\mathsf{f}(\mathsf{a}),\mathsf{f}(\mathsf{b}))$
$\to^*_{\Uopt(\TRSmarchiori)}$ $\mathsf{A}$ in
Example~\ref{ex:marchiori}, observing the reason why $\Uopt$ is not
sound for $\TRSmarchiori$ in Example~\ref{ex:marchiori}.

Recall the derivation
$\mathsf{h}(\mathsf{f}(\mathsf{a}),\mathsf{f}(\mathsf{b}))$
$\to^*_{\Uopt(\TRSmarchiori)}$ $\mathsf{A}$ in
Example~\ref{ex:marchiori}:
\[
 \begin{array}{l@{\>}c@{\>}l}
  \mathsf{h}(\mathsf{f}(\mathsf{a}),\mathsf{f}(\mathsf{b})) 
   & \to^*_{\Uopt(\TRSmarchiori)} & 
   \mathsf{h}(\Umarchiori(\mathsf{c},\mathsf{d}),\Umarchiori(\mathsf{c},\mathsf{d})) 
   \to_{\Uopt(\TRSmarchiori)} 
    \mathsf{g}(\Umarchiori(\mathsf{c},\mathsf{d}),\Umarchiori(\mathsf{c},\mathsf{d}),\mathsf{f}(\mathsf{k}))
    \\ 
   & \to^*_{\Uopt(\TRSmarchiori)} & 
    \mathsf{g}(\mathsf{d},\Umarchiori(\mathsf{l},\mathsf{m}),\Umarchiori(\mathsf{l},\mathsf{m})) 
    \to_{\Uopt(\TRSmarchiori)} 
    \mathsf{A} \\
 \end{array}
\]
To succeed in this derivation, the following subderivations are necessary:
 \begin{enumerate}[$\bullet$]
  \item to apply the rule $\mathsf{g}(\mathsf{d},x,x) \to \mathsf{A}$, the subterm
	$\mathsf{f}(\mathsf{a})$ in the initial term is reduced to $\mathsf{d}$,
  \item to apply the rule $\mathsf{h}(x,x) \to \mathsf{g}(x,x,\mathsf{f}(\mathsf{k}))$,
	both the subterms $\mathsf{f}(\mathsf{a})$ and $\mathsf{f}(\mathsf{b})$ in the initial
	term are reduced to the same term, and
  \item to apply the rule $\mathsf{g}(\mathsf{d},x,x) \to \mathsf{A}$, both the
	subterm $\mathsf{f}(\mathsf{b})$ in the initial term and the term
	$\mathsf{f}(\mathsf{k})$ derived from the application of $\mathsf{h}(x,x)
	\to \mathsf{g}(x,x,\mathsf{f}(\mathsf{k}))$ are reduced to the same term.
 \end{enumerate}
As a consequence, all of the terms $\mathsf{f}(\mathsf{a})$,
$\mathsf{f}(\mathsf{b})$, and 
$\mathsf{f}(\mathsf{k})$ have to be reduced to the same term $\mathsf{d}$.
However, this is impossible on the reduction of $\TRSmarchiori$.
Nevertheless, in the above derivation, $\mathsf{h}(x,x) \to
\mathsf{g}(x,x,\mathsf{f}(\mathsf{k}))$ is applied after reducing $\mathsf{f}(\mathsf{a})$ and
$\mathsf{f}(\mathsf{b})$ to $\Umarchiori(\mathsf{c},\mathsf{d})$:
the $\Umarchiori(\mathsf{c},\mathsf{d})$, that derives from
$\mathsf{f}(\mathsf{a})$, is reduced to $\mathsf{d}$, and the other
$\Umarchiori(\mathsf{c},\mathsf{d})$, that derives from
$\mathsf{f}(\mathsf{b})$, is reduced to 
$\Umarchiori(\mathsf{l},\mathsf{m})$ in order to be the same as
$\mathsf{f}(\mathsf{k})$. 
Finally, $\mathsf{g}(\mathsf{d},x,x) \to \mathsf{A}$ is applied.
These undesired subderivations must be caused by the non-RL
rule $\mathsf{h}(x,x) \to \mathsf{g}(x,x,\mathsf{f}(\mathsf{k}))$ and the erasing rule
$\mathsf{g}(\mathsf{d},x,x) \to \mathsf{A}$ in $\Uopt(\TRSmarchiori)$.
This stems from the following aspect:
\begin{enumerate}[$\bullet$]
 \item the application of $\mathsf{h}(x,x) \to \mathsf{g}(x,x,\mathsf{f}(\mathsf{k}))$ to
       $\mathsf{h}(\Umarchiori(\mathsf{c},\mathsf{d}),\Umarchiori(\mathsf{c},\mathsf{d}))$
       keeps two occurrences of $\Umarchiori(\mathsf{c},\mathsf{d})$
       that are intermediate states of evaluating
       $\mathsf{f}(\mathsf{a})$ 
       and $\mathsf{f}(\mathsf{b})$, respectively, and each of occurrence has a
       capability to be reduced to a different term later although they
       should be the same, and 
 \item $\mathsf{g}(\mathsf{d},x,x) \to \mathsf{A}$ erases the two occurrences of
       $\Umarchiori(\mathsf{l},\mathsf{m})$ as if they derive from the same term
       (in fact, they derive from the terms $\mathsf{f}(\mathsf{b})$ and
       $\mathsf{f}(\mathsf{k})$, respectively, although
       $\mathsf{f}(\mathsf{b})$ and $\mathsf{f}(\mathsf{k})$ should be
       reduced to different terms).  
\end{enumerate}
Viewed in this light, we conjecture that the RLNE property of the
unraveled TRSs is a sufficient condition for soundness of $\Uopt$.
Note that the above issue does not arise in the case of ultra-LL DCTRSs
since the LL property does not require equivalence at all between
subterms in redexes.

In the next subsection, we will prove the conjecture above,
 by reducing soundness for a $\Uopt$-RLNE DCTRS to
that for a DCTRS obtained by simply inverting.
The key feature is that if a DCTRS is $\Uopt$-NE, then,
\begin{enumerate}[$\bullet$]
 \item the unraveled TRS of the inverted one is equivalent
       to the inverted unraveled TRS of the DCTRS, and
 \item the inverted one is $\Uopt$-LL iff the DCTRS is $\Uopt$-RL.
\end{enumerate}
The converse of this approach is impossible since
 the first property above needs the $\Uopt$-NE property and not all
 $\Uopt$-LL DCTRSs have the $\Uopt$-NE property. 

\subsection{Soundness on Ultra-RLNE Property}
\label{subsec:RLNE-soundness}

In this subsection, we show that the optimized unraveling $\Uopt$ is
sound for $\Uopt$-RLNE DCTRSs.
To prove it, we reduce the soundness to that of $\Uopt$ for ultra-LL DCTRSs.
Moreover, we provide examples showing that neither $\Uopt$-RL nor
$\Uopt$-NE properties is sufficient on its own for soundness of $\Uopt$. 

We first define the operation to transform eDCTRSs into eDCTRSs that
define the inverse relation of the former eDCTRSs.
Note that the ``inverse'' here is slightly distinct from ``inverse'' in
the sense of {\em program inversion}. 
\begin{defi}
Let $\rho: l \to r \Leftarrow s_1 \tto t_1; \ldots; s_k \tto t_k$ be an
 (extended) conditional rewrite rule.
We define the operation $(\Blank)^{-1}$ as follows:
\[
 ( l \to r \Leftarrow s_1 \tto t_1; \ldots; s_k \tto t_k )^{-1}
 = r \to l \Leftarrow t_k \tto s_k ; \ldots; t_1 \tto s_1.
\]
This operation is extended to eDCTRSs as $(R)^{-1}$ $=$ $\{ (\rho)^{-1}
 \mid \rho \in R \}$. 
\end{defi}

For an eCTRS $R$, the inverse relation of $\to_R$ is equivalent to the
reduction of $(R)^{-1}$.
\begin{prop}
 \label{prop:inverse}
Let $R$ be an eCTRS.
Then, $\gets_R$ $=$ $\to_{(R)^{-1}}$.
\end{prop}
\proof[Proof (Sketch)]
It suffices to show that $\gets_{(n),R}$ $=$ $\to_{(n),(R)^{-1}}$ for
 all $n$ $\geq$ $0$.  
This can be proved by induction on $n$.
\qed

Regarding the operation $(\Blank)^{-1}$ and the $\Uopt$-NE property, the
unraveled TRSs are equivalent and we have dual relationships between the
$\Uopt$-LL and $\Uopt$-RL properties and between the non-LV and non-RV
properties. 
\begin{thm}
\label{th:RL-LL}
Let $R$ be an eDCTRS.
Then, all of the following hold:
\begin{enumerate}[\em(1)]
 \item $R$ is\/ $\Uopt$-NE iff\/ $(R)^{-1}$ is a 3-eDCTRS, 
 \item if $R$ is\/ $\Uopt$-NE, then,  
       \begin{enumerate}[\em a.]
	\item $\Uopt((R)^{-1})$ $=$ $(\Uopt(R))^{-1}$ up to the renaming
	      of U symbols (\/\ie, $\Usym^\rho_i$ $=$
	      $\Usym^{(\rho)^{-1}}_{k-i+1}$ for all\/ $1$ $\leq$ $i$
	      $\leq$ $k$),  
	\item $R$ is\/ $\Uopt$-LL iff\/ $(R)^{-1}$ is\/ $\Uopt$-RL, and 
	\item $R$ is\/ $\Uopt$-RL iff\/ $(R)^{-1}$ is\/ $\Uopt$-LL, 
       \end{enumerate}
 \item $R$ is non-LV iff\/ $(R)^{-1}$ is non-RV, and
 \item $R$ is non-RV iff\/ $(R)^{-1}$ is non-LV.
\end{enumerate}
\end{thm}
\proof
The proof can be seen in Appendix~\ref{subsec:proof:th:RL-LL}.
\qed
\noindent
Note that the claim (2)-a in Theorem~\ref{th:RL-LL} does not
hold for $\U$ in general.
\begin{exa}\label{ex:NE-insufficient0}
Consider the following $\Uopt$-NE 3-DCTRS $\TRSne$ and its unraveled TRSs:
\[
 \TRSne = 
 \left\{
\quad
 \mathsf{f}(x) \to \mathsf{c}(x,y)
   \Leftarrow \mathsf{g}(x) \tto y;~y \tto \mathsf{h}(x) 
 \quad\quad
  \mathsf{a} \to \mathsf{b} 
 \quad\quad
  \mathsf{g}(\mathsf{a}) \to \mathsf{h}(\mathsf{b}) 
 \quad
 \right\}
\]
\[
 \Uopt(\TRSne) = \U(\TRSne) =
 \left\{
 \begin{array}{r@{\>}c@{\>}l}
  \mathsf{f}(x) & \to & \Uneg(\mathsf{g}(x),x) \\
  \Uneg(y,x) & \to & \Uney(y,x,y) \\
  \Uney(\mathsf{h}(x),x,y) & \to & \mathsf{c}(x,y) \\
  & \vdots \\  
 \end{array}
 \right\}
\]
The following TRS is obtained from $\TRSne$ by applying $(\Blank)^{-1}$:
\[
 (\TRSne)^{-1} = 
 \left\{
 \quad
  \mathsf{c}(x,y) \to \mathsf{f}(x) 
   \Leftarrow \mathsf{h}(x) \tto y;
   ~ y \tto \mathsf{g}(x) 
 \quad\quad
  \mathsf{b} \to \mathsf{a} 
 \quad\quad
  \mathsf{h}(\mathsf{b}) \to \mathsf{g}(\mathsf{a}) 
 \quad
 \right\}
\]
The DCTRS $(\TRSne)^{-1}$ is unraveled by $\Uopt$ and $\U$ as follows:
\[
 \Uopt((\TRSne)^{-1}) =
 \left\{
 \begin{array}{@{\>}r@{\>}c@{\>}l@{\>}}
  \mathsf{c}(x,y) & \to & \Uney(\mathsf{h}(x),x,y) \\
   \Uney(y,x,y) & \to & \Uneg(y,x)  \\
  \Uneg(\mathsf{g}(x),x) & \to & \mathsf{f}(x) \\
  & \vdots \\  
 \end{array}
 \right\}
\]
\[
 \U((\TRSne)^{-1}) =
 \left\{
 \begin{array}{@{\>}r@{\>}c@{\>}l@{\>}}
  \mathsf{c}(x,y) & \to & \Uney(\mathsf{h}(x),x,y) \\
   \Uney(y,x,y) & \to & \Uneg(y,x,y)  \\
  \Uneg(\mathsf{g}(x),x,y) & \to & \mathsf{f}(x) \\
  & \vdots \\  
 \end{array}
 \right\}
\]
We have that $\Uopt((\TRSne)^{-1})$ $=$ $(\Uopt(\TRSne))^{-1}$, but
 $\U((\TRSne)^{-1})$ $\ne$ $(\U(\TRSne))^{-1}$.
\end{exa}

Finally, we show soundness of $\Uopt$ for a $\Uopt$-RLNE
eDCTRS $R$ by reducing it to soundness for the $\Uopt$-LL
eDCTRS $(R)^{-1}$.
\begin{thm}
 \label{th:RLNE-soundness}
$\Uopt$ is sound for\/ $\Uopt$-RLNE eDCTRSs that are non-LV or non-RV. 
\end{thm}
\proof
Let $R$ be a $\Uopt$-RLNE eDCTRS over a signature $\cF$.
Then, it follows from Theorem~\ref{th:RL-LL} that $(R)^{-1}$
 is a $\Uopt$-LL 3-eDCTRS which is non-RV or non-LV.
Thus, it follows from Theorem~\ref{th:LL-soundness} that $\Uopt$ is
sound for $(R)^{-1}$, \ie,
 $\to^*_{\Uopt((R)^{-1})}$ $\subseteq$ $\to^*_{(R)^{-1}}$ on terms in
 $\cT(\cF,\cV)$. 
 It follows from Theorem~\ref{th:RL-LL} that $\Uopt((R)^{-1})$ $=$
 $(\Uopt(R))^{-1}$, and hence $\to^*_{\Uopt((R)^{-1})}$ $=$
 $\to^*_{(\Uopt(R))^{-1}}$. 
 It follows from Proposition~\ref{prop:inverse} that 
 $\to^*_{(\Uopt(R))^{-1}}$ $=$ $\gets^*_{\Uopt(R)}$ and
 $\to^*_{(R)^{-1}}$ $=$ $\gets^*_R$. 
 Therefore, we have that $\to^*_{\Uopt(R)}$ $\subseteq$ $\to^*_R$ on
 terms in $\cT(\cF,\cV)$.
\qed

\begin{exa}
 \label{ex:quad}
 Consider the following TRS defining a function $\mathsf{quad}$ that
 computes the quadruple of input natural numbers:
 \[
 \TRSquad = 
 \left\{
 \begin{array}{r@{\>}c@{\>}l}
  \mathsf{quad}(x) & \to & \mathsf{twice}(\mathsf{twice}(x)) \\
  \mathsf{twice}(x) & \to & \mathsf{add}(x,x) \\
 \end{array}
 \right\}
 \cup \TRSmult
 \]
The inversion method in~\cite{Nishida04phd} 
 inverts this TRS to the following DCTRS $\TRSinvquad$:
 \[
 \TRSinvquad =
 \left\{
 \begin{array}{r@{\>}c@{\>}l}
  \mathsf{quad}^{-1}(y) & \to & \mathsf{tp}_1(x)
   \Leftarrow \mathsf{twice}^{-1}(y) \tto \mathsf{tp}_1(z);
   ~ \mathsf{twice}^{-1}(z) \tto \mathsf{tp}_1(x) \\
  \mathsf{twice}^{-1}(y) & \to & \mathsf{tp}_1(x)
   \Leftarrow \mathsf{add}^{-1}(y) \tto \mathsf{tp}_2(x,x) \\
 \end{array}
 \right\}
 \cup \TRSinvmult
 \]
 This DCTRS is $\Uopt$-RLNE, and thus, we recognize from
 Theorem~\ref{th:RLNE-soundness} that $\Uopt$ is sound for
 $\TRSinvquad$, while soundness of $\Uopt$ for the resulting EV-TRSs
 (\eg, $\TRSinvquad$) of  the inversion
 method~\cite{Nishida04phd} has already been shown
 (\cf,~\cite{NSS05rta,NSS05ieice}).  
 On the other hand, soundness of $\Uopt$ for DCTRSs obtained by removing
 the unary tuple symbol $\mathsf{tp}_1$ that seems meaningless:
 \[
 \TRSinvquad' =
 \left\{
 \begin{array}{r@{\>}c@{\>}l}
  \mathsf{quad}^{-1}(y) & \to & x 
   \Leftarrow \mathsf{twice}^{-1}(y) \tto z;
   ~ \mathsf{twice}^{-1}(z) \tto x \\
  \mathsf{twice}^{-1}(y) & \to & x 
   \Leftarrow \mathsf{add}^{-1}(y) \tto \mathsf{tp}_2(x,x) \\
 \end{array}
 \right\}
 \cup \TRSinvmult
 \]
 The soundness results in~\cite{Nishida04phd,NSS05rta,NSS05ieice} cannot
 guarantee that $\Uopt$ is sound for $\TRSinvquad'$. 
 However, since this DCTRS $\TRSinvquad'$ is also $\Uopt$-RLNE, we
 recognize from Theorem~\ref{th:RLNE-soundness} that $\Uopt$ is sound
 for $\TRSinvquad'$. 
\end{exa}

The open problem mentioned in~\cite{Nishida04phd} that $\Uopt$ is
sound for $\Uopt$-NE eDCTRSs does not hold in general.
This indicates that the ultra-NE property on its own is not a soundness
condition for either $\Uopt$ or $\U$. 
\begin{exa}\label{ex:NE-insufficient}
Consider the 3-DCTRS $\TRSne$ and the unraveled TRS $\Uopt(\TRSne)$ in
 Example~\ref{ex:NE-insufficient0} again.
$\TRSne$ is $\Uopt$-NE and $\U$-NE, but not $\Uopt$-RL or $\U$-RL, either.
We have the derivation $\mathsf{f}(\mathsf{a})$ $\to^*_{\Uopt(\TRSne)}$
 $\mathsf{c}(\mathsf{b},\mathsf{h}(\mathsf{b}))$, but
 $\mathsf{f}(\mathsf{a})$ cannot be reduced by $\TRSne$ to
 $\mathsf{c}(\mathsf{b},\mathsf{h}(\mathsf{b}))$.  
Therefore, $\Uopt$ is not sound for every $\Uopt$-NE DCTRS. 
By the same token, $\U$ is not sound for every $\U$-NE DCTRS since
 $\Uopt(\TRSne)$ $=$ $\U(\TRSne)$.
\end{exa}
\noindent
Moreover, the ultra-RL property on its own is not a soundness condition for
$\Uopt$. 
\begin{exa}\label{ex:feh}
Consider the following DCTRS $\TRSfeh$ and its unraveled TRSs:
\[
 \TRSfeh = 
 \left\{
  \begin{array}{r@{\>}c@{\>}l}
   \mathsf{f}(x) & \to & \mathsf{e} \Leftarrow \mathsf{d} \tto \mathsf{l}\\ 
   \mathsf{h}(x,x) & \to & \mathsf{A} \\
  \end{array}
 \right\}
\]
\[
 \Uopt(\TRSfeh) = 
 \left\{
  \begin{array}{r@{\>}c@{\>}l}
   \mathsf{f}(x) & \to & \Ufeh(\mathsf{d}) \\
   \Ufeh(\mathsf{l}) & \to & \mathsf{e} \\
   & \vdots \\  
  \end{array}
 \right\}
\quad\quad
 \U(\TRSfeh) = 
 \left\{
  \begin{array}{r@{\>}c@{\>}l}
   \mathsf{f}(x) & \to & \Ufeh(\mathsf{d},x) \\
   \Ufeh(\mathsf{l},x) & \to & \mathsf{e} \\
   & \vdots \\
  \end{array}
 \right\}
\]
The DCTRS $\TRSfeh$ is $\Uopt$-RL, but not $\Uopt$-NE.
Although we have the derivation
 $\mathsf{h}(\mathsf{f}(\mathsf{a}),\mathsf{f}(\mathsf{b}))$
 $\to^2_{\Uopt(\TRSfeh)}$
 $\mathsf{h}(\Ufeh(\mathsf{d}),\Ufeh(\mathsf{d}))$
 $\to_{\Uopt(\TRSfeh)}$ $\mathsf{A}$, the term 
 $\mathsf{h}(\mathsf{f}(\mathsf{a}),\mathsf{f}(\mathsf{b}))$
 cannot be reduced by $\TRSfeh$ to $\mathsf{A}$.
Therefore, $\Uopt$ is not sound for $\TRSfeh$ while $\U$ is sound for
 $\TRSfeh$.  
\end{exa}

It is possible to prove Theorem~\ref{th:RLNE-soundness}
directly~\cite{NSS04ss}, by using the feature that every reduction
sequence of RL TRSs can be transformed to a {\em basic}\/ reduction
sequence~\cite{MH94}. 
As stated at the end of Subsection~\ref{subsec:observation}, however,
Theorem~\ref{th:LL-soundness} cannot be proved by using  
Theorem~\ref{th:RLNE-soundness} since $\Uopt((R)^{-1})$ $=$
$(\Uopt(R))^{-1}$ does not hold for every $\Uopt$-LL DCTRS $R$ (see
$\Uopt(\TRSfeh)$ in Example~\ref{ex:feh}).  

In the proof of Theorem~\ref{th:LL-soundness} (and also the direct proof of
Theorem~\ref{th:RLNE-soundness}), linearity plays a very important role
and finding other soundness conditions by means of a similar proof
scheme is quite difficult. 

Finally, we revisit the resulting system of the program inversion
mentioned in Section~\ref{sec:intro}. 
\begin{exa}\label{ex:inversion}
 Consider the EV-TRS $\Uopt(\TRSinvmult)$ in Example~\ref{ex:U}
 again. 
 The original TRS $\TRSmult$ is left-linear, and thus,
 $\TRSinvmult$ is $\Uopt$-RLNE~\cite{Nishida04phd,NSS05ieice}.
 Theorem~\ref{th:RLNE-soundness} guarantees that $\Uopt(\TRSinvmult)$ is
 an {\em inverse}\/ system of $\TRSmult$. 
\end{exa}

\subsection{Soundness of Unraveled TRSs with Extra Variables}
\label{subsec:soundness_of_LL-evb}

As we stated in Section~\ref{sec:intro}, the optimized unraveling
$\Uopt$ is used in the program inversion method proposed
in~\cite{Nishida04phd,NSS05rta,NSS05ieice} and DCTRSs obtained by the
inversion method are of Type~4 (not of Type~3) in general.  
For this reason, in this subsection, we extend
Theorem~\ref{th:LL-soundness} to 4-eDCTRSs.
More precisely, we show that $\Uopt$ is sound for $\Uopt$-LL DCTRSs if
reduction sequences of the unraveled TRSs are restricted to {\em
EV-safe}\/ reduction sequences (see Definition~\ref{def:EV-safe}). 
Roughly speaking, in an EV-safe reduction sequence, any redex introduced
via extra variables at the application of rewrite rules is never reduced
anywhere. 
In practical cases (\eg, {\em inverse}\/
TRSs~\cite{Nishida04phd,NSS05rta,NSS05ieice,NSS03entcs}), extra
variables are instantiated with constructor terms. 
However, at the application of rewrite rules, extra variables in the
unraveled eTRSs may introduce undesired terms, \eg, terms rooted by U
symbols that are not reachable from terms over the original signature. 
As a consequence, $\Uopt$ is not always sound w.r.t.\ non-EV-safe
reduction sequences of the unraveled eTRSs (see Example~\ref{ex:efh}).  

We first define the notion of {\em EV-safe}\/ reduction sequences of
eTRSs~\cite{NSS03entcs,Nishida04phd,NSS04ss}.
This notion can be formalized by extending the notion of {\em basic}\/
reduction sequences~\cite{Hul80,MH94}.
\begin{defi}[EV-safe reduction~\cite{NSS03entcs}]
\label{def:EV-safe}
Let $R$\/ be an eTRS and $\rho_i: l_i \to r_i$ $\in$ $R$\/ for all $i$
 $\geq$ $1$.
Let $t_0$ $\to_{p_1,\rho_1}$ $t_2$ $\to_{p_2,\rho_2} \cdots$
be a reduction sequence of $R$, and $B_0$
$\subseteq$ $\Pos_\cF(t_0)$ such that $B_0$ is prefix closed (\ie, if
$p$ $<$ $q$ and $q$ $\in$ $B_0$, then $p$ $\in$ $B_0$). 
We define the sets $B_1,B_2,\ldots$ of positions from the sequence and $B_0$
inductively as
\[
 \begin{array}{l@{\>}l}
 B_i = & (B_{i-1} \setminus \{ q \in B_{i-1} \mid q \geq p_i \}) 
  \cup \{ p_iq \mid q \in \Pos_\cF(r_i) \} \\
  & \cup \{ p_ip'q \mid p_ipq \in B_{i-1}, ~ p \in \Pos_\cV(l_i), ~
   l_i|_p = r_i|_{p'} \} \\
 \end{array}
\]
for all $i$ $\geq$ $1$.
Note that $B_1,B_2,\ldots$ are also prefix closed.
For all $i$ $\geq$ $0$, positions in $B_i$ are referred as
{\em basic positions of $t_i$ w.r.t.\ extra variables}.
The reduction sequence above is said to be {\em based on $B_0$ w.r.t.\
extra variables}\/ if $p_i$ $\in$ $B_{i-1}$ for all $i$ $\geq$ $1$.
If the sequence is finite with length $n$, then we denote it by $B_0: t_0$
 $\evbto^*_R$ $B_n: t_n$ or $B_0: t_0$ $\evbto^*_R$ $t_n$. 
In particular, the reduction sequence is called {\em safe w.r.t.\ extra
variables}\/ (EV-safe) if $B_0$ $=$ $\Pos_\cF(t_0)$.
If the EV-safe sequence is finite with length $n$, then we denote it by
 $t_0$ $\evbto^*_R$ $t_n$.
\end{defi}
\noindent
Note that EV-safeness is different from {\em
basicness}~\cite{Hul80,MH94} in the sense that all the basic positions
are propagated at the application of rewrite rules, but none of the positions
for extra variables are added to basic positions. 
A typical instance of EV-safe reduction sequences is a reduction sequence
obtained by substituting a normal form for each extra variable when
applying rewrite rules.

To specify a set of terms that extra variables are possibly
instantiated at the rule application, 
we introduce the notion of {\em EV-instantiation on sets of terms}.
Let $R$\/ be an eTRS and $T$ be a set of terms.
A derivation $t_0$ $\to_{p_1,\rho_1}$ $t_1$ $\to_{p_2,\rho_2}
\cdots$ of $R$\/ is called {\em EV-instantiated on $T$}\/ if any extra
variable of $\rho_i: l_i \to r_i$ is instantiated by a term in $T$, 
\ie, $t_i|_{p_iq}$ $\in$ $T$ for any $q$ $\in$ $\Pos_\cV(r_i)$ such that
$r_i|_q$ $\in$ $\EVar(\rho_i)$.
By the same token, the notion of the EV-instantiation property is
defined for the parallel reduction of eTRSs. 
For any of the unraveled eTRSs, their EV-safe reduction sequences have
the following property related to EV-instantiation on the set of terms
over the original signature. 
\begin{lem}
\label{lem:EV-instantiated}
Let $R$ be a\/ $\Uopt$-LL eDCTRS over a signature $\cF$, and $s,t$ be terms
 in $\cT(\cF,\cV)$.
If $s$ $\evbto^*_{\Uopt(R)}$ $t$, then there exists a derivation $s$
 $\evbto^*_{\Uopt(R)}$ $t$ that is EV-instantiated on $\cT(\cF,\cV)$.
\end{lem}
\proof
The proof can be seen in
 Appendix~\ref{subsec:proof:lem:EV-instantiated}.
\qed

Lemma~\ref{lem:LL-soundness}, the key lemma for the case of $\Uopt$-LL
3-DCTRSs, is adapted to 4-eDCTRSs as follows. 
\begin{lem}
 \label{lem:LL-soundness-evb}
Let $R$ be a\/ $\Uopt$-LL eDCTRS over a signature $\cF$, $s$ be a term in
 $\cT(\cF,\cV)$, $t$ be a linear term in $\cT(\cF_{\Uopt(R)},\cV)$, and
 $\sigma$ be a substitution in $\Subst(\cF_{\Uopt(R)},\cV)$. 
Suppose that $R$ is non-LV or non-RV.
 If $s$ $\evbpto^n_{\Uopt(R)}$ $t\sigma$ for some $n$ $\geq$
 $0$ and the derivation is EV-instantiated on $\cT(\cF,\cV)$, then there
 exists a substitution $\theta$ in $\Subst(\cF,\cV)$ such that
 \begin{enumerate}[$\bullet$]
  \item $s$ $\to^*_R$ $t\theta$ $\evbpto^{n'}_{\geq \Pos_\cV(t),\Uopt(R)}$
	$t\sigma$ for some $n'$ $\leq$ $n$, 
  \item the derivation $t\theta$ $\evbpto^{n'}_{\geq \Pos_\cV(t),\Uopt(R)}$ 
	$t\sigma$ is EV-instantiated on $\cT(\cF,\cV)$, and
  \item if $t\sigma$ $\in$ $\cT(\cF,\cV)$, then $t\theta$ $=$
	$t\sigma$. 
 \end{enumerate}
\end{lem}
\proof
This lemma can be proved similarly to Lemma~\ref{lem:LL-soundness}.
\qed

As a consequence of Lemma~\ref{lem:LL-soundness-evb}, we extend
Theorem~\ref{th:LL-soundness} to 4-eDCTRSs. 
\begin{thm}
\label{th:LL-soundness-evb}
$\Uopt$ is sound for a\/ $\Uopt$-LL eDCTRSs $R$ over a signature $\cF$
 w.r.t.\ $\evbto_{\Uopt(R)}$
 if  $R$ is non-LV or non-RV.
\end{thm}
\proof
Suppose that $s$ $\evbto^*_{\Uopt(R)}$ $t$ and $s,t$ $\in$
 $\cT(\cF,\cV)$.
Then, it follows from Lemma~\ref{lem:EV-instantiated} that there is a
 derivation $s$ $\evbto^*_{\Uopt(R)}$ $t$ that is EV-instantiated on
 $\cT(\cF,\cV)$. 
 Since a single step of $\evbto_{\Uopt(R)}$ can be considered a
 single step of the parallel reduction, we have the derivation $s$
 $\evbpto^*_{\Uopt(R)}$ $t$ that is EV-instantiated on $\cT(\cF,\cV)$.  
 Let $x$\/ be a variable and $\sigma$ be a substitution such that
 $x\sigma$ $=$ $t$.
 Then, it follows from Lemma~\ref{lem:LL-soundness-evb} that $s$ $\to^*_R$
 $x\sigma$ $=$ $t$.
\qed
\noindent
Note that Lemma~\ref{lem:LL-soundness-evb} and
Theorem~\ref{th:LL-soundness-evb} are strict extensions of
Lemma~\ref{lem:LL-soundness} and Theorem~\ref{th:LL-soundness},
respectively. 

Finally, we show a counterexample against Theorem~\ref{th:LL-soundness}
without the EV-safe property. 
\begin{exa}\label{ex:efh}
Consider the DCTRS $\TRSfeh$ and its unraveled TRSs in
 Example~\ref{ex:feh} again.
Their inverted systems are as follows:
\[
 (\TRSfeh)^{-1} = 
 \left\{
  \begin{array}{r@{\>}c@{\>}l}
   \mathsf{e} & \to &  \mathsf{f}(x) \Leftarrow \mathsf{l} \tto \mathsf{d} \\
   \mathsf{A} & \to & \mathsf{h}(x,x) \\
  \end{array}
 \right\}
\]
\[
 \Uopt((\TRSfeh)^{-1}) = \U((\TRSfeh)^{-1}) = (\Uopt(\TRSfeh))^{-1} = 
 \left\{
  \begin{array}{r@{\>}c@{\>}l}
   \mathsf{e} & \to & \Ufeh(\mathsf{l}) \\
   \Ufeh(\mathsf{d}) & \to & \mathsf{f}(x) \\
   & \vdots \\  
  \end{array}
 \right\}
\]
We have the derivation $\mathsf{A}$ $\to_{\Uopt((\TRSfeh)^{-1})}$
 $\mathsf{h}(\Ufeh(\mathsf{d}),\Ufeh(\mathsf{d}))$
 $\to^*_{\Uopt((\TRSfeh)^{-1})}$
 $\mathsf{h}(\mathsf{f}(\mathsf{a}),\mathsf{f}(\mathsf{b}))$ that is not
 EV-safe:  
 the term $\Ufeh(\mathsf{d})$ introduced by instantiating the extra
 variable $x$\/ in the applied rule $\mathsf{A} \to \mathsf{h}(x,x)$ is
 reduced. 
 However, $\mathsf{A}$ cannot be reduced by $(\TRSfeh)^{-1}$ to
 $\mathsf{h}(\mathsf{f}(\mathsf{a}),\mathsf{f}(\mathsf{b}))$.  
 Therefore, $\Uopt$ is not sound for $(\TRSfeh)^{-1}$.
 Note that $\U$ is not sound for $(\TRSfeh)^{-1}$, either.
\end{exa}

Note that if an unraveling $U$ is sound for an eCTRS $R$, then $U$ is
sound for $R$ w.r.t.\ $\evbto_{U(R)}$.
For this reason, we need not discuss soundness of $U$ for $R$ w.r.t.\
$\evbto_{U(R)}$ when soundness of $U$ for $R$ w.r.t.\ $\to_{U(R)}$
has already been shown.  

\section{Soundness of Other Unravelings}  
\label{sec:soundness_of_unravelings}

In this section, we show that soundness of $\Uopt$ 
implies that of $\U$, 
and then we revisit soundness of the unravelings for join and normal
CTRSs. 

We first recall the notion of {\em tree homomorphisms}. 
Let $\cF$\/ and $\cG$\/ be signatures and $\phi_{\cF}$ be a mapping
which, for an $n$-ary function symbol $f$ $\in$ $\cF$, associates a term
in $\cT(\cG,\{x_1,\ldots,x_n\})$ where $x_1,\ldots,x_n$ $\in$ $\cV$.  
The {\em tree homomorphism}\/ $\phi: \cT(\cF,\cV) \to \cT(\cG,\cV)$
determined by $\phi_{\cF}$ is defined as follows~\cite{Tha73,tata2007}:
\begin{enumerate}[$\bullet$]
 \item $\phi(x)$ $=$ $x$\/ for $x$ $\in$ $\cV$, and 
 \item $\phi(f(t_1,\ldots,t_n))$ $=$ $\phi_{\cF}(f)\{ x_i \mapsto \phi(t_i) \mid
       1 \leq i \leq n \}$ for an $n$-ary function symbol $f$ $\in$
       $\cF$. 
\end{enumerate}
When $\phi_{\cF}(f)$ is not specified explicitly for an $n$-ary function
symbol $f$, we let $\phi_{\cF}(f)$ $=$ $f(x_1,\ldots,x_n)$ with assuming
that $f$ $\in$ $\cG$. 
To declare $\phi_{\cF}$ intelligibly, we may use the notation
``\,$\phi_{\cF}(f(x_1,\ldots,x_n)) = t$\,'' instead of  
``\,$\phi_{\cF}(f) = t$\,'', 
\eg, $\phi_{\cF}(\mathsf{f}(x,y,z))$ $=$ $\mathsf{g}(y,\mathsf{g}(x,z))$.
The tree homomorphism $\phi$\/ is called {\em linear}\/ if $\phi_{\cF}(f)$
is linear for any function symbol $f$ $\in$ $\cF$, and called
{\em non-erasing}\/ if $\Var(\phi(f))$ $=$ $\{x_1,\ldots,x_n\}$ for any
$n$-ary function symbol $f$ $\in$ $\cF$. 
The tree homomorphism $\phi$\/ is extended to eCTRSs as follows:
$\phi(R)$ $=$ $\{ \phi(l) \to \phi(r) \Leftarrow \phi(s_1) = \phi(t_1);
 \ldots; \phi(s_k) = \phi(t_k) \mid l \to r \Leftarrow
 \Condition{s_1}{t_1}; \ldots; \Condition{s_k}{t_k} \in R \}$.
We extend it to a set of term pairs $T$ (\eg, a binary relation) as
follows: 
$\phi(T)$ $=$ $\{ (\phi(s),\phi(t)) \mid (s,t) \in T \}$.
For a substitution $\sigma$ $\in$ $\Subst(\cF,\cV)$, $\sigma_\phi$
denotes the substitution $\{ x \mapsto \phi(x\sigma) \mid x \in
\Dom(\sigma) \}$.
Tree homomorphisms have the following properties.
\begin{lem}\label{lem:TH}
Let $\phi$ be a tree homomorphism.
\begin{enumerate}[\em(1)]
 \item Let $t$ be a term and $\sigma$ be a substitution in $\Subst(\cF,\cV)$.
       Then, $\phi(t\sigma)$ $=$ $(\phi(t))\sigma_\phi$.
 \item Let $t$ be a term and $C[~]$ be a one-hole context.
       Then, all of the following hold:
       \begin{enumerate}[$\bullet$]
	\item $\phi(C[t])$ $=$
	      $\phi(C)[\overbrace{\phi(t),\ldots,\phi(t)}^n]$,%
	      \footnote{ $\phi(C)$ has no hole (\ie, $n$ $=$ $0$) if 
	      $\phi$\/ removes the hole from $C[~]$.} 
	\item if $\phi$ is non-erasing, then $\phi(C)$ has at least one
	      hole, 
	\item if $\phi$ is linear, then $\phi(C)$ has at most one hole.
	      That is, if $\phi$ is linear, then, for any term $t$ and
	      any one-hole context $C[~]$, either $\phi(C[t])$ $=$
	      $\phi(C)$ or $\phi(C[t])$ $=$ $\phi(C)[\phi(t)]$.
       \end{enumerate}
 \item\label{lem:TH:phi-simulation}
      Let $R$ be an eCTRS and $s,t$ be terms in $\cT(\cF,\cV)$.
      If $s$ $\to^*_R$ $t$, then $\phi(s)$ $\to^*_{\phi(R)}$ $\phi(t)$.
      That is, $\phi(\to^*_R)$ $\subseteq$ $\to^*_{\phi(R)}$.
\end{enumerate}
\end{lem}
\noindent
The proof of Lemma~\ref{lem:TH} is omitted since it can be easily proved
by induction.

\subsection{Abstract Comparison Method for Soundness of Unravelings} 
\label{subsec:idea_of_comparison}

Before we discuss the relationship between soundness of two or more
unravelings, we present a sufficient condition of two unravelings under
which soundness of the first implies soundness of the other.  

To show soundness of an unraveling $U_2$ by means of a sound unraveling
$U_1$, it suffices to show that all the derivations of $U_2$ on terms
over the original signature are included in the derivations of $U_1$,
\ie, $\to^*_{U_2(R)}$ $\subseteq$ $\to^*_{U_1(R)}$ on terms in $\cT(\cF,\cV)$.
Suppose that $U_1$ is sound for a CTRS $R$, \ie, $\to^*_{U_1(R)}$ $\subseteq$
$\to^*_R$ on terms in $\cT(\cF,\cV)$.
Then, it follows from  $\to^*_{U_2(R)}$ $\subseteq$ $\to^*_{U_1(R)}$
that $\to^*_{U_2(R)}$ $\subseteq$ $\to^*_R$ on terms in $\cT(\cF,\cV)$.
Therefore, $U_2$ is sound for $R$.	      

To show that $\to^*_{U_2(R)}$ $\subseteq$ $\to^*_{U_1(R)}$ on terms in
$\cT(\cF,\cV)$, it suffices to show the existence of a tree homomorphism
$\phi$\/ for an extended signature of $\cF$ such that 
\begin{enumerate}[$\bullet$]
 \item $U_1(R)$ $=$ $\phi(U_2(R))$, and
 \item $\phi(t)$ $=$ $t$\/ for any term $t$ $\in$ $\cT(\cF,\cV)$.
\end{enumerate}
Moreover, since we consider soundness w.r.t.\ $\evbto$, we are
interested in a sufficient condition under which $\evbto^*_{U_2(R)}$
$\subseteq$ $\evbto^*_{U_1(R)}$. 
To show that $\evbto^*_{U_2(R)}$ $\subseteq$ $\evbto^*_{U_1(R)}$, it suffices
to additionally show that $\EVar(\phi(l)\to \phi(r))$ $\subseteq$
$\EVar(l\to r)$ for any rule $l\to r$ $\in$ $U_2(R)$. 

For a set $\cG$ $\subseteq$ $\cF$\/ of function symbols, a tree
homomorphism $\phi$\/ determined by a mapping $\phi_\cF$ is called {\em
$\cG$-identical}\/ if $\phi_\cF(f)$ $=$ $f(x_1,\ldots,x_n)$ for
any $n$-ary function symbol $f$ $\in$ $\cG$. 
Moreover, $\phi$\/ is called {\em EV-preserving for an eTRS $R$}\/ if
$\EVar(\phi(l)\to \phi(r))$ $=$ $\EVar(l\to r)$ for any rule
$l\to r$ $\in$ $R$.  
\begin{lem}
\label{lem:pi-evb}
Let $R$ be an eTRS, $s,t$ be terms, and $\phi$ be a tree homomorphism
 that is EV-preserving for $R$.
If $s$ $\evbto^*_R$ $t$, then $\phi(s)$ $\evbto^*_{\phi(R)}$ $\phi(t)$.
\end{lem}
\proof
We first define the mapping $\phi_u$ from a position of a term $u$\/
 to a set of positions of $\phi(u)$, and extend it to sets of positions
 of $u$\/:
\begin{enumerate}[$\bullet$]
 \item $\phi_x(\varepsilon)$ $=$ $\{\varepsilon\}$ for $x$ $\in$ $\cV$,
 \item $\phi_{f(u_1,\ldots,u_n)}(\varepsilon)$ $=$ $\{\varepsilon\}$ for
       an $n$-ary function symbol $f$, 
 \item $\phi_{f(u_1,\ldots,u_n)}(ip)$ $=$ $\{ qp' \mid q \in \Pos_\cV(\phi_{\cF}(f)), ~
       \phi_{\cF}(f)|_q = x_i, ~ p' \in \phi_{u_i}(p) \}$ for an
       $n$-ary function symbol $f$, where $1$ $\leq$ $i$ $\leq$ $n$ and
       $p$ $\in$ $\Pos(u_i)$, and 
\item $\phi_u(P)$ $=$ $\bigcup_{p \in P} \phi_u(p)$.
\end{enumerate}
For a position $p$\/ of $u$, we mean by $p'$ $\in$ $\phi_u(p)$ that the
 application of $\phi$\/ to $u$\/ maps $u|_p$ to $\phi(u)|_{p'}$.
 Note that positions in $\phi_u(p)$ are parallel since variable
 positions of $\phi_{\cF}(f)$ are parallel.
 We prove that if $B: s$ $\evbto^*_R$ $B': t$\/ and $\phi_s(B)$ $\subseteq$
 $B_1$ $\subseteq$ $\Pos(\phi(s))$, then $B_1: \phi(s)$
 $\evbto^*_{\phi(R)}$ $B_1': \phi(t)$ and $\phi_t(B')$ $\subseteq$ $B_1'$.
To prove this claim by induction on the length $n$\/ of the derivation $B:
 s$ $\evbto^*_R$ $B': t$, it suffices to show that if $B: s$
 $\evbto_{p,R}$ $B': t$, then $\phi_s(p)$ is defined and 
 $\phi_s(B): \phi(s)$ $\evbto_{q_1,\phi(R)} \cdots \evbto_{q_m,\phi(R)}$
 $\phi_t(B'): \phi(t)$ where $\phi_s(p)$ $=$ $\{q_1,\ldots,q_m\}$.
This follows from the assumption and the definitions of $\evbto$ and $\phi_t$.
\qed

Accordingly, to show soundness of $U_2$ by means of
soundness of $U_1$, we obtain the following useful lemma. 
\begin{lem}
 \label{lem:soundness_of_two-unravelings}
 Let $U_1$ and $U_2$ be unravelings, $R$ be an eCTRS over a signature
 $\cF$, and $\cG$ be an extended signature of $\cF$ such that $U_2(R)$
 is defined over $\cG$. 
 Let $\phi$ be an $\cF$-identical tree homomorphism determined by
 $\phi_\cG$ such that $U_1(R)$ $=$ $\phi(U_2(R))$. 
 Then, all of the following hold:
 \begin{enumerate}[\em(1)]
  \item $\to^*_{U_2(R)}$ $\subseteq$ $\to^*_{U_1(R)}$ on terms in
	$\cT(\cF,\cV)$,
  \item if $\phi$ is non-erasing, then $\phi$ is EV-preserving for any
	eTRS, and 
  \item if $\phi$ is EV-preserving for $U_2(R)$, then
	$\evbto^*_{U_2(R)}$ $\subseteq$ $\evbto^*_{U_1(R)}$ on terms in
	$\cT(\cF,\cV)$.
 \end{enumerate}
That is, all of the following holds:
 \begin{enumerate}[\em(1)]
  \setcounter{enumi}{3}
  \item if $U_1$ is sound for $R$, then so is $U_2$, 
  \item if $\phi$ is non-erasing and $U_1$ is sound for $R$ w.r.t.\
	$\evbto_{U_1(R)}$, then $U_2$ is sound for $R$ w.r.t.\
	$\evbto_{U_2(R)}$, and 
 \item if $\phi$ is EV-preserving for $U_2(R)$ and $U_1$ is sound for
       $R$ w.r.t.\ $\evbto_{U_1(R)}$, then $U_2$ is sound for $R$
       w.r.t.\ $\evbto_{U_2(R)}$.
 \end{enumerate}
\end{lem}
\proof
We first prove the first claim $\to^*_{U_2(R)}$ $\subseteq$
$\to^*_{U_1(R)}$ on terms in $\cT(\cF,\cV)$. 
It follows from Lemma~\ref{lem:TH} and the assumption $U_1(R)$ $=$
$\phi(U_2(R))$ that $\phi(\to^*_{U_2(R)})$ $\subseteq$ $\to^*_{U_1(R)}$.
Since $\phi$\/ is $\cF$-identical, we have that $\to^*_{U_2(R)}$
$\subseteq$ $\to^*_{U_1(R)}$ on terms in $\cT(\cF,\cV)$. 

Let $\phi$\/ be determined by a mapping $\phi_\cG$.
To prove the second claim, it suffices to show that 
$\Var(t)$ $=$ $\Var(\phi(t))$ for any term $t$.
We prove this claim by induction on the structure of $t$.
Since the case that $t$\/ is a variable is trivial, we only consider the
remaining case that $t$\/ is of the form $f(t_1,\ldots,t_n)$.
By the induction hypothesis, $\Var(t_i)$ $=$ $\Var(\phi(t_i))$ for all
$1$ $\leq$ $i$ $\leq$ $n$. 
It follows from the non-erasingness of $\phi$\/ that
$\Var(\phi(f(t_1,\ldots,t_n)))$ $=$ $\Var(\phi_\cG(f)\{ x_i \mapsto
\phi(t_i)\ \mid 1 \leq i \leq n \})$ $=$ $\bigcup_{i=1}^n
\Var(\phi(t_i))$, and hence $\Var(\phi(t))$ $=$ $\Var(t)$. 

The third claim follows from the first claim and
Lemma~\ref{lem:pi-evb}.
The remaining claims follow from the first, second, and third claims, and
 soundness of $U_1$.
\qed

Due to Lemma~\ref{lem:soundness_of_two-unravelings}, to show soundness
of $U_2$ by soundness of $U_1$, it suffices to show
the existence of an $\cF$-identical tree homomorphism $\phi$\/ satisfying
that $U_1(R)$ $=$ $\phi(U_2(R))$.
Moreover, for the case of soundness w.r.t.\ $\evbto$, it suffices to
additionally show that the tree homomorphism $\phi$\/ is non-erasing or
EV-preserving for $U_2(R)$.

\subsection{On Ohlebusch's Unraveling for DCTRSs}
\label{subsec:soundness_of_U}

As stated in Section~\ref{sec:unraveling}, the optimized unraveling
$\Uopt$ is a variant of the unraveling $\U$, in 
the sense that variables carried by U symbols are optimized.
For this reason, for a DCTRS $R$, it is easy to find a tree homomorphism
$\phi$\/ such that $\Uopt(R)$ $=$ $\phi(\U(R))$.
In the following, we assume that for every rule $\rho$ $\in$ $R$, the
 same U symbols $\Usym^\rho_1,\ldots,\Usym^\rho_k$ are introduced 
for $\Uopt(\rho)$ and $\U(\rho)$. 
\begin{lem}
 \label{lem:U-TH}
Let $R$ be an eDCTRS over a signature $\cF$.
 There exists an $\cF$-identical tree homomorphism $\phi$ such that
 $\phi(\U(R))$ $=$ $\Uopt(R)$ and $\phi$ is EV-preserving for $\U(R)$. 
\end{lem}
\proof
 Let $\phi$\/ be a tree homomorphism determined by $\phi_{\cF_{\U(R)}}$
 such that 
 \[
 \phi_{\cF_{\U(R)}}(\Usym^\rho_i(x_i,\overrightarrow{\AppearedVar_i}))
 = \Usym^\rho_i(x_i,\overrightarrow{\AppearedLaterUsedVar_i})
 \] 
 where $x_i$ is a fresh variable such that $x_i$ $\not\in$
 $\AppearedVar_i$. 
 Then, it is clear that $\phi(\U(\rho))$ $=$ $\Uopt(\rho)$, and hence
 $\phi(\U(R))$ $=$ $\Uopt(R)$.

 Next, we show that $\phi_{\cF_{\U(R)}}$ is EV-preserving for $\U(R)$. 
 For unconditional rules $l_0 \to r_0$ in $R$, it is clear that
 $\EVar(l_0 \to r_0)$ $=$ $\EVar(\phi(l_0) \to \phi(r_0))$, since
 $l_0,r_0$ $\in$ $\cT(\cF,\cV)$ and $\phi(t)$ $=$ $t$\/ for all 
 $t$ $\in$ $\cT(\cF,\cV)$.
 Thus, we only consider the case of conditional rules $\rho: l \to r
 \Leftarrow s_1 \tto t_1 ; \ldots; s_k \tto t_k$ $\in$ $R$. 
 Rules in $\Uopt(\rho)$ and $\U(\rho)$ that may contain extra variables
 are rules of $\Usym^\rho_k$, that is, 
 $\rho_k: \Usym^\rho_k(t_k,\overrightarrow{\AppearedLaterUsedVar_k}) \to
 r$ $\in$ $\Uopt(R)$, 
 $\rho_k': \Usym^\rho_k(t_k,\overrightarrow{\AppearedVar_k}) \to r$
 $\in$ $\U(R)$.  
 It follows from $\AppearedVar_k \cap \LaterUsedVar_k$ $\subseteq$
 $\AppearedVar_k$ and $\Var(r) \cap \AppearedLaterUsedVar_k$ $=$
 $\Var(r) \cap \AppearedVar_k$ that $\EVar(\rho_k)$ $=$ $\Var(r)
 \setminus (\Var(t_k) \cup (\Var(l,t_1,\ldots,t_{k-1}) \cap
 \Var(r,t_k)))$ $=$ $\Var(r) \setminus \Var(l,t_1,\ldots,t_k)$ $=$
 $\EVar(\rho_k')$.   
 Therefore, $\phi_{\cF_{\U(R)}}$ is EV-preserving for $\U(R)$. 
\qed

As a consequence, we conclude that soundness of $\Uopt$ 
implies soundness of $\U$. 
\begin{cor}
 \label{cor:U-soundness_from_Uopt}
Let $R$ be an eDCTRS over a signature $\cF$.
 $\U$ is sound for $R$ (w.r.t.\ $\evbto_{\U(R)}$) if\/ $\Uopt$ is
 sound for $R$ (w.r.t.\ $\evbto_{\Uopt(R)}$).
\end{cor}
\begin{exa}\label{ex:invquad-soundness}
Consider the DCTRS $\TRSinvquad'$ in Example~\ref{ex:quad} again. 
As stated in Example~\ref{ex:invquad-soundness}, $\Uopt$ is sound for
 $\TRSinvquad'$, and thus, we recognize from
 Corollary~\ref{cor:U-soundness_from_Uopt} that $\U$ is also sound for
 $\TRSinvquad'$. 
\end{exa}

The converse of Corollary~\ref{cor:U-soundness_from_Uopt} does not hold
in general since, for a DCTRS $R$\/ over a signature $\cF$,  
$\to^*_{\U(R)}$ $\ne$ $\to^*_{\Uopt(R)}$ on terms in $\cT(\cF,\cV)$ in
general
(see Example~\ref{ex:feh}).
The reason why the converse of Corollary~\ref{cor:U-soundness_from_Uopt}
does not hold must be that the U symbols introduced via the application of
$\U$ have more variables (\ie, information) than the corresponding U
symbols introduced by $\Uopt$.   
Thus, $\U$ is sufficient to produce TRSs that can be used instead of the
original DCTRSs.
Nonetheless, $\Uopt$ will be useful in investigating soundness
of $\U$ since the unraveled TRSs obtained by $\Uopt$
are simpler than those obtained by $\U$.

\subsection{On Unravelings for Join and Normal CTRSs}
\label{subsec:soundness_of_Uj-Un}

Join CTRSs can be converted into equivalent normal CTRSs that are special
cases of DCTRSs, and normal CTRSs are join CTRSs since the conditions $s_i
\tto n_i$ and $s_i \downarrow n_i$ are identical:
\[
 \mbox{join CTRS} = \mbox{normal CTRS} \subset \mbox{DCTRS}
\]
In this subsection, we show that the unraveling $\Ujl$ for join
CTRSs~\cite{Mar96} is sound for join CTRSs if the unraveling $\Unl$ for
normal CTRSs~\cite{Ohl02,GGS10} is sound for the corresponding normal
CTRSs.
Then, by using this result and the existing soundness condition of
$\Unl$~\cite{GGS10}, we show that $\Ujl$ is sound for LL join CTRSs.
We also show that $\Ujl$ is sound for join CTRSs that can be considered
normal CTRSs. 
Moreover, we show that $\Unl$ is sound for a normal CTRS if $\Ujl$ is
sound for the normal CTRS that is considered as a join CTRS.
Finally, we show that soundness of $\Unl$ implies soundness of $\U$. 
As far as we know, soundness of $\Ujl$ has never been discussed, 
whereas soundness of $\Unl$ has been investigated in some
papers~\cite{Mar96,GGS10}. 
For this reason, we show the soundness condition for $\Ujl$ and compare
soundness of $\Ujl$ with $\Unl$.

A CTRS $R$\/ is called {\em join}\/ if the symbol $\Condition{}{}$ 
in the conditions of rewrite rules is interpreted as {\em joinability}\/:
the {\em reduction relation}\/ of $R$\/ is defined as $\to_R$ $=$
$\bigcup_{n \geq 0} \to_{(n),R}$ where 
\begin{enumerate}[$\bullet$]
 \item $\to_{(0),R}$ $=$ $\emptyset$, and
 \item $\to_{(i+1),R}$ $=$ $\to_{(i),R}$ $\cup$ $\{
       (C[l\sigma]_p,C[l\sigma]_p) \mid \rho: l \to r \Leftarrow
       \Condition{s_1}{t_1}; \ldots; \Condition{s_k}{t_k} \in R, ~
       s_1\sigma \downarrow_{(i),R} 
       t_1\sigma, ~\ldots,~ s_k\sigma \downarrow_{(i),R} t_k\sigma \}$
       for $i$ $\geq$ $0$.
\end{enumerate}
From now on, rewrite rules $l \to r \Leftarrow \Condition{s_1}{t_1};
\ldots; \Condition{s_k}{t_k}$ of join CTRSs are written as $l \to r
\Leftarrow s_1 \downarrow 
t_1; \ldots; s_k \downarrow t_k$. 

We first recall the definition of the unravelings $\Ujl$ and $\Unl$ for
join and normal CTRSs, that are variants of unravelings proposed by
Marchiori~\cite{Mar96}. 
\begin{defi}[$\Ujl$, $\Unl$~\cite{Ohl02,GGS10}]
 \label{def:Uj-Un}
Let $R$\/ be a join or normal eCTRS over a signature $\cF$.
Introducing a U symbol $\Usym^\rho$, we transform $\rho: l \to r
 \Leftarrow \Condition{s_1}{t_1};\ldots; \Condition{s_k}{t_k}$ into sets 
 $\Ujl(\rho)$ and $\Unl(\rho)$ of two unconditional rules as follows:
 \begin{enumerate}[$\bullet$]
  \item $\Ujl(\rho)$ $=$ $\{
	l \to
	\Usym^\rho(s_1,t_1,\ldots,s_k,t_k,\overrightarrow{\Var(l)}),
	~
	\Usym^\rho(x_1,x_1,\ldots,x_k,x_k,\overrightarrow{\Var(l)}) \to
	r \}$
	if $R$\/ is join, and
  \item $\Unl(\rho)$ $=$ $\{
       l \to \Usym^\rho(s_1,\ldots,s_k,\overrightarrow{\Var(l)}),
       ~
       \Usym^\rho(t_1,\ldots,t_k,\overrightarrow{\Var(l)}) \to r \}$
	if $R$\/ is normal. 
\end{enumerate}
where $x_1,\ldots,x_k$ are different fresh variables.
Note that $\Ujl(l' \to r')$ $=$ $\Unl(l'\to r')$ $=$ $\{ l' \to r' \}$. 
$\Ujl$ and $\Unl$ are extended to join and normal CTRSs, respectively,
 \ie, $\Ujl(R)$ $=$ $\bigcup_{\rho \in R} \Ujl(\rho)$
 and  $\Unl(R)$ $=$ $\bigcup_{\rho \in R} \Unl(\rho)$. 
We define the extended signatures $\cF_{\Ujl(R)}$ and $\cF_{\Unl(R)}$ of
 $\cF$ as $\cF_{\Ujl(R)}$ $=$ $\cF_{\Unl(R)}$ $=$ $\cF \cup \{ \Usym^\rho
 \mid \rho \in R \}$. 
\end{defi}
\noindent
Note that $\Ujl$ and $\Unl$ are tidy unravelings for join and normal
CTRSs, respectively.%
\footnote{ Compared with sequential unravelings (\eg, $\U$ and
$\Uopt$), $\Ujl$ and $\Unl$ are called {\em simultaneous
unravelings}~\cite{GGS12}. }
The difference from the original definition 
in~\cite{Mar96} is the replacement of $\overrightarrow{\Var(r)}$ by
$\overrightarrow{\Var(l)}$. 
We denote the original unravelings for join and normal CTRSs by $\Uj$
and $\Un$, respectively.   
$\Uj$ and $\Un$ can be considered optimized variants of $\Ujl$ and
$\Unl$, respectively, as well as the optimized variant $\Uopt$ of $\U$.
The relationship between $\Uj$ and $\Ujl$ and between $\Un$ and $\Unl$
is similar to that between $\Uopt$ and $\U$, 
\ie, if $\Uj$ ($\Un$) is sound for a join (normal) CTRS $R$, 
then so is $\Ujl$ ($\Unl$) 
 (\cf, Corollary~\ref{cor:U-soundness_from_Uopt}).
Thus, in the following, we deal with $\Ujl$ and $\Unl$. 
\begin{exa}
\label{ex:oddeven}
 Consider the following join CTRS defining $\mathsf{odd}$ and
 $\mathsf{even}$ that, given a natural number
 $\mathsf{s}^n(\mathsf{0})$, return $\mathsf{true}$ and
 $\mathsf{false}$, respectively, if $n$\/ is odd, and return
 $\mathsf{false}$ and $\mathsf{true}$, respectively, otherwise:
 \[
 \TRSoddeven = 
 \left\{
 \begin{array}{r@{\>}c@{\>}l@{\quad\quad}r@{\>}c@{\>}l}
  \mathsf{odd}(\mathsf{0}) & \to & \mathsf{false} 
   & \mathsf{even}(\mathsf{0}) & \to & \mathsf{true} \\
  \mathsf{odd}(\mathsf{s}(x)) & \to & \mathsf{true}
   \Leftarrow \mathsf{even}(x) \downarrow \mathsf{true}  
   & \mathsf{even}(\mathsf{s}(x)) & \to & \mathsf{true} 
   \Leftarrow \mathsf{odd}(x) \downarrow \mathsf{true} \\
  \mathsf{odd}(\mathsf{s}(x)) & \to & \mathsf{false}
   \Leftarrow \mathsf{even}(x) \downarrow \mathsf{false}    
   & \mathsf{even}(\mathsf{s}(x)) & \to & \mathsf{false} 
   \Leftarrow \mathsf{odd}(x) \downarrow \mathsf{false} \\
 \end{array}
 \right\}
\]
This join CTRS is unraveled by $\Ujl$ into the following TRS:
\[
 \Ujl(\TRSoddeven) = \\
 \left\{
 \begin{array}{r@{\>}c@{\>}l@{\quad}r@{\>}c@{\>}l}
  & \vdots \\
  \mathsf{odd}(\mathsf{s}(x)) & \to &
   \Uodda(\mathsf{even}(x),\mathsf{true},x) 
   &
  \Uodda(y,y,x) & \to & \mathsf{true} \\
  \mathsf{odd}(\mathsf{s}(x)) & \to &
   \Uoddb(\mathsf{even}(x),\mathsf{false},x) 
   &
  \Uoddb(y,y,x) & \to & \mathsf{false} \\
  & \vdots \\
  \mathsf{even}(\mathsf{s}(x)) & \to &
   \Uevena(\mathsf{odd}(x),\mathsf{true},x) 
   &
  \Uevena(y,y,x) & \to & \mathsf{true} \\
  \mathsf{even}(\mathsf{s}(x)) & \to &
   \Uevenb(\mathsf{odd}(x),\mathsf{false},x) 
   & 
  \Uevenb(y,y,x) & \to & \mathsf{false} \\
 \end{array}
 \right\} 
\]
 When we consider $\TRSoddeven$ as a normal CTRS by replacing
 $\downarrow$ by $\tto$, the CTRS, denoted by $\TRSoddeven'$ below, is
 unraveled by $\Unl$ as follows: 
\[
 \Unl(\TRSoddeven') = \\
 \left\{
 \begin{array}{r@{\>}c@{\>}l@{\quad}r@{\>}c@{\>}l}
  & \vdots \\
  \mathsf{odd}(\mathsf{s}(x)) & \to & \Uodda(\mathsf{even}(x),x) 
   &
  \Uodda(\mathsf{true},x) & \to & \mathsf{true} \\
  \mathsf{odd}(\mathsf{s}(x)) & \to & \Uoddb(\mathsf{even}(x),x) 
   &
  \Uoddb(\mathsf{false},x) & \to & \mathsf{false} \\
  & \vdots \\
  \mathsf{even}(\mathsf{s}(x)) & \to & \Uevena(\mathsf{odd}(x),x) 
   &
  \Uevena(\mathsf{true},x) & \to & \mathsf{true} \\
  \mathsf{even}(\mathsf{s}(x)) & \to & \Uevenb(\mathsf{odd}(x),x) 
   &
  \Uevenb(\mathsf{false},x) & \to & \mathsf{false} \\
 \end{array}
 \right\} 
\]
\end{exa}

Next, we define a variant of a transformation from join CTRSs to normal
ones, that is proposed in~\cite{DO90} (\cf,~\cite{Ohl02}).
\begin{defi}
 \label{def:Tjn}
Let $R$\/ be a join CTRS over a signature $\cF$.
Introducing a fresh binary function symbol $\mathsf{eq}$ and a fresh
 constant $\top$, we define a transformation $\Tjn$ as follows: 
 \[
 \begin{array}{l}
  \Tjn(l \to r \Leftarrow s_1 \downarrow t_1; \ldots; s_k
 \downarrow t_k) \\
  \hspace{30mm}
   =  l \to r \Leftarrow \mathsf{eq}(s_1,t_1) \tto \mathsf{eq}(\top,\top); 
 \ldots; \mathsf{eq}(s_k,t_k) \tto \mathsf{eq}(\top,\top) \\
 \end{array}
 \]
and
 \[
  \Tjn(R) = \{ \mathsf{eq}(x,x) \to \mathsf{eq}(\top,\top) \} \cup \{ 
 \Tjn(\rho) \mid 
 \rho \in R \}
 \]
\end{defi}
\noindent
The added rule $\mathsf{eq}(x,x) \to \mathsf{eq}(\top,\top)$
results in non-termination, but non-termination does not affect the following
discussion.%
\footnote{ To avoid non-termination caused by the added rule
$\mathsf{eq}(x,x) \to \mathsf{eq}(\top,\top)$, we may introduce a unary
constructor $\mathsf{c}_1$ as follows:
 \[
 \begin{array}{l}
\Tjn'(l \to r \Leftarrow s_1 \downarrow t_1; \ldots; s_k
 \downarrow t_k) = \\
  \hspace{30mm}
 l \to r \Leftarrow \mathsf{eq}(\mathsf{c}_1(s_1),\mathsf{c}_1(t_1)) \tto
 \mathsf{eq}(\top,\top); 
 \ldots; \mathsf{eq}(\mathsf{c}_1(s_k),\mathsf{c}_1(t_k)) \tto
 \mathsf{eq}(\top,\top) \\
 \end{array}
 \]
and
 \[
  \Tjn'(R) = \{ \mathsf{eq}(\mathsf{c}_1(x),\mathsf{c}_1(x)) \to
 \mathsf{eq}(\top,\top) \} \cup \{ 
 \Tjn'(\rho) \mid 
 \rho \in R \}
 \]
This variant can substitute for $\Tjn$\/ in the following discussion.}
The difference from the original transformation~\cite{DO90} is the use
of $\mathsf{eq}(\top,\top)$ instead of and $\top$.
The reason of this difference is to make it simple to prove a theorem 
shown later (Theorem~\ref{th:UnTjn-simulation})|the original
transformation, denoted by $n$\/ in~\cite{Ohl02}, can substitute for
$\Tjn$\/ since $\to_{n(R)}$ $=$ $\to_{\Tjn(R)}$ on terms in
$\cT(\cF,\cV)$. 
It is clear that $\Tjn(R)$ is a normal CTRS over the signature $\cF \cup
\{ \mathsf{eq}, \top \}$, $\to^*_R$ $\subseteq$ $\to^*_{\Tjn(R)}$, and
especially $\to_R$ $=$ $\to_{\Tjn(R)}$ on terms in  $\cT(\cF,\cV)$.
Note that the composed transformation $\Unl\circ\Tjn$\/  
is an unraveling for join CTRSs. 
\begin{exa}
 The join CTRS $\TRSoddeven$ in Example~\ref{ex:oddeven} is transformed
 by $\Tjn$\/ as follows: 
 \[
 \Tjn(\TRSoddeven) = 
 \left\{
 \begin{array}{r@{\>}c@{\>}l}
  & \vdots \\
  \mathsf{odd}(\mathsf{s}(x)) & \to & \mathsf{true}
   \Leftarrow \mathsf{eq}(\mathsf{even}(x),\mathsf{true}) \tto
   \mathsf{eq}(\top,\top) \\   
  \mathsf{odd}(\mathsf{s}(x)) & \to & \mathsf{false}
   \Leftarrow \mathsf{eq}(\mathsf{even}(x),\mathsf{false}) \tto
   \mathsf{eq}(\top,\top) \\    
  & \vdots \\
  \mathsf{even}(\mathsf{s}(x)) & \to & \mathsf{true} 
   \Leftarrow \mathsf{eq}(\mathsf{odd}(x),\mathsf{true}) \tto
   \mathsf{eq}(\top,\top) \\ 
  \mathsf{even}(\mathsf{s}(x)) & \to & \mathsf{false} 
   \Leftarrow \mathsf{eq}(\mathsf{odd}(x),\mathsf{false}) \tto
   \mathsf{eq}(\top,\top) \\ 
  \mathsf{eq}(x,x) & \to & \mathsf{eq}(\top,\top) \\
 \end{array}
 \right\}
\]
\end{exa}

Here, we recall the notion of weak left-linearity. 
A normal 1-CTRS $R$\/ is called {\em weakly left-linear}\/
(WLL)~\cite{GGS10} if any 
conditional rule with a non-empty condition 
in $R$\/ is $\Uopt$-LL and any unconditional rule in $R$\/ 
is LL w.r.t.\ non-erasing variables (\ie, for $l \to r$, the
left-hand side is linear w.r.t.\ variables in $\Var(l) \cap \Var(r)$).
For example, the normal CTRS $\TRSfeh$ in Example~\ref{ex:feh} is WLL.
$R$\/ is called {\em ground conditional}\/ if, for any rule $l \to r
\Leftarrow \Condition{s_1}{t_1}; \ldots; \Condition{s_k}{t_k}$ in $R$,
the terms $s_1,t_1,\ldots,s_k,t_k$ in the conditional part are ground. 
As mentioned before, some soundness conditions for $\Unl$ and $\Un$ are
known, that are related to the (W)LL property, while no soundness
condition for either $\Ujl$ or $\Uj$ is known. 
\begin{thm}[\cite{GGS10}]
\label{th:Un-soundness}\ 
 $\Unl$ is sound for WLL normal 1-CTRSs. 
\end{thm}

The WLL property and Theorem~\ref{th:Un-soundness} lead to the following
soundness condition of the composed unraveling $\Unl\circ\Tjn$. 
\begin{lem}
 \label{lem:Tjn}\ 
\begin{enumerate}[$\bullet$]
 \item 
 If a join CTRS $R$ is LL, then $\Tjn(R)$ is WLL.
\item $\Unl \circ \Tjn$ is sound for LL join CTRSs.
\end{enumerate}
\end{lem}
\proof
The first claim is trivial by definition.
The second claim follows from the first claim and Theorem~\ref{th:Un-soundness}.
\qed

TRSs obtained by $\Unl \circ \Tjn$\/ can completely 
 derive reduction sequences of the corresponding TRSs obtained by $\Ujl$,
 \ie, if $\Unl\circ\Tjn$ is sound for a join CTRS $R$, then so is $\Ujl$. 
\begin{thm}
 \label{th:UnTjn-simulation}
 Let $R$ be a join CTRS over a signature $\cF$. 
 Then, there exists an $\cF$-identical and non-erasing tree
 homomorphism $\phi$ such that $\Unl\circ\Tjn(R)$ $=$ $\phi(\Ujl(R))$.
 That is, if $\Unl\circ\Tjn$\/ is sound for $R$ (w.r.t.\
 $\evbto_{\Unl\circ\Tjn(R)}$), then $\Ujl$ is sound for $R$ (w.r.t.\
 $\evbto_{\Ujl(R)}$). 
\end{thm}
\proof
 Let $\phi$\/ be an $\cF$-identical and non-erasing tree homomorphism
 determined by $\phi_{\cF_{\Ujl(R)}}$ such that 
 \[
 \phi_{\cF_{\Ujl(R)}}(\Usym^\rho(x_1,x'_1,\ldots,x_k,x'_k,\overrightarrow{\Var(l)})) 
 =
 \Usym^\rho(\mathsf{eq}(x_1,x'_1),\ldots,\mathsf{eq}(x_k,x'_k),\overrightarrow{\Var(l)})
 \]
 where $\rho: l \to r \Leftarrow s_1 \downarrow t_1; \ldots; s_k
 \downarrow t_k$ $\in$ $R$\/ and $x_1,x'_1,\ldots,x_k,x'_k$ are fresh
 different variables. 
 Then, it is clear that $\Unl\circ\Tjn(R)$ $=$ $\phi(\Ujl(R))$. 
\qed
\noindent
Theorem~\ref{th:UnTjn-simulation} indicates that, for a join CTRS $R$,
soundness conditions of $\Unl$ for $\Tjn(R)$ are soundness conditions of
$\Ujl$ for $R$. 
For example, as a consequence of Lemma~\ref{lem:Tjn} and
Theorem~\ref{th:UnTjn-simulation}, we conclude the following result on
soundness of $\Ujl$.  
\begin{cor}
 \label{cor:Ujl-soundness}
$\Ujl$ is sound for LL join 3-CTRSs.
\end{cor}
\noindent
We recognize from Corollary~\ref{cor:Ujl-soundness} that $\Ujl$ is sound for
$\TRSoddeven$ in Example~\ref{ex:oddeven}.

As we mentioned before, normal CTRSs can be considered join CTRSs
because the conditions $s_i \tto n_i$ and $s_i \downarrow n_i$ with a
ground normal form $n_i$ are identical.
Thus, soundness of $\Ujl$ implies soundness of $\Unl$.
\begin{thm}
 \label{th:Un-soundness_by_Uj}
 $\Unl$ is sound for a normal CTRS $R$ (w.r.t.\ $\evbto_{\Unl(R)}$)
 if\/ $\Ujl$ is sound for the the corresponding join CTRS $R'$ $=$ $\{ l
 \to r \Leftarrow s_1 \downarrow n_1; \ldots; s_k \downarrow n_k \mid l
 \to r \Leftarrow s_1 \tto n_1; \ldots; s_k \tto n_k \in R \}$ (w.r.t.\
 $\evbto_{\Ujl(R')}$). 
\end{thm}
\proof
Let $R$\/ be over a signature $\cF$\/ and $\phi$\/ be an $\cF$-identical and
non-erasing tree homomorphism determined by $\phi_{\Unl(R)}$ such that 
\[
 \phi_{\cF_{\Unl(R)}}(\Usym^\rho(x_1,\ldots,x_k,\overrightarrow{\Var(l)}))
 = \Usym^\rho(x_1,n_1,\ldots,x_k,n_k,\overrightarrow{\Var(l)})
 \]
 where $\rho: l \to r \Leftarrow s_1 \downarrow n_1; \ldots; s_k
 \downarrow n_k$ $\in$ $R$\/ and $x_1,x'_1,\ldots,x_k,x'_k$ are different
 fresh variables.  
 Then, it is clear that $\phi(\Unl(R))$ $=$ $\Ujl(R')$. 
\qed
\begin{exa}
 Consider the join CTRS $\TRSoddeven$ and the corresponding normal CTRS
 $\TRSoddeven$ in Example~\ref{ex:oddeven} again.
 Let $\phi$\/ be a tree homomorphism determined by the mapping $\phi_\cF$
 such that $\phi_\cF(\Uodda(y,x))$ $=$ $\Uodda(y,\mathsf{true},x)$, 
 $\phi_\cF(\Uoddb(y,x))$ $=$
 $\Uoddb(y,\mathsf{false},x)$,
 $\phi_\cF(\Uevena(y,x))$ $=$ $\Uevena(y,\mathsf{true},x)$,
 and $\phi_\cF(\Uevenb(y,x))$ $=$
 $\Uevenb(y,\mathsf{false},x)$.
 Then, $\phi(\Unl(\TRSoddeven'))$ $=$ $\Ujl(\TRSoddeven)$.
 Since $\Ujl$ is sound for $\TRSoddeven$, we recognize from
 Theorem~\ref{th:Un-soundness_by_Uj} that $\Unl$ is sound for
 $\TRSoddeven'$. 
\end{exa}

By the same token, a join CTRS $R$\/ can be considered a normal CTRS
if, for any rule $l \to r \Leftarrow s_1 \downarrow t_1; \ldots; s_k
\downarrow t_k$ $\in$ $R$ and for all $1$ $\leq$ $i$ $\leq$ $k$, at
least one $s_i$ and $t_i$ is a ground normal form w.r.t.\ $R_u$. 
We call such a join CTRS $R$ {\em normal}\/ and assume w.l.o.g.\ that $t_i$ is
a ground normal form w.r.t.\ $R_u$.
Then, we obtain the following soundness condition of $\Ujl$.
\begin{thm}
 \label{th:Uj-soundness_by_Un}
 $\Ujl$ is sound for a normal join CTRS $R$ (w.r.t.\ $\evbto_{\Ujl(R)}$)
 if\/ $\Unl$ is sound for the corresponding normal CTRS $R'$ $=$ $\{ l \to
 r \Leftarrow s_1 \tto n_1; \ldots; s_k \tto n_k \mid l \to r \Leftarrow
 s_1 \downarrow n_1; \ldots; s_k \downarrow n_k \in R \}$ (w.r.t.\
 $\evbto_{\Unl(R')}$). 
\end{thm}
\proof
Let $R$\/ be over a signature $\cF$\/ and $\phi$\/ be an $\cF$-identical and
 non-erasing tree homomorphism determined by $\phi_{\Ujl(R)}$ such that  
 \[
  \phi_{\cF_{\Ujl(R)}}(\Usym^\rho(x_1,x'_1,\ldots,x_k,x'_k,\overrightarrow{\Var(l)}))
 = \Usym^\rho(x_1,\ldots,x_k,\overrightarrow{\Var(l)})
\]
 where $\rho: l \to r \Leftarrow s_1 \downarrow n_1; \ldots; s_k
 \downarrow n_k$ $\in$ $R$\/ and $x_1,x'_1,\ldots,x_k,x'_k$ are different
 fresh variables.  
 Then, it is clear that $\phi(\Ujl(R))$ $=$ $\Unl(R')$.  
\qed

It is possible to transform join CTRSs into DCTRSs without adding the
rule $\mathsf{eq}(x,x) \to \mathsf{eq}(\top,\top)$.
\begin{defi}
 \label{def:Tjd}
Let $R$\/ be a join CTRS over a signature $\cF$.
Introducing a fresh $2k$-ary constructor $\mathsf{eq}_k$ for each $\rho:
 l \to r \Leftarrow s_1 \downarrow t_1; \ldots; s_k \downarrow t_k$ $\in$
 $R$, we define a transformation $\Tjd$\/ as follows: 
\begin{enumerate}[$\bullet$]
 \item $\Tjd(l \to r \Leftarrow s_1 \downarrow t_1; \ldots; s_k
       \downarrow t_k)$ $=$ $l \to r \Leftarrow
       \mathsf{eq}_k(s_1,t_1,\ldots,s_k,t_k) \!\tto\!
       \mathsf{eq}_k(x_1,x_1,\ldots,x_k,x_k)$ where $x_1,\ldots,x_k$ are
       different fresh 
       variables, and
 \item $\Tjd(R)$ $=$ $\{ \Tjd(\rho) \mid \rho \in R \}$.
\end{enumerate}
\end{defi}
\noindent
The reason why we introduced $\mathsf{eq}_k$ instead of $s_1 \tto x_1;
t_1 \tto x_1; \ldots; s_k \tto x_k; t_k \tto x_k$ is to make the number
of the conditions in each rule of $\Tjd(R)$ at most one. 
It is clear that 
$\U\circ\Tjd$\/ is an unraveling for join CTRSs.
It is also clear that $\to_{\Tjd(R)}$ $=$ $\to_R$ on terms in
$\cT(\cF,\cV)$.  
TRSs obtained by $\U \circ \Tjn$\/ can completely 
 derive reduction sequences of the corresponding TRSs obtained by
 $\Ujl$.
 This indicates that if $\U\circ\Tjd$\/ is sound for a join CTRS $R$,
 then so is $\Ujl$. 
\begin{thm}
 \label{th:Uj-soundness_from_U-Tjd}
 Let $R$ be a join CTRS over a signature $\cF$.
 Then, there exists an $\cF$-identical and non-erasing tree homomorphism
 $\phi$ such that $\U\circ\Tjd(R)$ $=$ $\phi(\Ujl(R))$.
 That is, if\/ $\U\circ\Tjd$ is sound for $R$ (w.r.t.\
 $\evbto_{\U\circ\Tjd(R)}$),  then\/ $\Ujl$ is sound for $R$ (w.r.t.\
 $\evbto_{\Ujl(R)}$).  
\end{thm}
\proof
Let $\phi$\/ be an $\cF$-identical and non-erasing tree homomorphism
determined by $\phi_{\cF_{\Uj(R)}}$ such that
 \[
 \phi_{\cF_{\Ujl(R)}}(\Usym^\rho(x_1,x'_1,\ldots,x_k,x'_k,\overrightarrow{\Var(l)}))
 =
 \Usym^\rho_1(\mathsf{eq}_k(x_1,x'_1,\ldots,x_k,x'_k),\overrightarrow{\Var(l)})
 \]
 where $\rho: l \to r \Leftarrow s_1 \downarrow t_1; \ldots; s_k
 \downarrow t_k$ $\in$ $R$\/ and $x_1,x'_1,\ldots,x_k,x'_k$ are fresh
 different variables. 
 Then, it is clear that $\phi(\Ujl(R))$ $=$ $\U\circ\Tjd(R)$.
\qed
\noindent
Note that it is easy to adapt Theorem~\ref{th:Uj-soundness_from_U-Tjd}
to $\Unl$ and normal CTRSs.

Normal CTRSs are special cases of DCTRSs, and thus, the unravelings $\U$
and $\Uopt$ for DCTRSs are applicable to normal CTRSs.
Moreover, by definition, $\Unl$ 
 can be considered a special variant of $\U$ 
while there is a slight difference: 
$\Unl$ 
introduces at most one U symbol for each rewrite rule, and $\U$ 
introduces $k$ U symbols for each rewrite rule with $k$ conditions.
This difference prevents us from using
Lemma~\ref{lem:soundness_of_two-unravelings} to prove that if
$\Unl$ is sound for $R$, then so is $\U$. 
For this reason, we extend
Lemma~\ref{lem:soundness_of_two-unravelings} as follows.
\begin{lem}
 \label{lem:ex-soundness_of_two-unravelings}
 Let $U_1$ and $U_2$ be unravelings, $R$ be an eCTRS over a signature
 $\cF$, and $\cG_1,\cG_2$ be extended signatures of $\cF$ such that
 $U_1(R)$ and $U_2(R)$ are defined over $\cG_1$ and $\cG_2$,
 respectively. 
 Let $\phi$ be an $\cF$-identical tree homomorphism determined by
 $\phi_\cG$ such that $U_1(R)$ $=$ $\phi(U_2(R)) \setminus \{ t \to t
 \mid t \in \cT(\cG_1,\cV) \setminus \cT(\cF,\cV) \}$. 
 Then, all of the following hold:
 \begin{enumerate}[\em(1)]
  \item $\to^*_{U_2(R)}$ $\subseteq$ $\to^*_{U_1(R)}$ on terms in
	$\cT(\cF,\cV)$,
  \item if $\phi$ is EV-preserving for $U_2(R)$, then
	$\evbto^*_{U_2(R)}$ $\subseteq$ $\evbto^*_{U_1(R)}$ on terms in
	$\cT(\cF,\cV)$.
 \end{enumerate}
 That is, all of the following hold:
 \begin{enumerate}[\em(1)]
  \setcounter{enumi}{2}
  \item if $U_1$ is sound for $R$, then so is $U_2$, and,
  \item if $\phi$ is EV-preserving for $U_2(R)$ and $U_1$ is sound for
	$R$ w.r.t.\ $\evbto_{U_1(R)}$, then $U_2$ is sound for $R$
	w.r.t.\ $\evbto_{U_2(R)}$.
 \end{enumerate}
\end{lem}
\proof
We first prove the first claim $\to^*_{U_2(R)}$ $\subseteq$
$\to^*_{U_1(R)}$ on terms in $\cT(\cF,\cV)$. 
It follows from the assumption that $U_1(R)$ $=$ $(\phi(U_2(R))
 \setminus \{ t \to t \mid t \in \cT(\cG_1,\cV) \} \cup R'$
for some eTRS $R'$ $\subseteq$ $\{ t \to t \mid t \in \cT(\cG_1,\cV)
\}$. 
Then, it follows from Lemma~\ref{lem:TH} that $\phi(\to^*_{U_2(R)})$
$\subseteq$ $\to^*_{U_1(R) \cup R'}$.
Since $\to_{R'}$ is the identity relation, we have that $\to^*_{U_1(R)
\cup R'}$ $=$ $\to^*_{U_1(R)}$, and hence $\phi(\to^*_{U_2(R)})$ $\subseteq$
$\to^*_{U_1(R)}$. 
Since $\phi$\/ is $\cF$-identical, we have that $\to^*_{U_2(R)}$
$\subseteq$ $\to^*_{U_1(R)}$ on terms in $\cT(\cF,\cV)$. 

The second claim follows from the first claim and
 Lemma~\ref{lem:pi-evb}. 
The third and fourth claims follow from the first and second claims, and
 soundness of $U_1$.
\qed

Due to Lemma~\ref{lem:ex-soundness_of_two-unravelings}, we obtain the
following theorem.
\begin{thm}
 \label{th:soundness_by_Un-full}
Let $R$ be a normal CTRS over a signature $\cF$.
 Then, there exists an $\cF$-identical and non-erasing tree homomorphism
 $\phi$ such that $\Unl(R)$ $=$ $\phi(\U(R)) \setminus \{ t \to t \mid
 t \in \cT(\cF_{\U(R)},\cV) \setminus \cT(\cF,\cV) \}$.
 That is, if\/ $\Unl$ is sound for a normal CTRS $R$ (w.r.t.\
 $\evbto_{\Unl(R)}$), then\/ $\U$ is sound for $R$ (w.r.t.\
 $\evbto_{\U(R)}$). 
\end{thm}
\proof
Let $\phi$\/ be an $\cF$-identical tree homomorphism
 determined by $\phi_{\cF_{\U(R)}}$ such that  
\[
 \phi_{\cF_{\U(R)}}(\Usym^\rho_i(x_i,\overrightarrow{\AppearedVar_i}))
 =
 \Usym^\rho(n_1,\ldots,n_{i-1},x_i,s_{i+1},\ldots,s_k,\overrightarrow{\Var(l)})
\]
 where $\rho: l \to r \Leftarrow s_1 \tto n_1; \ldots; s_k \tto n_k$
 $\in$ $R$\/ and $x_i$ is a fresh variable. 
 It is clear that $\Unl(R)$ $=$ $\phi(\U(R)) \setminus \{ t \to t
 \mid t \in \cT(\cF_{\U(R)},\cV) \setminus \cT(\cF,\cV) \}$:
 \begin{enumerate}[$\bullet$]
  \item $\phi(l \to \Usym^\rho_1(s_1,\overrightarrow{\AppearedVar_1}))$ $=$ $l
	\to \Usym^\rho(s_1,\ldots,s_k,\overrightarrow{\Var(l)})$ $\in$ $\Unl(R)$, 
  \item $\phi(\Usym^\rho_i(n_i,\overrightarrow{\AppearedVar_i})
	\to
	\Usym^\rho_{i+1}(s_{i+1},\overrightarrow{\AppearedVar_{i+2}}))$
	$=$
	$\Usym^\rho(n_1,\ldots,n_i,s_{i+1},\ldots,s_k,\overrightarrow{\Var(l)})
	\to
	\Usym^\rho(n_1,\ldots,n_i,$ $s_{i+1},\ldots,s_k,\overrightarrow{\Var(l)})$
	$\in$ $\{ t \to t \mid t \in \cT(\cF_{\Unl(R)},\cV) \setminus
	\cT(\cF,\cV) \}$, and
  \item $\phi(\Usym^\rho_k(n_k,\overrightarrow{\AppearedVar_k}) \to r)$ $=$
 $\Usym^\rho(n_1,\ldots,n_k,\overrightarrow{\Var(l)}) \to r$ $\in$
	$\Unl(R)$.
 \end{enumerate}
Since $R$\/ is normal, we have that $\Var(n_1,\ldots,n_k)$ $=$
 $\emptyset$ and $\Var(s_1,\ldots,s_k)$ $\subseteq$ $\Var(l)$, and hence
 $\AppearedVar_i$ $=$ $\Var(l)$ for all $1$ $\leq$ $i$ $\leq$ $k$. 
 Thus, 
 $\Var(\phi_{\cF_{\U(R)}}(\Usym^\rho_i(x_i,\overrightarrow{\AppearedVar_i})))$
 $=$
 $\{x_i\} \cup \AppearedVar_i$
 $=$
 $\{x_i\} \cup \Var(l)$
 $=$
 $\Var(\Usym^\rho(n_1,\ldots,n_{i-1},x_i,s_{i+1},\ldots,s_k,\overrightarrow{\Var(l)}))$,
 and hence $\phi$ is non-erasing.
\qed
\noindent
It is not known whether the converse of
Theorem~\ref{th:soundness_by_Un-full} (\ie, $\to^*_{\Unl(R)}$ $\subseteq$
$\to^*_{\U(R)}$ on terms in $\cT(\cF,\cV)$) holds or not.
In other words, it is not known whether the following claim holds or not:
if $\U$ is sound for a normal CTRS, then so is $\Unl$.
As we mentioned before, to show soundness of $\Unl$ by means of $\U$, we would
like to show that, for any normal CTRS $R$, all the derivations of
$\Unl(R)$ can be derived by $\U(R)$.
However, this is not true in general.
\begin{exa}\label{ex:open1}
Consider the following variant $\TRSmarchiori'$ of the DCTRS
 $\TRSmarchiori$ in Example~\ref{ex:marchiori}, that is obtained by
 replacing the conditional part $x \tto \mathsf{e}$ of the first rule by
 $x \tto \mathsf{e};~x \tto \mathsf{e}'$ and by adding $\mathsf{c} \to
 \mathsf{e}'$ to the rules: 
\[
 \TRSmarchiori' = 
 \left\{
 \begin{array}{r@{\>}c@{\>}l}
  \mathsf{f}(x) & \to & x \Leftarrow x \tto \mathsf{e};~x \tto \mathsf{e}' \\
  \mathsf{g}(\mathsf{d},x,x) & \to & \mathsf{A} \\
  \mathsf{h}(x,x) & \to & \mathsf{g}(x,x,\mathsf{f}(\mathsf{k})) \\
  \mathsf{c} & \to & \mathsf{e}' \\
 \end{array}
 \right\}
 \cup \TRSmarchioriconstants
\]
The CTRS $\TRSmarchiori'$ is unraveled by $\Unl$ and $\U$ as follows:
\[
  \Unl(\TRSmarchiori') = 
 \left\{
 \begin{array}{@{\>}r@{\>}c@{\>}l@{\>}}
  \mathsf{f}(x) & \to & \Umarchiori(x,x,x) \\
  \Umarchiori(\mathsf{e},\mathsf{e}',x) & \to & x \\
  & \vdots \\
 \end{array}
 \right\} 
 \cup \TRSmarchioriconstants
 \quad\quad
  \U(\TRSmarchiori') = 
 \left\{
 \begin{array}{@{\>}r@{\>}c@{\>}l@{\>}}
  \mathsf{f}(x) & \to & \Umarchiori'(x,x) \\
  \Umarchiori'(\mathsf{e},x) & \to & \Umarchiori''(x,x) \\
  \Umarchiori''(\mathsf{e}',x) & \to & x \\
  & \vdots \\
 \end{array}
 \right\} 
 \cup \TRSmarchioriconstants
\]
We have that $\mathsf{h}(\mathsf{f}(\mathsf{a}),\mathsf{f}(\mathsf{b}))$
 $\to^*_{\Unl(\TRSmarchiori')}$ $\mathsf{A}$, but
 $\mathsf{h}(\mathsf{f}(\mathsf{a}),\mathsf{f}(\mathsf{b}))$
 $\not\to^*_{\U(\TRSmarchiori')}$ $\mathsf{A}$.  
This means that $\U(\TRSmarchiori')$ cannot derive 
 every reduction sequence of $\Unl(\TRSmarchiori')$ that starts from
 terms over the original signature of $\TRSmarchiori'$.
$\U$ seems sound for $\TRSmarchiori'$.
However, we have no sufficient condition to prove soundness of $\U$ for
 $\TRSmarchiori'$, and thus, it is not known whether $\U$ is sound for
 $\TRSmarchiori'$ or not.

The symbols $\mathsf{e}$ and $\mathsf{e}'$ are used for the same role;
therefore, this distinction is meaningless.
Thus, the replacement of $x \tto \mathsf{e}$ with $x \tto \mathsf{e};~x
 \tto \mathsf{e}$ is sufficient for the purpose of this example.
For the original CTRS $\TRSmarchiori'$, this duplication of $x \tto
 \mathsf{e}$ is quite meaningless, but this greatly affects the reduction
 of $\U(\TRSmarchiori')$.
For this reason, this would be an interesting example for investigating
 soundness conditions of unravelings.
\end{exa}

A trivial sufficient condition for the converse of
Theorem~\ref{th:soundness_by_Un-full} is that any rule of $R$\/ has at
most one condition:
by considering $\Usym^\rho$ $=$ $\Usym^\rho_1$, we have that $\Unl(R)$
$=$ $\U(R)$. 

As stated above, the relationship between $\Uj$, $\Un$, and $\Uopt$ is
similar to that between $\Ujl$, $\Unl$, and $\U$.
For this reason,
Theorems~\ref{th:Un-soundness_by_Uj},~\ref{th:Uj-soundness_by_Un},~\ref{th:soundness_by_Un-full}
also hold for $\Uj$, $\Un$, and $\Uopt$. 

\section{Comparison with {\Serbanuta}-{\Rosu} Transformation}
\label{sec:comparison}

In this section, we compare the unraveling $\U$ with the SR
transformation, in terms of soundness, operational termination,
confluence, computational equivalence, and so on.

\subsection{Formalization of Transformations for CTRSs}
\label{subsec:CTRS-transformations}

In this subsection, to make it easier to compare unravelings with other
transformations, we first formalize transformations of CTRSs into TRSs,
and also generalize the notion of  soundness and completeness for unravelings.
Then, we present relationship between soundness of two transformations
by generalizing Lemma~\ref{lem:soundness_of_two-unravelings}.   

We first formalize transformations of CTRSs and the notions of
soundness and completeness.
\begin{defi}[CTRS transformations]
\label{def:CTRS-transformations}
A {\em CTRS transformation}\/ is a computable transformation $T$\/ from
 eCTRSs into eTRSs with injective mappings as follows: 
 for an eCTRS $R$\/ over a signature $\cF$, the 
 transformed eTRS $R_T$ over a signature $\cG$\/ is defined and
 the corresponding mapping $\phi_{T(R)}$ from $\cT(\cF,\cV)$ to 
 $\cT(\cG,\cV)$ is also defined, \ie, $T(R)$ $=$ $(R_T,\phi_{T(R)})$. 
 The mapping $\phi_{T(R)}$ is called a {\em translation related to
 $T(R)$}.%
 \footnote{ The mapping $\phi$\/ can be considered a
 translation from original terms for $R$\/ into the corresponding ones for
 $T(R)$.} 
 We extend $\phi_{T(R)}$ to pairs of terms in $\cT(\cF,\cV)$:
 for $S$ $\subseteq$ $\cT(\cF,\cV) \times \cT(\cF,\cV)$, $\phi(S)$ $=$
 $\{ (\phi(s),\phi(t)) \mid (s,t) \in S\}$.
 Moreover, $T$\/ is called {\em simple}\/ if the related translation
 $\phi_{T(R)}$ is the identity mapping (\ie, $\cF$ $\subseteq$ $\cG$\/ and
 $\phi_{T(R)}(t)$ $=$ $t$\/ for all $t$ $\in$ $\cT(\cF,\cV)$), and we
 abuse notation and write $T(R)$ as the transformed system $R_T$. 
 
 Let $\Rightarrow_{R_T}$ be a subrelation of $\to_{R_T}$. 
\begin{enumerate}[$\bullet$]
 \item $T$\/ is called {\em sound for $R$ w.r.t.\ $\Rightarrow_{R_T}$}\/ if
       $\Rightarrow^*_{R_T}$ $\subseteq$ $\phi_{T(R)}(\to^*_R)$ on terms
       in $\cT(\cF,\cV)$ (\ie, for 
       all terms $s,t$ in $\cT(\cF,\cV)$, if $\phi_{T(R)}(s)$
       $\Rightarrow^*_{R_T}$ $\phi_{T(R)}(t)$, then $s$ $\to^*_R$ $t$).
 \item $T$\/ is called {\em complete for $R$ w.r.t.\
       $\Rightarrow_{R_T}$}\/ if $\phi_{T(R)}(\to^*_R)$ $\subseteq$
       $\Rightarrow^*_{R_T}$ (\ie, for all terms $s,t$
       $\in$ $\cT(\cF,\cV)$, if $s$ $\to^*_R$ $t$, then $\phi_{T(R)}(s)$
       $\Rightarrow^*_{R_T}$ $\phi_{T(R)}(t)$). 
\end{enumerate}
 When $T$\/ is sound and complete for $R$\/ w.r.t.\ $\to_{R_T}$,
we simply say that $T$\/ 
 is {\em sound}\/ and {\em complete for $R$}, respectively. 
 Moreover, $T$\/ is called {\em sound}\/ ({\em complete}\/) if $T$\/ is sound
 (complete) for any eCTRS $R$\/ such that $T(R)$ is defined. 
\end{defi}
\noindent
Note that unravelings are complete simple CTRS transformations.

Next, we generalize Lemma~\ref{lem:soundness_of_two-unravelings} to two
CTRS transformations, one of which is simple. 
\begin{thm}
 \label{th:CTRS-transformations}
 Let $T$ be a CTRS transformation, $U$ be a simple CTRS transformation,
 $R$ be an eCTRS over a signature $\cF$ such that $T(R)$ and $U(R)$ are
 defined, $R_T$ is over a signature $\cG_T$, and $U(R)$ is over a
 signature $\cG_U$. 
 Then, all of the following hold:
 \begin{enumerate}[$\bullet$]
  \item\label{itm:soundness-preservation}
       if $T$ is sound for $R$ and $\phi(\to^*_{U(R)})$
       $\subseteq$ $\to^*_{R_T}$,%
       \footnote{ Note that $\phi(\to^*_{U(R)})$ $=$ $\{
       (\phi(s),\phi(t)) \mid s, t \in \cT(\cF,\cV), ~ s
       \to^*_{U(R)} t \}$ since $\phi$\/ is not defined for any term
       containing a function symbol in $\cG_U \setminus \cF$.}
       then $U$ is sound for $R$, 
  \item\label{itm:soundness-preservation2}
       if $U$ is sound for $R$ and 
       $\to^*_{R_T}$ $\subseteq$ $\phi(\to^*_{U(R)})$,
       then $T$ is sound for $R$, 
  \item\label{itm:completeness-preservation}
       if $T$ is complete for $R$ and 
       $\to^*_{R_T}$ $\subseteq$ $\phi(\to^*_{U(R)})$, then $U$ is
       complete for $R$, and
  \item\label{itm:completeness-preservation2}
       if $U$ is complete for $R$ and $\phi(\to^*_{U(R)})$ $\subseteq$
       $\to^*_{R_T}$, then $T$ is complete for $R$.
 \end{enumerate}
\end{thm}
\proof
We only prove the first claim since the other claims can be proved
 similarly to the first one.
Let $s,t$ be terms in $\cT(\cF,\cV)$.
	      Suppose that $s$ $\to^*_{U(R)}$ $t$.
	      Then, it follows from $\phi(\to^*_{U(R)})$ $\subseteq$
	      $\to^*_{R_T}$ that $\phi(s)$ $\to^*_{R_T}$ $\phi(t)$.
	      It follows from soundness of $T$\/ for $R$\/ that $s$ $\to^*_R$
	      $t$, and hence $U$\/ is sound for $R$.	      
\qed

\subsection{{\Serbanuta}-{\Rosu} Transformation}
\label{subsec:SR-transformation}

In this subsection, we recall the definition of the SR transformation
proposed by {\Serbanuta} and {\Rosu}~\cite{SR06,SR06b}, which is
basically applied to {\em strongly}\/ or {\em syntactically}\/ DCTRSs.
We also recall some of its properties. 

Let $R$\/ be an eDCTRS.
A term $t$\/ is called {\em strongly irreducible w.r.t.\ $R$}\/ if
 $t\sigma$\/ is a normal form w.r.t.\ $R$\/ for every normalized
 substitution $\sigma$. 
$R$\/ is called {\em strongly deterministic}\/ (strongly DCTRS) if, for
every rule $l \to r \Leftarrow s_1 \tto t_1; \ldots; s_k 
\tto t_k$ $\in$ $R$, every term $t_i$ is strongly irreducible w.r.t.\
$R$.
$R$\/ is called {\em syntactically deterministic}\/ (syntactically DCTRS)
if, for every rule $l \to r \Leftarrow s_1 \tto t_1; 
\ldots; s_k \tto t_k$ $\in$ $R$, every term $t_i$ is a constructor term
or a ground normal form w.r.t.\ $R_u$.
Note that normal CTRSs are syntactically DCTRSs, and syntactically
DCTRSs are also strongly DCTRSs.

In the following, we assume that for each defined symbol $f$ of $R$,
there are $n_f$ many $f$-rules in $R$\/ that have non-empty conditions
and are ordered. 
We denote the $i$-th conditional rewrite rule of $f$\/ with a non-empty
condition by $\rho_{f,i}$. 

In the SR transformation $\SRTd$ below, a fresh unary function symbol
$\{\Blank\}$, a fresh constant $\bot$, and fresh $k$-ary constructors
$[\Blank]_k$ are introduced and for a defined symbol $f$, a fresh
function symbol $\overline{f}$ is introduced by adding $n_f$ arguments
to $f$.  
The ``\,$n+i$\,''-th argument of $\overline{f}$ is used for evaluating the
conditions of the $i$-th conditional rule $\rho_{f,i}: l
\to r \Leftarrow s_1 \tto t_1; \ldots; s_k \tto t_k$, by initializing
with $\bot$ and by replacing $\bot$ with an instance of
$[\{\overline{s_1}\},\bot,\ldots,\bot]_k$ to start the evaluation, 
where $\overline{s_1}$ is the term obtained by replacing each defined
symbol $f$\/ by $\overline{f}$ with filling extra arguments with $\bot$.
The $k$-ary symbol $[\Blank]_k$ is used as a stack with $k$\/ elements, 
\eg, when $[\{\overline{s_1}\sigma\},\bot,\ldots,\bot]_k$ is reduced to
$[\{\overline{t_1}\theta\},\bot,\ldots,\bot]_k$, the evaluation of the
second condition $s_2 \tto t_2$ with $\theta$\/ starts from
$[\{\overline{s_2}\theta\},\overline{t_1}\theta,\bot,\ldots,\bot]_k$.
\begin{defi}[SR transformation $\SRTd$~\cite{SR06b}]
 \label{def:SRTd}
Let $R$\/ be a strongly or syntactically DCTRS over a signature $\cF$.%
\footnote{ In~\cite{SR06b}, it is assumed that any
 deterministic conditional rule $l \to r \Leftarrow s_1 \tto
 t_1;\ldots;s_k \tto t_k$ satisfies $\Var(s_i)$ $\not\subseteq$
 $\Var(l,t_1,\ldots,t_{i-2})$, \ie, the $i$-th condition $s_i \tto t_i$
 cannot be evaluated before finishing the evaluation of the ``\,$i-1$\,''-th
 condition $s_{i-1} \tto t_{i-1}$. 
 However, this is not essential for the definition of $\SRTd$.}
For $f$ $\in$ $\cD_R$, we prepare a function symbol $\overline{f}$ with
 $\Arity(\overline{f})$ $=$ $\Arity(f) + n_f$.
Introducing a fresh unary function symbol $\{\Blank\}$, a fresh constant
 $\bot$ and fresh $j$-ary constructors $[\Blank]_j$ with $j$ $>$ $0$
 ($[\Blank]_1,[\Blank]_2,\ldots$ are sometimes abbreviated to $[\Blank]$)
 into the signature, the DCTRS $R$\/ is transformed into the
 following TRS $\SRTd^\to(R)$:
\[
 \begin{array}{@{}l@{}}
  \SRTd^{\rm rule}(\rho_{f,i}: f(w_1,\ldots,w_n) \to r \Leftarrow s_1 \tto t_1;
   \ldots; s_k \tto t_k) = \\
  \quad
   \left\{
    \begin{array}{r@{\hspace{18ex}}c@{\>}l}
     \lefteqn{\overline{f}(\overline{w_1},\ldots,\overline{w_n},z_1,\ldots,z_{i-1},\bot,z_{i+1},\ldots,z_{n_f})} \\
      & \to & \overline{f}(\overline{w_1},\ldots,\overline{w_n},z_1,\ldots,z_{i-1},[\{\overline{s_1}\},\bot,\ldots,\bot]_k,z_{i+1},\ldots,z_{n_f})
      \\ 
     \lefteqn{\overline{f}(\overline{w_1},\ldots,\overline{w_n},z_1,\ldots,z_{i-1},[\{\overline{t_1}\},\bot,\ldots,\bot]_k,z_{i+1},\ldots,z_{n_f})} \\
     & \to & 
      \overline{f}(\overline{w_1},\ldots,\overline{w_n},z_1,\ldots,z_{i-1},[\{\overline{s_2}\},\overline{t_1},\ldots,\bot]_k,z_{i+1},\ldots,z_{n_f})
 \\ 
     & \vdots & \\
\lefteqn{\overline{f}(\overline{w_1},\ldots,\overline{w_n},z_1,\ldots,z_{i-1},[\{\overline{t_{k-1}}\},\overline{t_{k-2}},\ldots,\overline{t_1},\bot]_k,z_{i+1},\ldots,z_{n_f})} \\
     & \to & 
      \overline{f}(\overline{w_1},\ldots,\overline{w_n},z_1,\ldots,z_{i-1},[\{\overline{s_k}\},\overline{t_{k-1}},\ldots,\overline{t_1}]_k,z_{i+1},\ldots,z_{n_f})
 \\ 
     \lefteqn{\overline{f}(\overline{w_1},\ldots,\overline{w_n},z_1,\ldots,z_{i-1},[\{\overline{t_k}\},\overline{t_{k-1}},\ldots,\overline{t_1}]_k,z_{i+1},\ldots,z_{n_f}) 
   \to \{\overline{r}\}} \\
    \end{array}
 \right\} \\
  \SRTd^{\rm rule}(f(w_1,\ldots,w_n) \to r) = 
   \{~ \overline{f}(\overline{w_1},\ldots,\overline{w_n},z_1,\ldots,z_{n_f})
   \to \{\overline{r}\} ~\}
 \end{array}
\]
where $z_1,\ldots,z_{n_f}$ are fresh different
 variables and the operation $\overline{\Blank}$ is a linear
 non-erasing tree homomorphism determined by $\overline{\phi}$ such that
 \begin{enumerate}[$\bullet$]
 \item $\overline{\phi}(c(x_1,\ldots,x_n)$ $=$
       $c(x_1,\ldots,x_n)$ for an $n$-ary constructor $c$ $\in$ $\Cc_R$, and
 \item $\overline{\phi}(f(x_1,\ldots,x_n)$ $=$
       $\overline{f}(x_1,\ldots,x_n,\overbrace{\bot,\ldots,\bot}^{n_f})$ 
       for an $n$-ary defined symbol $f\in\cD_R$.
\end{enumerate}
Note that the operation $\overline{\Blank}$ is injective.
The transformed TRS $\SRTd^\to(R)$ is defined as follows:
\[
 \begin{array}{@{}l@{\>}l@{}}
  \lefteqn{\SRTd^\to(R) = 
   \textstyle \bigcup_{ \rho \in R} \SRTd^{\rm
   rule}(\rho)} \\ 
  & \cup
   \{
   \overline{f}(x_1,\ldots,x_{i-1},\{x_i\},x_{i+1},\ldots,x_n,z_1,\ldots,z_{n_f})
   \to 
   \{\overline{f}(x_1,\ldots,x_n,\bot,\ldots,\bot)\} \mid f \in \cD_R \} \\
  & \cup
   \{ c(x_1,\ldots,x_{i-1},\{x_i\},x_{i+1},\ldots,x_n) \to
   \{c(x_1,\ldots,x_n)\} \mid c \in \Cc_R \}  \\
   & \cup
   \{~\{\{x\}\} \to \{x\}~\} \\
 \end{array}
\]
where $x_1,\ldots,x_n,z_1,\ldots,z_{n_f}$ are variables.
Note that $\SRTd^\to(R)$ is a TRS over $\cF_{\SRTd(R)}$ $=$ $\{ \bot,
 \{\Blank\} \} \cup \{ \overline{f} \mid f \in \cD_R \} \cup \Cc_R$.
Moreover, a partial mapping $\,\widehat{\Blank}\,$ from
 $\cT(\cF_{\SRTd(R)},\cV)$ to $\cT(\cF,\cV)$ is defined as follows:
\begin{enumerate}[$\bullet$]
 \item $\widehat{x}$ $=$ $x$ for $x$ $\in$ $\cV$,
 \item $\widehat{\{t\}}$ $=$ $\widehat{t}$,
 \item $\widehat{c(t_1,\ldots,t_n)}$ $=$
       $c(\widehat{t_1},\ldots,\widehat{t_n})$ for an $n$-ary
       constructor $c$ $\in$ $\Cc_R$,
       and
 \item $\widehat{\overline{f}(t_1,\ldots,t_n,u_1,\ldots,u_{n_f})}$ $=$
       $f(\widehat{t_1},\ldots,\widehat{t_n})$ for an $n$-ary defined
       symbol $f$ $\in$ $\cD_R$.
\end{enumerate}
Note that the operation $\,\widehat{\Blank}\,$ partially translates
 terms in $\cT(\cF_{\SRTd(R)},\cV)$ back into terms in $\cT(\cF,\cV)$.
 The {\em SR transformation}\/ $\SRTd$ is defined as $\SRTd(R)$ $=$
 $(\SRTd^\to(R),\phi_{\SRTd(R)})$, where the translation
 $\phi_{\SRTd(R)}$ 
 is defined as $\phi_{\SRTd(R)}(t)$ $=$ $\{\overline{t}\}$.  
 Moreover, a term $t$\/ in $\cT(\cF_{\SRTd(R)},\cV)$ is called {\em
 reachable}\/ if there exists a term $s$\/ in $\cT(\cF,\cV)$ such that
 $\phi_{\SRTd(R)}(s)$ $\to^*_{\SRTd^\to(R)}$ $\{t\}$.
\end{defi}
\noindent
Note that $\SRTd$ is a complete CTRS transformation~\cite{SR06b}.
By definition, it is clear that $R$\/ is $\Uopt$-LL iff $\SRTd^\to(R)$ is LL.
A reachable term $s$\/ has the following property:
\begin{enumerate}[$\bullet$]
 \item every subterm of $s$, rooted by $\overline{f}$, is of the form
       $\overline{f}(s_1,\ldots,s_n,u_1,\ldots,u_{n_f})$ such that,
       for all $1$ $\leq$ $j$ $\leq$ $n_f$, 
       $u_j$ is either $\bot$ or of the form
       $[\{\overline{t_i}\},\overline{t_{i-1}},\ldots,\overline{t_1},\bot,\ldots,\bot]_k$ 
       for some $i$,
       where $\rho_{f,j}: f(w_1,\ldots,w_n) \to r \Leftarrow s_1 \tto
       t_1; \ldots; s_k \tto t_k$ $\in$ $R$,
       and
 \item both the symbols $\bot$ and $[\Blank]_k$ appear only as in the form
       mentioned in the previous case.
\end{enumerate} 

For a DCTRS $R$, the transformed TRS $\SRTd^\to(R)$ is overlapping (not
only at root position, but also at properly inner positions), thus not a
constructor system, and all non-constant constructors of $R$\/ are defined
symbols of $\SRTd^\to(R)$. 
However, critical pairs generated from rules to push out the special
constructor $\{\Blank\}$ are joinable and they are not so critical in terms of
confluence. 
\begin{exa}[\cite{SR06b}]\label{ex:SRTd-qsort}
Consider the $\Uopt$-LL DCTRS $\TRSqsort$ in Example~\ref{ex:qsort} again. 
 $\TRSqsort$ is a syntactically DCTRS and it is transformed by
 $\SRTd^\to$ into the following TRS:
\[\small
 \begin{array}{@{}l@{}}
  \SRTd^\to(\TRSqsort) = \\
  ~
 \left\{
\begin{array}{@{}c@{}}
 \begin{array}{@{\,}r@{\>}c@{\>}l@{\,}}
  \overline{\mathsf{split}}(x,\mathsf{nil},z_1,z_2) & \to &
   \{\mathsf{tp}_2(\mathsf{nil},\mathsf{nil})\} \\
  \overline{\mathsf{split}}(x,\mathsf{cons}(y,ys),\bot,z_2) & \to &
   \overline{\mathsf{split}}(x,\mathsf{cons}(y,ys),[\{\overline{\mathsf{split}}(x,ys,\bot,\bot)\},\bot],z_2) \\
   \overline{\mathsf{split}}(x,\mathsf{cons}(y,ys),[\{\mathsf{tp}_2(zs_1,zs_2)\},\bot],z_2)
    & \to &
    \overline{\mathsf{split}}(x,\mathsf{cons}(y,ys),[\{\overline{\mathsf{le}}(x,y)\},\mathsf{tp}_2(zs_1,zs_2)],z_2) \\
\overline{\mathsf{split}}(x,\mathsf{cons}(y,ys),[\{\mathsf{true}\},\mathsf{tp}_2(zs_1,zs_2)],z_2) 
 & \to & \{\mathsf{tp}_2(zs_1,\mathsf{cons}(y,zs_2)) \} \\
  \overline{\mathsf{split}}(x,\mathsf{cons}(y,ys),z_1,\bot) & \to &
\overline{\mathsf{split}}(x,\mathsf{cons}(y,ys),z_1,[\{\overline{\mathsf{split}}(x,ys,\bot,\bot)\},\bot]) \\
\overline{\mathsf{split}}(x,\mathsf{cons}(y,ys),z_1,[\{\mathsf{tp}_2(zs_1,zs_2)\},\bot])
 & \to &
 \overline{\mathsf{split}}(x,\mathsf{cons}(y,ys),z_1,[\{\overline{\mathsf{le}}(x,y)\},\mathsf{tp}_2(zs_1,zs_2)]) \\
\overline{\mathsf{split}}(x,\mathsf{cons}(y,ys),z_1,[\{\mathsf{false}\},\mathsf{tp}_2(zs_1,zs_2)])
 & \to & \{\mathsf{tp}_2(\mathsf{cons}(y,zs_1),zs_2)\} \\
  \overline{\mathsf{le}}(\mathsf{0},y) & \to & \{\mathsf{true}\} \\
 \end{array}
 \\
 \begin{array}{r@{\>}c@{\>}l@{\quad\quad}r@{\>}c@{\>}l}
   \overline{\mathsf{le}}(\mathsf{s}(x),\mathsf{0}) & \to & \{\mathsf{false}\} 
   & \overline{\mathsf{le}}(\mathsf{s}(x),\mathsf{s}(y)) & \to &
   \{\overline{\mathsf{le}}(x,y)\} \\
  \overline{\mathsf{split}}(\{x\},ys,z_1,z_2) & \to &
   \{\overline{\mathsf{split}}(x,ys,\bot,\bot)\} 
  & \overline{\mathsf{le}}(\{x\},y) & \to &
   \{\overline{\mathsf{le}}(x,y)\} \\
   \overline{\mathsf{split}}(x,\{ys\},z_1,z_2) & \to &
   \{\overline{\mathsf{split}}(x,ys,\bot,\bot)\}
   & \overline{\mathsf{le}}(x,\{y\}) & \to &
   \{\overline{\mathsf{le}}(x,y)\} \\
  \mathsf{cons}(\{x\},xs) & \to & \{\mathsf{cons}(x,xs)\} 
   & \mathsf{tp}_2(\{x\},y) & \to & \{\mathsf{tp}_2(x,y)\} \\
   \mathsf{cons}(x,\{xs\}) & \to & \{\mathsf{cons}(x,xs)\} 
   & \mathsf{tp}_2(x,\{y\}) & \to & \{\mathsf{tp}_2(x,y)\} \\
  \mathsf{s}(\{x\}) & \to & \{\mathsf{s}(x)\} 
   & \{\{x\}\} & \to & \{x\} \\
  \end{array}\\
 \end{array}
 \right\} 
 \\
 \end{array}
\]
 Consider the term
 $\mathsf{split}(\mathsf{s}(\mathsf{0}),\mathsf{cons}(\mathsf{0},\mathsf{cons}(\mathsf{s}(\mathsf{s}(\mathsf{0})),\mathsf{nil})))$.
 Starting from its translated term, we have the following derivation of
 $\SRTd^\to(\TRSqsort)$ under the leftmost innermost strategy that
 selects the topmost rules of applicable ones: 
 \[
 \begin{array}{@{}l@{\quad}l@{}}
  \lefteqn{\overline{\mathsf{split}(\mathsf{s}(\mathsf{0}),\mathsf{cons}(\mathsf{0},\mathsf{cons}(\mathsf{s}(\mathsf{s}(\mathsf{0})),\mathsf{nil})))}} \\
  & = 
     \overline{\mathsf{split}}(\mathsf{s}(\mathsf{0}),\mathsf{cons}(\mathsf{0},\mathsf{cons}(\mathsf{s}(\mathsf{s}(\mathsf{0})),\mathsf{nil})),\bot,\bot) \\
  & \to_{\SRTd^\to(\TRSqsort)} 
   \overline{\mathsf{split}}(\mathsf{s}(\mathsf{0}),\mathsf{cons}(\mathsf{0},
   \ldots
   ),[\{\overline{\mathsf{split}}(\mathsf{s}(\mathsf{0}),\mathsf{cons}(\mathsf{s}(\mathsf{s}(\mathsf{0})),\mathsf{nil}),\bot,\bot)\},\bot],\bot)
   \\
  & \to^*_{\SRTd^\to(\TRSqsort)} 
   \overline{\mathsf{split}}(\mathsf{s}(\mathsf{0}),\mathsf{cons}(\mathsf{0},
   \ldots
   ),[\{\mathsf{tp}_2(\mathsf{nil},\mathsf{cons}(\mathsf{s}(\mathsf{s}(\mathsf{0})),\mathsf{nil}))\},\bot],\bot) 
   \\
  & \to_{\SRTd^\to(\TRSqsort)} 
   \overline{\mathsf{split}}(\mathsf{s}(\mathsf{0}),\mathsf{cons}(\mathsf{0},
   \ldots
   ),[\{\overline{\mathsf{le}}(\mathsf{s}(\mathsf{0}),\mathsf{0})\},\mathsf{tp}_2(\mathsf{nil},\mathsf{cons}(\mathsf{s}(\mathsf{s}(\mathsf{0})),\mathsf{nil}))],\bot) 
   \\
  & \to_{\SRTd^\to(\TRSqsort)} 
   \overline{\mathsf{split}}(\mathsf{s}(\mathsf{0}),\mathsf{cons}(\mathsf{0},
   \ldots
   ),[\{\{\mathsf{false}\}\},\mathsf{tp}_2(\mathsf{nil},\mathsf{cons}(\mathsf{s}(\mathsf{s}(\mathsf{0})),\mathsf{nil}))],\bot) 
   \\
  & \to_{\SRTd^\to(\TRSqsort)} 
   \overline{\mathsf{split}}(\mathsf{s}(\mathsf{0}),\mathsf{cons}(\mathsf{0},
   \ldots
   ),[\{\mathsf{false}\},\mathsf{tp}_2(\mathsf{nil},\mathsf{cons}(\mathsf{s}(\mathsf{s}(\mathsf{0})),\mathsf{nil}))],\bot) 
   \\
  & \to_{\SRTd^\to(\TRSqsort)} 
   \overline{\mathsf{split}}(\mathsf{s}(\mathsf{0}),\mathsf{cons}(\mathsf{0},
   \ldots
   ),[\{\mathsf{false}\},\ldots
],[\{\overline{\mathsf{split}}(\mathsf{s}(\mathsf{0}),\mathsf{cons}(\mathsf{s}(\mathsf{s}(\mathsf{0})),\mathsf{nil}),\bot,\bot)\},\bot]) 
   \\
  & \to^*_{\SRTd^\to(\TRSqsort)} 
   \overline{\mathsf{split}}(\mathsf{s}(\mathsf{0}),\mathsf{cons}(\mathsf{0},
   \ldots
   ),[\{\mathsf{false}\},\ldots
],[\{\mathsf{tp}_2(\mathsf{nil},\mathsf{cons}(\mathsf{s}(\mathsf{s}(\mathsf{0})),\mathsf{nil}))\},\bot])
   \\
  & \to_{\SRTd^\to(\TRSqsort)} 
   \overline{\mathsf{split}}(\mathsf{s}(\mathsf{0}),\mathsf{cons}(\mathsf{0},
   \ldots
   ),[\{\mathsf{false}\},\ldots
],[\{\overline{\mathsf{le}}(\mathsf{s}(\mathsf{0}),\mathsf{0})\},\mathsf{tp}_2(\mathsf{nil},\mathsf{cons}(\mathsf{s}(\mathsf{s}(\mathsf{0})),\mathsf{nil}))])
   \\
  & \to_{\SRTd^\to(\TRSqsort)} 
   \overline{\mathsf{split}}(\mathsf{s}(\mathsf{0}),\mathsf{cons}(\mathsf{0},
   \ldots
   ),[\{\mathsf{false}\},\ldots
],[\{\{\mathsf{false}\}\},\mathsf{tp}_2(\mathsf{nil},\mathsf{cons}(\mathsf{s}(\mathsf{s}(\mathsf{0})),\mathsf{nil}))])
   \\
  & \to_{\SRTd^\to(\TRSqsort)} 
   \overline{\mathsf{split}}(\mathsf{s}(\mathsf{0}),\mathsf{cons}(\mathsf{0},
   \ldots
   ),[\{\mathsf{false}\},\ldots
],[\{\mathsf{false}\},\mathsf{tp}_2(\mathsf{nil},\mathsf{cons}(\mathsf{s}(\mathsf{s}(\mathsf{0})),\mathsf{nil}))])
   \\
   & \to_{\SRTd^\to(\TRSqsort)} 
   \{\{\mathsf{tp}_2(\mathsf{cons}(\mathsf{0},\mathsf{nil}),\mathsf{cons}(\mathsf{s}(\mathsf{s}(\mathsf{0})),\mathsf{nil}))\}\}
   \\
  & \to_{\SRTd^\to(\TRSqsort)} 
   \{\mathsf{tp}_2(\mathsf{cons}(\mathsf{0},\mathsf{nil}),\mathsf{cons}(\mathsf{s}(\mathsf{s}(\mathsf{0})),\mathsf{nil}))\}
   \\
 \end{array}
\]
By applying the translation-back mapping $\,\widehat{\Blank}\,$ to 
 $\{\mathsf{tp}_2(\mathsf{cons}(\mathsf{0},\mathsf{nil}),\mathsf{cons}(\mathsf{s}(\mathsf{s}(\mathsf{0})),\mathsf{nil}))\}$,
 we obtain 
 $\mathsf{tp}_2(\mathsf{cons}(\mathsf{0},\mathsf{nil}),\mathsf{cons}(\mathsf{s}(\mathsf{s}(\mathsf{0})),\mathsf{nil}))$,
 a normal form of
 $\mathsf{split}(\mathsf{s}(\mathsf{0}),\mathsf{cons}(\mathsf{0},\mathsf{cons}(\mathsf{s}(\mathsf{s}(\mathsf{0})),\mathsf{nil})))$
 w.r.t.\ $\TRSqsort$.
\end{exa}

The SR transformation $\SRTd$ has the following properties.
\begin{thm}[\cite{SR06b}]
 \label{th:SRTd-properties}
Let $R$ be a strongly or syntactically DCTRS.
Then, all of the following hold:
\begin{enumerate}[$\bullet$]
 \item $\SRTd$ is sound for $R$ if $R$ is confluent~%
       \footnote{ In~\cite{SR06,SR06b}, soundness and completeness are
       discussed on ground reduction sequences only.
       In the proof of soundness and completeness, groundness of terms
       in derivations is only used with groundness in ``ground
       confluence''.  
       For this reason, confluence is a soundness condition for the 
       case of arbitrary reduction sequences. } 
       or\/ $\Uopt$-LL, 
 \item if $R$ is\/ $\Uopt$-LL and $\SRTd^\to(R)$ is confluent
       on reachable terms, then $R$ is confluent, and
 \item if $R$ is\/ $\Uopt$-LL and confluent, then
       $\SRTd^\to(R)$ is confluent on reachable terms.
\end{enumerate}
\end{thm}
We recognize from the second statement of
Theorem~\ref{th:SRTd-properties} that confluence of $\SRTd^\to(R)$ is a
sufficient condition for confluence of $R$. 
\begin{exa}
 \label{ex:SRTd-qsort-CR}
Consider the DCTRS $\TRSqsort$ and the transformed TRS
 $\SRTd(\TRSqsort)$ in Examples~\ref{ex:qsort},~\ref{ex:SRTd-qsort}
 again. 
 The DCTRS $\TRSqsort$ is operationally terminating since
 $\SRTd^\to(\TRSqsort)$ is terminating~\cite{SR06b}. 
 We have only a critical pair of $\TRSqsort$ between the second and
 third rules.
 The critical pair is infeasible since there exists no terms $s,t$\/ such
 that $\mathsf{le}(s,t)$ $\to^*_{\TRSqsort}$ $\mathsf{true}$ and
 $\mathsf{le}(s,t)$ $\to^*_{\TRSqsort}$ $\mathsf{false}$.  
 Thus, we can see that $\TRSqsort$ is confluent~\cite{ALS94} (\cf,~\cite{Ohl02}).
 Though, we have no formal method for proving confluence of
 $\TRSqsort$. 
 On the other hand, all the critical pairs of $\SRTd(\TRSqsort)$ are
 joinable and $\SRTd^\to(\TRSqsort)$ is terminating, and hence
 $\SRTd^\to(\TRSqsort)$ is confluent. 
 Due to Theorem~\ref{th:SRTd-properties}, confluence of
 $\SRTd^\to(\TRSqsort)$ guarantees confluence of $\TRSqsort$.

 Consider the unraveled TRS $\U(\TRSqsort)$:
\[
 \U(\TRSqsort) =
 \left\{
 \begin{array}{r@{\>}c@{\>}l}
  & \vdots \\
  \mathsf{split}(x,\mathsf{cons}(y,ys)) & \to &
   \Usplita(\mathsf{split}(x,ys),x,y,ys) \\
  \Usplita(\mathsf{tp}_2(zs_1,zs_2),x,y,ys) & \to & 
   \Usplitb(\mathsf{le}(x,y),x,y,ys,zs_1,zs_2) \\
  \Usplitb(\mathsf{true},x,y,ys,zs_1,zs_2) & \to & \mathsf{tp}_2(zs_1,\mathsf{cons}(y,zs_2)) \\
  \mathsf{split}(x,\mathsf{cons}(y,ys)) & \to &
   \Usplitc(\mathsf{split}(x,ys),x,y,ys) \\
  \Usplitc(\mathsf{tp}_2(zs_1,zs_2),x,y,ys) & \to & 
   \Usplitd(\mathsf{le}(x,y),x,y,ys,zs_1,zs_2) \\
  \Usplitd(\mathsf{false},x,y,ys,zs_1,zs_2) & \to & 
   \mathsf{tp}_2(\mathsf{cons}(y,zs_1),zs_2) \\
  & \vdots \\ 
 \end{array}
 \right\}
\]
 Unlike $\SRTd^\to(\TRSqsort)$, this unraveled TRS $\U(\TRSqsort)$ is not
 confluent since we have a {\em critical}\/ peak, \eg,
 $\mathsf{tp}_2(\mathsf{nil},\mathsf{cons}(\mathsf{0},\mathsf{nil}))$
 $\gets^*_{\U(\TRSqsort)}$
 $\mathsf{split}(\mathsf{0},\mathsf{cons}(\mathsf{0},\mathsf{nil}))$
 $\to^*_{\U(\TRSqsort)}$ 
 $\Usplitd(\mathsf{false},\mathsf{0},\mathsf{0},\mathsf{nil},\mathsf{nil},\mathsf{nil})$
 that is not joinable. 
 In this case, we can solve this non-confluence by replacing $\Usplitd$ with
 $\Usplitb$ since the only difference between the second and third rules
 of $\TRSqsort$ is whether $\mathsf{le}(x,y)$ reduces to $\mathsf{true}$ or $\mathsf{false}$.
 However, this simple solution is not possible in general.  
\end{exa}
\begin{exa}
 \label{ex:snoc}
Consider the following TRS defining $\mathsf{snoc}$ that appends the
 element to the end of the list, \eg, $\mathsf{snoc}([1,2,3],4)$ $=$
 $[1,2,3,4]$:
 \[
 \TRSsnoc = 
 \left\{
 \begin{array}{r@{\>}c@{\>}l}
  \mathsf{snoc}(\mathsf{nil},y) & \to & \mathsf{cons}(y,\mathsf{nil}) \\
  \mathsf{snoc}(\mathsf{cons}(x,xs),y) & \to &
   \mathsf{cons}(x,\mathsf{snoc}(xs,y)) \\
 \end{array}
 \right\}
 \]
The inversion method in~\cite{Nishida04phd} 
 inverts this TRS to the following DCTRS $\TRSinvsnoc$:
 \[
 \TRSinvsnoc = 
 \left\{
 \begin{array}{r@{\>}c@{\>}l}
  \mathsf{snoc}^{-1}(\mathsf{cons}(y,\mathsf{nil})) & \to &
   \mathsf{tp}_2(\mathsf{nil},y) \\ 
  \mathsf{snoc}^{-1}(\mathsf{cons}(x,ys)) & \to &
   \mathsf{tp}_2(\mathsf{cons}(x,xs),y) 
   \Leftarrow \mathsf{snoc}^{-1}(ys) \tto \mathsf{tp}_2(xs,y) \\
 \end{array}
 \right\}
 \]
This DCTRS $\TRSinvsnoc$ is unraveled by $\U$ as follows:
 \[
 \U(\TRSinvsnoc) = 
 \left\{
 \begin{array}{r@{\>}c@{\>}l}
  & \vdots \\
  \mathsf{snoc}^{-1}(\mathsf{cons}(x,ys)) & \to &
   \Uinvsnoc(\mathsf{snoc}^{-1}(ys),x,ys) \\
  \Uinvsnoc(\mathsf{tp}_2(xs,y),x,ys) & \to &
   \mathsf{tp}_2(\mathsf{cons}(x,xs),y) \\ 
 \end{array}
 \right\}
 \]
The DCTRS $\TRSinvsnoc$ is confluent, but the unraveled TRS
 $\U(\TRSinvsnoc)$ is not since we have a {\em critical}\/ peak
 $\Uinvsnoc(\mathsf{snoc}^{-1}(\mathsf{nil}),x,\mathsf{nil})$
 $\gets_{\U(\TRSinvsnoc)} \cdot \to_{\U(\TRSinvsnoc)}$
 $\mathsf{tp}_2(\mathsf{nil},x)$ that is not joinable. 
 The simple solution described in Example~\ref{ex:SRTd-qsort-CR} cannot
 solve non-confluence of $\U(\TRSinvsnoc)$.  
\end{exa}

Finally, we show some properties and a notion related to reachable terms
that are helpful to compare the SR transformation with unravelings.
\begin{defi}[\cite{SR06,SR06b}]
 \label{def:structural-positions}
Let $R$\/ be a DCTRS over a signature $\cF$.
For a reachable term $s$\/ in $\cT(\cF_{\SRTd(R)},\cV)$, we define the set
$\Pos_{\rm str}(s)$ of {\em structural positions}\/ for $s$\/ as follows:
\begin{enumerate}[$\bullet$]
 \item $\Pos_{\rm str}(x)$ $=$ $\{\varepsilon\}$ for $x$ $\in$ $\cV$,
 \item $\Pos_{\rm str}(\{t\})$ $=$ $\{ 1p \mid p \in \Pos_{\rm str}(t) \}$, 
 \item $\Pos_{\rm str}(c(t_1,\ldots,t_n))$ $=$ $\{ ip \mid 1 \leq i \leq
       n,~p \in \Pos_{\rm str}(t_i)\}$ for $c$ $\in$ $\Cc_R$, and
 \item $\Pos_{\rm str}(\overline{f}(t_1,\ldots,t_n,u_1,\ldots,u_{n_f}))$
       $=$ $\{\varepsilon\} \cup \{ ip \mid 1 \leq i \leq n, ~p \in
       \Pos_{\rm str}(t_i)\}$ 
       for an $n$-ary defined symbol $f$ $\in$ $\cD_R$.
\end{enumerate}
\end{defi}
\noindent
Note that $\Pos_{\rm str}$ is well-defined for reachable terms while it is
not defined for the symbols $\bot$ and $[\Blank]$.
\begin{exa}
Consider the following term related to $\SRTd^\to(\TRSqsort)$ in
 Example~\ref{ex:SRTd-qsort-CR}:
\[
 \{\overline{\mathsf{split}}(\mathsf{s}(\mathsf{0}),\mathsf{cons}(\mathsf{0},\mathsf{cons}(\mathsf{s}(\mathsf{s}(\mathsf{0})),\mathsf{nil})),[\{\overline{\mathsf{split}}(\mathsf{s}(\mathsf{0}),\mathsf{cons}(\mathsf{s}(\mathsf{s}(\mathsf{0})),\mathsf{nil}),\bot,\bot)\},\bot],\bot)\}
\]
The structural positions of this terms are 
 $1$, $1.1$, $1.1.1$, $1.2$, $1.2.1$, $1.2.2$, $1.2.2.1$, $1.2.2.1.1$,
 $1.2.2.1.1.1$, and $1.2.2.2$. 
\end{exa}
By definition, structural positions have the following property related to
contexts.
\begin{lem}
 \label{lem:structural-positions}
Let $R$ be a DCTRS over a signature $\cF$, $t$ be a term in
 $\cT(\cF_{\SRTd(R)},\cV)$, and $C[~]_p$ be a one-hole context over
 $\cF_{\SRTd(R)}$ such that $p$ $\in$ $\Pos_{\rm str}(C[~])$.
 Then, $C[\{t\}]$ $\to^*_{\SRTd^\to(R)}$ $\{C[t]\}$.
\end{lem}
\noindent
The proof of Lemma~\ref{lem:structural-positions} is omitted since it
can be easily proved by induction.

\subsection{Relationship between Soundness}
\label{subsec:soundness_of_SR}

In this subsection, we show that if $\SRTd$ is sound for a DCTRS, then so
is $\U$.
To this end, as in Section~\ref{sec:soundness_of_unravelings}, we show
that all the derivations of $\U$ on  terms over the original signature
are included in the derivations of $\SRTd$.

In rewrite rules obtained from $\SRTd$, the conditional parts related to
the same defined symbol are evaluated in parallel, and thus, the system
$\SRTd(R)$ is more reasonable than the system $\U(R)$.
Due to the parallel evaluation of conditional parts, $\SRTd(R)$ can
derive all the reduction sequences of $\U(R)$, and thus, soundness of $\SRTd$
implies that of $\U$.
\begin{lem}
\label{lem:SRTd-simulates-U}
Let $R$ be a DCTRS over a signature $\cF$.
 Then, $\phi_{\SRTd(R)}(\to^*_{\U(R)})$ $\subseteq$
 $\to^*_{\SRTd^\to(R)}$ on terms in $\cT(\cF,\cV)$.
\end{lem}
\proof
 The proof can be seen in
 Appendix~\ref{subsec:proof:lem:SRTd-simulates-U}.
\qed

Due to Lemma~\ref{lem:SRTd-simulates-U}, we obtain the following theorem.
\begin{thm}
 \label{th:SRTd-soundness}
 If\/ $\SRTd$ is sound for a syntactically or strongly DCTRS, then
 so is\/ $\U$. 
\end{thm}
\proof
Suppose that $\SRTd$ is sound for a syntactically or strongly DCTRS $R$.
Then, it follows from Lemma~\ref{lem:SRTd-simulates-U} that
$\phi_{\SRTd(R)}(\to^*_{\U(R)})$ $\subseteq$ $\to^*_{\SRTd^\to(R)}$.  
Therefore, it follows from Theorem~\ref{th:CTRS-transformations} that
 $\U$ is sound for $R$.
\qed
\noindent
It is not known whether the converse of Theorem~\ref{th:SRTd-soundness}
holds or not. 

Similarly to $\U$, the LL property of DCTRSs is not a soundness
condition of $\SRTd$;
Suppose that $\SRTd$ is sound for LL DCTRSs;
Then, it follows from Theorem~\ref{th:SRTd-soundness} that $\U$ is sound
for LL DCTRSs, but $\U$ is not sound for every LL DCTRS (see
Example~\ref{ex:marchiorill}).  
\begin{exa}\label{ex:SRTd-marchiorill}
The DCTRS $\TRSmarchiorill$ in Example~\ref{ex:marchiorill} is
 transformed by $\SRTd^\to$ 
 into the following TRS:
 \[
 \SRTd^\to(\TRSmarchiorill) = 
 \left\{
 \begin{array}{@{}r@{\>}c@{\>}l@{\hspace{5ex}}r@{\>}c@{\>}l@{}}
  \overline{\mathsf{f}}(x,\bot) & \to & \overline{\mathsf{f}}(x,[\{x\}]) 
   &
  \overline{\mathsf{f}}(x,[\{\mathsf{e}\}]) & \to & \{x\} \\
  \overline{\mathsf{g}}(\mathsf{d},x,y,\bot) & \to &
   \overline{\mathsf{g}}(\mathsf{d},x,y,[\{y\}]) 
   & 
  \overline{\mathsf{g}}(\mathsf{d},x,y,[\{x\}]) & \to & \{\mathsf{A}\} \\
  \overline{\mathsf{h}}(x,y,\bot) & \to &
   \overline{\mathsf{h}}(x,y,[\{y\}]) 
   &
  \overline{\mathsf{h}}(x,y,[\{x\}]) & \to & 
   \{\overline{\mathsf{g}}(x,y,\overline{\mathsf{f}}(\overline{\mathsf{k}},\bot),\bot)\}
   \\ 
  \overline{\mathsf{a}} & \to & \{\overline{\mathsf{c}}\} 
   &
  \overline{\mathsf{a}} & \to & \{\mathsf{d}\} \\
  \overline{\mathsf{b}} & \to & \{\overline{\mathsf{c}}\} 
   & 
  \overline{\mathsf{b}} & \to & \{\mathsf{d}\} \\
  \overline{\mathsf{c}} & \to & \{\mathsf{e}\} 
   &
  \overline{\mathsf{c}} & \to & \{\mathsf{l}\} \\
  \overline{\mathsf{d}} & \to & \{\mathsf{m}\} 
   &
  \overline{\mathsf{k}} & \to & \{\mathsf{l}\} \\
  \overline{\mathsf{k}} & \to & \{\mathsf{m}\} 
   &
  \{\{x\}\} & \to & \{x\} \\
  \overline{\mathsf{f}}(\{x\},z_1) & \to &
   \{\overline{\mathsf{f}}(x,\bot)\} 
   & 
  \overline{\mathsf{g}}(\{x\},y,z,z_1) & \to & \{\overline{\mathsf{g}}(x,y,z,\bot)\} \\
  \overline{\mathsf{g}}(x,\{y\},z,z_1) & \to &
   \{\overline{\mathsf{g}}(x,y,z,\bot)\} 
   &
  \overline{\mathsf{g}}(x,y,\{z\},z_1) & \to &
   \{\overline{\mathsf{g}}(x,y,z,\bot)\} \\
  \overline{\mathsf{h}}(\{x\},y,z_1) & \to &
   \{\overline{\mathsf{h}}(x,y,\bot)\} 
   & 
  \overline{\mathsf{h}}(x,\{y\},z_1) & \to & \{\overline{\mathsf{h}}(x,y,\bot)\} \\
 \end{array}
\right\}
\]
 We have the derivation $\overline{\mathsf{h}}(\overline{\mathsf{f}}(\overline{\mathsf{a}},\bot),\overline{\mathsf{f}}(\overline{\mathsf{a}},\bot))$
 $\to^*_{\SRTd^\to(\TRSmarchiorill)}$ $\{\mathsf{A}\}$, but
 $\mathsf{h}(\mathsf{f}(\mathsf{a}),\mathsf{f}(\mathsf{a}))$
 $\not\to_{\TRSmarchiorill}$ $\mathsf{A}$. 
Thus, the LL property is not a sufficient condition for soundness of
 $\SRTd$.
\end{exa}

\subsection{A Comparison from Several Viewpoints}
\label{subsec:comparison}

Finally, we compare unravelings with the SR transformation, in terms of
the following points.  

\begin{enumerate}[$\bullet$]
 \item{\it Proving Operational Termination.}
      Both the unravelings and the SR transformation can be used
      for proving operational termination:   
      if the transformed TRS is terminating, then the
      original CTRS is operationally
      terminating~\cite{LMM05,SR06,SR06b,SG07,SG10}.  

 \item{\it Soundness.}
      As shown in Theorem~\ref{th:SRTd-soundness}, for strongly
      or syntactically DCTRSs, soundness of 
      $\SRTd$ implies soundness of $\U$.
      The known soundness conditions are the $\Uopt$-LL property
      and confluence only.
      These conditions are also the ones for unravelings and
      more soundness conditions for unravelings are known 
      than those for $\SRTd$ (see Table~\ref{tbl:all-results} in
      Section~\ref{sec:summary_and_related-work}). 

 \item{\it Strong Soundness.}
      A CTRS transformation $T$\/ is called {\em strongly sound}\/
      for an eCTRS $R$\/ over a signature $\cF$\/ if there exists a
      (partial)~%
      \footnote{ The mapping $\psi$\/ only needs to translate
      resulting terms (terms reachable from $\phi(s)$ for some
      original term $s$) for $T(R)$ back into the corresponding
      terms for $R$. }  
      mapping $\psi$\/ as an inverse to $\phi$\/ (\ie,
      $\psi(\phi(t))$ $=$ $t$\/ for $t$ $\in$ $\cT(\cF,\cV)$)
      such that, for all terms $s$ $\in$ $\cT(\cF,\cV)$ and $t$
      $\in$ $\cT(\cG,\cV)$, $\phi(s)$ $\to^*_{R_T}$ $t$\/
      implies $s$ $\to^*_R$ $\psi(t)$,
      where $\cG$\/ is a signature over which $R_T$ is defined.
      The well-designed rules obtained by 
      the SR transformation provide strong soundness from soundness,
      that plays an important role in the points below.  
      On the other hand, strong soundness of unravelings has never been
      discussed, and soundness of unravelings does not imply strong
      soundness of the unravelings in general. 

 \item{\it Proving Confluence.}
      As stated in Theorem~\ref{th:SRTd-properties}, 
      the SR transformation 
      provides a method for proving confluence of strongly or
      syntactically $\U$-LL DCTRSs. 
      For unravelings, this has never been discussed, and furthermore,
      for any overlapping confluent DCTRS, usual unravelings (\eg, $\U$
      and $\Uopt$) do not preserve confluence, \ie, the unraveled TRS is
      not confluent (see Examples~\ref{ex:SRTd-qsort-CR},~\ref{ex:snoc}).

 \item{\it Computing Normal Forms.}
      For a strongly or syntactically DCTRS $R$, the normal forms of
      $\SRTd^\to(R)$ can be converted to the corresponding
      normal forms of $R$\/ if $\SRTd$ is strongly sound for $R$. 
      Thus, $\SRTd(R)$ can be used for the normalizing reduction of
      $R$.  
      Moreover, the obtained normal form is a unique one if $R$\/
      is confluent.  
      In general, this is impossible for unravelings. 
      
 \item{\it Computational Equivalence.}
      For a CTRS transformation $T$\/ and an eCTRS $R$, the
      transformed eTRS $R_T$ is called {\em computationally
      equivalent to $R$}\/ if, whenever $R$\/ terminates on $s$\/
      admitting a unique normal form $t$\/ (\ie, $t'$ $=$ $t$\/
      for all normal forms $t'$ of $s$), $R_T$ also terminates
      on $\phi(s)$ and for any of its normal forms $t'$, we have
      that $\psi(t')$ $=$ $t$~\cite{SR06,SR06b}.
      $\SRTd^\to(R)$ is computationally equivalent to $R$\/ if $R$\/
      is finite, confluent, and operationally terminating~\cite{SR06b}.  
      Thus, for such a DCTRS $R$, $\SRTd(R)$ can be used as 
      a rewriting engine 
      for $R$\/ in terms of reduction. 
      This is the main advantage of 
      $\SRTd$ and has never been discussed for unravelings. 
\end{enumerate}

In summary, 
when the SR transformation is sound for a strongly or syntactically
DCTRS with confluence and operational termination, the SR transformation
seems better to use as a reasonable rewriting engine for the DCTRS than
the unravelings mentioned in this paper.
On the other hand, unravelings are good tools for investigating
soundness conditions of CTRS transformations, which is required for
computational equivalence. 
Moreover, as stated in Section~\ref{sec:intro}, unravelings are useful
in order to analyze or modify DCTRSs.
Currently, for DCTRSs that are neither strongly nor syntactically
DCTRSs, unravelings are more useful than the SR transformation
since it is not known whether $\SRTd$ provides computational equivalence
(and even soundness) for such DCTRSs or not.

\section{Summary and Related Work of Soundness Conditions}
\label{sec:summary_and_related-work}

In this section, we briefly describe related work on soundness of
unravelings and we summarize positive and negative results on soundness
conditions of unravelings and the SR transformation.

First, we briefly describe a comparison with related work, in terms of the
approach to the proof of soundness related to the $\Uopt$-LL property
(Subsection~\ref{subsec:LL-soundness}).   
For an LL normal CTRS $R$ over a signature $\cF$, the approach
to the proof of soundness in~\cite{GGS10} is the use of the
transformation $\nabla$ from 
$\cT(\cF_{\Unl(R)},\cV)$ to $\cT(\cF,\cV)$, proving that 
for any term $s$ $\in$ $\cT(\cF,\cV)$ and term $t$ $\in$
$\cT(\cF_{\Unl(R)},\cV)$, if $s$ $\to^*_{\Unl(R)}$ $t$, then $s$ $\to^*_R$
$\nabla(t)$~\cite{Ohl02}.
Note that the transformation $\nabla$ has been extended to
$\U$~\cite{Ohl01} (\cf,~\cite{Ohl02}). 
The transformation $\nabla$ was introduced in~\cite{Ohl01} to
discuss innermost termination.
Unlike the case of normal CTRSs, however, $\nabla$ has never been used
to show soundness. 
The transformation $\nabla$ cannot be defined well for $\Uopt$ since not
all the variables in $l$\/ appear in
$\Usym^\rho_i(t_i,\overrightarrow{\AppearedLaterUsedVar_i})$. 
For this reason, the proof in this paper takes a direct approach to the
proof of soundness for $\Uopt$-LL DCTRSs (\cf, Lemma~\ref{lem:LL-soundness}). 

Extending the results in~\cite{GGS10}, Gmeiner et al.\ have shown that
$\U$ is sound for confluent and {\em right-stable}\/
 3-DCTRSs w.r.t.\ the reduction to
normal forms,%
\footnote{ A syntactically DCTRS $R$\/ is called {\em
right-stable}~\cite{SMI95,GGS12} if for every rule $l \to r \Leftarrow s_1
\tto t_1;\ldots;s_k \tto t_k$ and for all $1$ $\leq$ $i$ $\leq$ $n$,
$t_i$ is linear and $\AppearedVar_i \cap \Var(t_i)$ $=$ $\emptyset$.}
 and $\U$ is sound for $\U$-RL or WLL 3-DCTRSs~\cite{GGS12}.%
\footnote{ A 3-DCTRS $R$\/ is called {\em weakly left-linear}\/
(WLL)~\cite{GGS12} if
for every rule $\rho: l \to r \Leftarrow s_1 \tto t_1; \ldots; s_k \tto t_k$
$\in$ $R$\/ and all variables $x$ $\in$ $\Var(\rho)$, 
$x$\/ does not appear in any of $r,s_1,\ldots,s_k$ whenever $x$\/
appears at least twice in $l,t_1,\ldots,t_k$.
Note that this WLL property for 3-DCTRSs is an extension of the WLL
property for normal 1-CTRSs.
}
For the case of $\U$-RL 3-DCTRSs, this result is
incompatible with Theorem~\ref{th:RLNE-soundness} since $\U$-RL is
strictly more restrictive than $\Uopt$-RL.
For example, the DCTRS $\TRSinvquad'$ is $\Uopt$-RLNE, but not $\U$-RL.
This indicates that soundness of $\U$ for $\TRSinvquad'$ cannot be
proved by using the result in~\cite{GGS12}, while the soundness can be
proved by the results in this paper (see
Example~\ref{ex:invquad-soundness}).  
For the case of WLL 3-DCTRS, this result strictly contains the
combination of Theorem~\ref{th:LL-soundness} and
Corollary~\ref{cor:U-soundness_from_Uopt} since $\Uopt$-LL 3-DCTRSs are
WLL. 
On the other hand, the WLL property is not a soundness condition of
$\Uopt$ since $\Uopt$ is not sound for the WLL DCTRS
$\TRSfeh$ shown in Example~\ref{ex:feh}.
Furthermore, Gmeiner et al.\ have also shown that
$\Uopt$ is sound for $\Uopt$-NE and {\em right-separated}\/
 2-DCTRSs~\cite{GGS12}.%
\footnote{ A DCTRS $R$\/ is called {\em right-separated}~\cite{GGS12} if
for every rule $l \to r \Leftarrow s_1 \tto t_1;\ldots;s_k \tto t_k$ and all
$1$ $\leq$ $i$ $\leq$ $k$, $\Var(t_i)\cap\AppearedVar_i$ $=$
$\emptyset$.}
The $\Uopt$-RL property is incompatible with the right-separated
property even if the DCTRSs are $\Uopt$-NE and of Type~2.%
\footnote{ The rule $\mathsf{f}(x,y,z) \to x \Leftarrow \mathsf{g}(y) \tto
\mathsf{w};~\mathsf{g}(z) \tto y;~ \mathsf{h}(y) \tto \mathsf{a}$
is $\Uopt$-RLNE and of Type~2, but not right-separated.
On the other hand, the rule $\mathsf{f}(x,y) \to x \Leftarrow
\mathsf{g}(y) \tto z;~ \mathsf{g}(z) \tto \mathsf{a};~\mathsf{h}(z) \tto
\mathsf{b}$ is $\Uopt$-NE, right-separated, and of Type~2, but not
$\Uopt$-RLNE. }
For this reason, the soundness result on $\Uopt$ in~\cite{GGS12} is
incomparable with Theorem~\ref{th:RLNE-soundness}.

Finally, we summarize the positive and negative results on soundness in
Table~\ref{tbl:all-results}, \ie, 
sufficient and insufficient conditions for soundness of the CTRS
transformations mentioned in this paper.
We can recognize from Example~\ref{ex:feh} that neither confluence nor
$\Uopt$-confluence is sufficient on its own for soundness of $\Uopt$.
As we have seen, soundness of the unraveling $\U$ is provided by the
other transformations $\Ujl$, $\Unl$, $\Uopt$, and $\SRTd$.
In summary, many sufficient and insufficient conditions for soundness of
unravelings are investigated and all the soundness conditions of
$\SRTd$ are soundness conditions of unravelings. 
 \begin{table}[ptb]
  \caption{Results on soundness conditions for CTRS transformations.}
  \label{tbl:all-results}

  \smallskip
  \begin{tabular}{|@{\,}r|c|c||c|c|}
   \hline
   \multicolumn{3}{|c||}{} & soundness & insufficient for soundness \\
   \hline
   \hline
   \multicolumn{2}{|c|}{J}
   &
   $\Ujl$ 
   & 
       \begin{tabular}{@{}c@{}}
	LL (Corollary~\ref{cor:Ujl-soundness}) \\
	soundness of $\Unl\circ\Tjn$ (Theorem~\ref{th:UnTjn-simulation}) \\
	soundness of $\Unl$ (Theorem~\ref{th:Uj-soundness_by_Un}) \\
	soundness of $\U \circ \Tjd$ (Theorem~\ref{th:Uj-soundness_from_U-Tjd}) \\
       \end{tabular}       
	   &
	   \\
   \hline
   &
       \raisebox{-2mm}[0mm][0mm]{N}
   &
       \begin{tabular}{@{}l@{}}
	$\Unl$ \\
       \end{tabular}
   &
       \begin{tabular}{@{}c@{}}
	LL (= $\Uopt$-LL)~\cite{GGS10} \\
	confluence~\cite{GGS10} \\
	NE~\cite{GGS10} \\
	groundness of all conditions~\cite{GGS10} \\
	WLL ($\supset$ LL)~\cite{GGS10} \\
	soundness of $\Ujl$ (Theorem~\ref{th:Un-soundness_by_Uj}) \\
       \end{tabular}
   &
       \begin{tabular}{@{}c@{}}
	constructor systems~\cite{GGS10} \\
	overlay systems~\cite{GGS10} \\
	non-RV~\cite{GGS10} \\
	RL~\cite{GGS10} \\
	overlappingness~\cite{GGS10} \\
	UN~\cite{GGS10} \\
	UN${}^{\to}$~\cite{GGS10} \\
       \end{tabular}
       \\
   \cline{2-5}
   &
   \raisebox{-3mm}[0mm][0mm]{S}
   &
   $\SRTd$ 
	   &
       \begin{tabular}{@{}c@{}}
	$\Uopt$-LL~\cite{SR06b} \\
	confluence~\cite{SR06b} \\
       \end{tabular}
       &
       \begin{tabular}{@{}c@{}}
	LL (Example~\ref{ex:SRTd-marchiorill}) \\
       \end{tabular}
   \\
   \cline{3-5}
   &    \raisebox{-20mm}[0mm][0mm]{D}
       & \raisebox{-15mm}[0mm][0mm]{$\U$}
	   &
       confluence and right-stability~\cite{GGS12}
       & 
	   confluence~\cite{GGS12} \\       
   \cline{2-2}\cline{4-5}
   \multicolumn{2}{|c|}{}
   &
   &
       \begin{tabular}{@{}c@{}}
	$\Uopt$-LL~\cite{Mar97} 
	\\
	soundness of $\Uopt$ (Corollary~\ref{cor:U-soundness_from_Uopt})
	\\
	soundness of $\Unl$ (Theorem~\ref{th:soundness_by_Un-full})
	\\
	soundness of $\SRTd$ (Theorem~\ref{th:SRTd-soundness}) \\
	$\U$-RL~\cite{GGS12} \\
	WLL ($\supset$ $\Uopt$-LL)~\cite{GGS12} \\
       \end{tabular}
       &
       \begin{tabular}{@{}c@{}}
	confluence~\cite{GGS12} \\
	$\U$-NE~\cite{GGS12} \\
       \end{tabular}
   \\
   \cline{3-5}
   \multicolumn{2}{|c|}{}
   &
   $\Uopt$ 
   &
       \begin{tabular}{@{}c@{}}
	$\Uopt$-LL (Theorem~\ref{th:LL-soundness})
	\\
	$\Uopt$-NE-RL (Theorem~\ref{th:RLNE-soundness})
	\\
	$\Uopt$-NE, right-separation, and Type 2~\cite{GGS12} \\
       \end{tabular}
       &
       \begin{tabular}{@{}c@{}}
	LL (Example~\ref{ex:marchiorill}) \\
	$\Uopt$-NE (Example~\ref{ex:NE-insufficient}) 
	\\
	WLL (Example~\ref{ex:feh}) \\
	$\Uopt$-RL (Example~\ref{ex:feh})
	\\
	soundness of $\U$ (Example~\ref{ex:feh}) 
	\\
	confluence (Example~\ref{ex:feh}) \\
	$\Uopt$-confluence (Example~\ref{ex:feh}) \\
       \end{tabular}
   \\
   \hline
  \end{tabular}
{\small
\begin{enumerate}[--]
 \item ``J'', ``N'', ``D'', and ``S'' in the first column represent
       ``join CTRSs'', ``normal CTRSs'', ``DCTRSs'', and ``strongly or
       syntactically DCTRSs'', respectively.
 \item ``soundness of $\Unl$'' means that the target is (or can be
       considered) a normal CTRS and $\Unl$
       is sound for the target.
 \item ``soundness of $\SRTd$'' means that the target is a strongly or
       syntactically DCTRS and $\SRTd$ is sound for the target.
\end{enumerate}
  }
 \end{table}

\section{Conclusion}
\label{sec:conclusion}

In this paper, we showed that the optimized unraveling for DCTRSs is
sound for ultra-LL or ultra-RLNE DCTRSs, and showed
that if the optimized unraveling is sound for a DCTRS, then so is
Ohlebusch's unraveling. 
We also presented necessary and sufficient syntactic conditions for
ultra-LL, ultra-RL, and ultra-NE, respectively, and soundness conditions of
unravelings for join and normal CTRSs.
Moreover, we showed that
soundness of the existing unravelings and the SR transformation
respectively implies soundness of Ohlebusch's unraveling. 

Our future work is to solve the remaining open problems, \eg, either to
show soundness of $\U$ and $\Uopt$ for $\TRSmarchiori'$ in
Example~\ref{ex:open1}, 
or to prove the converse of Theorem~\ref{th:soundness_by_Un-full}.
We are also interested in 
a study on strong soundness and
computational equivalence of unravelings. 

There seems to be room for discussing sufficient conditions of
unravelings related to confluence, \eg, under which confluence of the
unraveled TRSs implies that of the original CTRSs, or under which
confluence of the original CTRSs implies that of the unraveled TRSs. 
For a confluent DCTRS $R$, a trivial such condition is that $R$\/
is $\Uopt$-LL and non-overlapping, \ie, $\U(R)$ and $\Uopt(R)$ are LL
and non-overlapping.
In many cases, however, neither $\U(R)$ nor $\Uopt(R)$ is confluent even
if $R$\/ is confluent (see Example~\ref{ex:SRTd-qsort-CR}). 
Viewed in this light, an interesting further direction related to
confluence will be to improve unraveling transformations themselves,
\eg, to optimize introduction of U symbols as stated in
Example~\ref{ex:SRTd-qsort-CR}. 
Such an optimization has been already discussed in~\cite{SR06,SR06b}. 
For unravelings, however, it is not clear what the optimization leads
to. 
What has to be noticed in this direction is that the improvement is in
agreement with the SR transformation. 

As stated in the comparison with the SR transformation, 
soundness conditions of unravelings are better studied than soundness of
the SR transformation and it must be easier to investigate
soundness of unravelings than that of the SR transformation.
Thus, it is still worth investigating unravelings while the SR
transformation provides a reasonable rewriting engine in terms of
computational equivalence to the original CTRSs.
On the other hand, if the converse of Theorem~\ref{th:SRTd-soundness}
holds, then unravelings would be useful tools to show soundness of the
SR transformation.  
A further direction of this research will be to prove or disprove the
converse of Theorem~\ref{th:SRTd-soundness}. 

\section*{Acknowledgements}

We would like to thank the anonymous reviewers for their kind and useful
comments to improve this paper, especially the remark in
Footnote~\ref{fnt:unraveling-definition}. 



\appendix

\section{Proofs of Technical Results}

In this appendix, we show missing proofs of some technical results.

\subsection{Proof of Theorem~\ref{th:Uopt-P}}\ \\
\label{subsec:proof:th:Uopt-P}

\begin{againtheorem}[Theorem~\ref{th:Uopt-P}]
 Let $\rho: l \to r \Leftarrow s_1 \tto t_1; \ldots; s_k \tto t_k$ be an
 extended deterministic conditional rewrite rule.
 Then, all of the following hold:
\begin{enumerate}[\em(1)]
 \item $\rho$ is $\Uopt$-LL iff all of $l,t_1,\ldots,t_k$ are linear and
       $\Var(t_i) \cap \AppearedVar_i$ $=$ $\emptyset$ for
       all\/ $1$ $\leq$ $i$ $\leq$ $k$, 
 \item $\rho$ is $\Uopt$-RL iff all of $r,s_1,\ldots,s_k$ are linear and
       $\Var(s_i) \cap \LaterUsedVar_i$ $=$
       $\emptyset$ for all\/ $1$ $\leq$ $i$ $\leq$ $k$, and   
 \item $\rho$ is $\Uopt$-NE iff\/
       $\Var(l)$ $\subseteq$ $\Var(r,s_1,\ldots,s_k)$ and
       $\Var(t_i)$ $\subseteq$ $\Var(r,s_{i+1},\ldots,s_k)$
       for all\/ $1$ $\leq$ $i$ $\leq$ $k$.
\end{enumerate}
\end{againtheorem}

\proof
The case that $\rho$\/ is unconditional is trivial, so let $k$ $>$ $0$.
Recall that $\Uopt(\rho)$ $=$ $\{\> l \to
   \Usym^\rho_1(s_1,\overrightarrow{\AppearedLaterUsedVar_1}),
   ~
   \Usym^\rho_1(t_1,\overrightarrow{\AppearedLaterUsedVar_1}) \to 
   \Usym^\rho_2(s_2,\overrightarrow{\AppearedLaterUsedVar_2}), 
   ~\ldots,~
   \Usym^\rho_k(t_k,\overrightarrow{\AppearedLaterUsedVar_k}) \to r \>\}$.
\begin{enumerate}[(1)]
 \item Suppose that $\rho$\/ is $\Uopt$-LL.
       Then, by definition, 
       all of 
       $l,\Usym^\rho_1(t_1,\overrightarrow{\AppearedLaterUsedVar_1}),\ldots,\Usym^\rho_k(t_k,\overrightarrow{\AppearedLaterUsedVar_k})$
       are linear. 
       Thus, all of $l,t_1,\ldots,t_k$ are linear and $\Var(t_i) \cap
       \AppearedLaterUsedVar_i$ $=$ $\emptyset$ for all $1$ $\leq$ $i$ $\leq$
       $k$, and hence $\Var(t_i) \cap 
       \AppearedVar_i$ 
       $=$ $\emptyset$ for all $1$ $\leq$ $i$ $\leq$ $k$. 
       Therefore, the {\it only-if}\/ part holds.
 
       Suppose that $\rho$\/ is not $\Uopt$-LL, all of $l,t_1,\ldots,t_k$
       are linear, and $\Var(t_i) \cap \AppearedVar_i$ $=$ $\emptyset$
       for all $1$ $\leq$ $i$ $\leq$ $k$.  
       Then, $\Var(t_j) \cap \AppearedLaterUsedVar_j$ $\ne$
       $\emptyset$ for some $j$\/ since all of $l,t_1,\ldots,t_k$ are
       linear and the sequence
       $\overrightarrow{\AppearedLaterUsedVar_i}$ is linear w.r.t.\
       variable occurrences for all $1$ $\leq$ $i$ $\leq$ $k$.  
       Since $\AppearedLaterUsedVar_i$ $\subseteq$
       $\AppearedVar_i$ for all $1$ $\leq$ $i$ $\leq$ $k$\/ by definition,
       we have $\Var(t_j) \cap \AppearedVar_j$ $\ne$ $\emptyset$.
       This contradicts the assumption that 
       $\Var(t_i) \cap \AppearedVar_i$ $=$ $\emptyset$
       for all $1$ $\leq$ $i$ $\leq$ $k$. 
       Therefore, the {\it if}\/ part holds.
 \item Suppose that $\rho$\/ is $\Uopt$-RL.
       Then, by definition,
       all of
       $r,\Usym^\rho_1(s_1,\overrightarrow{\AppearedLaterUsedVar_1}),\ldots,\Usym^\rho_k(s_k,\overrightarrow{\AppearedLaterUsedVar_k})$
       are linear and $\Var(s_i) \cap \AppearedLaterUsedVar_i$ $=$
       $\emptyset$ for all $1$ $\leq$ $i$ $\leq$ $k$, and hence
       all of $r,s_1,\ldots,s_k$ are linear. 
       Since $\rho$\/ is deterministic, we have $\Var(s_i)$ $\subseteq$
       $\AppearedVar_i$, and hence $\Var(s_i) \cap \LaterUsedVar_i$ $=$
       $\emptyset$ for all $1$ $\leq$ $i$ $\leq$ $k$.
       Therefore, the {\it only-if}\/ part holds.

       Suppose that $\rho$ is not $\Uopt$-RL, all of $r,s_1,\ldots,s_k$ are
       linear, and $\Var(s_i) \cap \LaterUsedVar_i$ $=$
       $\emptyset$ for all $1$ $\leq$ $i$ $\leq$ $k$.
       Then, by definition,
       $\Var(s_j) \cap \AppearedLaterUsedVar_j$ $\ne$
       $\emptyset$ for some $j$\/ since all of $r,s_1,\ldots,s_k$ are
       linear and the variable sequence
       $\overrightarrow{\AppearedLaterUsedVar_i}$ is linear w.r.t.\
       variable occurrences for all $1$ $\leq$ $i$ $\leq$ $k$.  
       Since $\AppearedLaterUsedVar_i$ $\subseteq$ $\LaterUsedVar_i$ for
       all $1$ $\leq$ $i$ $\leq$ $k$\/ by definition, it follows from
       $\AppearedLaterUsedVar_j$ $\subseteq$ $\LaterUsedVar_j$ that $\Var(s_j) \cap 
       \LaterUsedVar_j$ $\ne$ $\emptyset$. 
       This contradicts the assumption that $\Var(s_i) \cap \LaterUsedVar_i$ $=$
       $\emptyset$ for all $1$ $\leq$ $i$ $\leq$ $k$.
       Therefore, the {\it if}\/ part holds.
 \item Suppose that $\rho$\/ is $\Uopt$-NE.
       Then, by definition,  $\Var(l)$ $\subseteq$
       $\Var(s_1) \cup \AppearedLaterUsedVar_1$, $\Var(t_i) \cup \AppearedLaterUsedVar_i$ $\subseteq$
       $\Var(s_{i+1}) \cup \AppearedLaterUsedVar_{i+1}$ for all $1$ $\leq$ $i$ $<$ $k$, and
       $\Var(t_k) \cup \AppearedLaterUsedVar_k$ $\subseteq$ $\Var(r)$, and hence
       $\Var(t_k)$ $\subseteq$ $\Var(r)$, $\AppearedLaterUsedVar_k$ $\subseteq$ $\Var(r)$, 
       $\Var(t_i)$ $\subseteq$ $\Var(s_{i+1}) \cup \AppearedLaterUsedVar_{i+1}$ and $\AppearedLaterUsedVar_i$
       $\subseteq$ $\Var(s_{i+1}) \cup \AppearedLaterUsedVar_{i+1}$ for all $1$ $\leq$ $i$
       $<$ $k$. 
       Thus, $\AppearedLaterUsedVar_i$ $\subseteq$ $\Var(s_{i+1}) \cup
       \AppearedLaterUsedVar_{i+1}$ $\subseteq$ $\Var(s_{i+1},s_{i+2})
       \cup \AppearedLaterUsedVar_{i+2}$ $\subseteq \cdots \subseteq$
       $\Var(s_{i+1},\ldots,s_k) \cup \Var(r)$ for all $1$ $\leq$ $i$
       $<$ $k$, and hence $\Var(t_i)$ $\subseteq$
       $\Var(r,s_{i+1},\ldots,s_k)$ for all $1$ $\leq$ $i$ $\leq$ $k$.  
       Moreover, $\Var(l)$ $\subseteq$ $\Var(s_1) \cup
       \AppearedLaterUsedVar_1$ $\subseteq$ $\Var(s_1) \cup
       \Var(r,s_2,\ldots,s_k)$ $=$ $\Var(r,s_1,\ldots,s_k)$. 
       Therefore, the {\it only-if}\/ part holds.

       Suppose that $\Var(l)$ $\subseteq$ $\Var(r,s_1,\ldots,s_k)$ and 
       $\Var(t_i)$ $\subseteq$ $\Var(r,s_{i+1},\ldots,s_k)$ for all $1$
       $\leq$ $i$ $\leq$ $k$.
       Then, by the definition of $\LaterUsedVar_i$, we have
       $\LaterUsedVar_i$ $=$ $\Var(r,s_{i+1},\ldots,s_k)$ for all $1$
       $\leq$ $i$ $\leq$ $k$.  
       \begin{enumerate}[$\bullet$]
	\item Consider the rule
	      $\Usym^\rho_k(t_k,\overrightarrow{\AppearedLaterUsedVar_k})
	      \to r$ $\in$ $\Uopt(\rho)$. 
	      It follows from $\Var(t_k)$ $\subseteq$ $\Var(r)$ and
	      $\LaterUsedVar_k$ $=$ $\Var(r)$ that $\Var(t_k) \cup
	      \AppearedLaterUsedVar_k$ $=$ $\Var(t_k) \cup
	      (\AppearedVar_k \cap \LaterUsedVar_k)$ $\subseteq$
	      $\Var(r)$. 
	      Thus,
	      $\Usym^\rho_k(t_k,\overrightarrow{\AppearedLaterUsedVar_k})
	      \to r$ $\in$ $\Uopt(\rho)$ is NE.  
	\item Consider the rule
	      $\Usym^\rho_i(t_i,\overrightarrow{\AppearedLaterUsedVar_i})
	      \to
	      \Usym^\rho_{i+1}(s_{i+1},\overrightarrow{\AppearedLaterUsedVar_{i+1}})$
	      $\in$ $\Uopt(\rho)$ with $1$ $\leq$ $i$ $<$ $k$.
	      Suppose that $\Var(t_i) \cup
	      \AppearedLaterUsedVar_i$ $\not\subseteq$ $\Var(s_{i+1})
	      \cup \AppearedLaterUsedVar_{i+1}$.
	      Then, there exists a variable $x$ $\in$ $\Var(t_i) \cup
	      \AppearedLaterUsedVar_i$ such that $x$ $\not\in$
	      $\Var(s_{i+1}) \cup \AppearedLaterUsedVar_{i+1}$, and
	      hence $x$ $\not\in$ $\Var(s_{i+1}) \cup \LaterUsedVar_{i+1}$.  
	      It follows from $\LaterUsedVar_{i+1}$ $=$
	      $\Var(r,s_{i+2},\ldots,s_k)$ that $x$ $\not\in$
	      $\Var(r,s_{i+1},\ldots,s_k)$. 
	      \begin{enumerate}[--]
	       \item Suppose that $x$ $\in$ $\Var(t_i)$.
		     Then, it follows from $\Var(t_i)$ $\subseteq$
		     $\Var(r,s_{i+1},\ldots,s_k)$ that $x$ $\in$
		     $\Var(r,s_{i+1},\ldots,s_k)$. 
		     This contradicts the fact that $x$ $\not\in$
		     $\Var(r,s_{i+1},\ldots,s_k)$. 
	       \item Suppose that $x$ $\not\in$ $\Var(t_i)$.
		     Then, $x$ $\in$ $\AppearedLaterUsedVar_i$ $=$
		     $\AppearedVar_i \cap \LaterUsedVar_i$, and hence
		     $x$ $\in$ $\AppearedVar_i$ and $x$ $\in$
		     $\LaterUsedVar_i$ $=$ $\Var(r,s_{i+1},\ldots,s_k) $.
		     This contradicts the fact that $x$ $\not\in$
		     $\Var(r,s_{i+1},\ldots,s_k)$. 		     
	      \end{enumerate}
	      Thus, $\Usym^\rho_i(t_i,\overrightarrow{\AppearedLaterUsedVar_i}) \to
	      \Usym^\rho_{i+1}(s_{i+1},\overrightarrow{\AppearedLaterUsedVar_{i+1}})$
	      is NE.
	\item Consider the remaining rule $l \to
	      \Usym^\rho_1(s_1,\overrightarrow{\AppearedLaterUsedVar_1}$
	      $\in$ $\Uopt(\rho)$.
	      Suppose that $\Var(l)$ $\not\subseteq$ $\Var(s_1) \cup
	      \AppearedLaterUsedVar_1)$. 
	      Then, there exists a variable $x$ $\in$ $\Var(l)$
	      such that $x$ $\not\in$ $\Var(s_1) \cup
	      \AppearedLaterUsedVar_1$, and hence $x$ $\not\in$
	      $\Var(s_1) \cup \LaterUsedVar_1$. 
	      It follows from $\LaterUsedVar_1$ $=$ $\Var(r,s_2\ldots,s_k)$ that $x$
	      $\not\in$ $\Var(r,s_1,\ldots,s_k)$.
	      This contradicts the fact that $\Var(l)$ $\subseteq$
	      $\Var(r,s_1,\ldots,s_k)$.
	      Thus, $l \to
	      \Usym^\rho_1(s_1,\overrightarrow{\AppearedLaterUsedVar_1})$
	      is NE.
       \end{enumerate}
       Therefore, $\Uopt(\rho)$ is NE, and hence the {\it if}\/ part holds.
\qed
\end{enumerate}

\subsection{Proof of Theorem~\ref{th:U-P}}\ \\
\label{subsec:proof:th:U-P}

\begin{againtheorem}[Theorem~\ref{th:U-P}]
Let $\rho: l \to r \Leftarrow s_1 \tto t_1; \ldots; s_k \tto t_k$ be an
 extended deterministic conditional rewrite rule.
Then, all of the following hold:
\begin{enumerate}[\em(1)]
 \item $\rho$ is $\U$-LL iff all of $l,t_1,\ldots,t_k$ are linear and
       $\Var(t_i) \cap \AppearedVar_i$ $=$ $\emptyset$ for all\/
       $1$ $\leq$ $i$ $\leq$ $k$, 
 \item $\rho$ is $\U$-RL iff\/ $r$ is linear and all of
       $s_1,\ldots,s_k$ are ground, and
 \item $\rho$ is $\U$-NE iff\/
       $\Var(l,t_1,\ldots,t_k)$ $\subseteq$ $\Var(r)$.
\end{enumerate}
\end{againtheorem}

\proof
The case that $\rho$\/ is unconditional is trivial, so let $k$ $>$ $0$.
Recall that $\U(\rho)$ $=$ $\{\> l \to
   \Usym^\rho_1(s_1,\overrightarrow{\AppearedVar_1}),
   ~
   \Usym^\rho_1(t_1,\overrightarrow{\AppearedVar_1}) \to 
   \Usym^\rho_2(s_2,\overrightarrow{\AppearedVar_2}),
   ~\ldots, ~
   \Usym^\rho_k(t_k,\overrightarrow{\AppearedVar_k}) \to r \>\}$.
\begin{enumerate}[(1)]
 \item This claim can be proved similarly to Theorem~\ref{th:Uopt-P} (1).
 \item Suppose that $\rho$\/ is $\U$-RL.
       Then, by definition, all of 
       $r,\Usym^\rho_1(s_1,\overrightarrow{\AppearedVar_1}),\ldots,\Usym^\rho_k(s_k,\overrightarrow{\AppearedVar_k})$   
       are linear.
       Suppose that $s_j$ is not ground for some $j$.
       Then, there exists a variable $x$ $\in$ $\Var(s_j)$.
       Since $\rho$\/ is deterministic, $x$\/ appears in
       any of $l,t_1,\ldots,t_{j-1}$, and hence $x$ $\in$
       $\AppearedVar_j$.
       Thus, $\Usym^\rho_i(s_i,\overrightarrow{\AppearedVar_j})$ is not
       linear, and hence $\U(\rho)$ is not RL, \ie, $\rho$\/ is not $\U$-RL.
       This contradicts the assumption that $\rho$\/ is $\U$-RL.
       Therefore, all of $s_1,\ldots,s_k$ are ground, and hence the {\it
       only-if} part holds. 

       Suppose that $\rho$\/ is not $\U$-RL, $r$\/ is linear, and all of
       $s_1,\ldots,s_k$ are ground. 
       Then, by definition, 
       $\Usym^\rho_j(s_j,\overrightarrow{\AppearedVar_j})$ is not linear
       for some $j$\/ since $r$\/ is linear. 
       It follows from groundness of $s_j$ that
       $\Var(s_j) \cap \AppearedVar_j$ $=$ $\emptyset$.
       Moreover, since the variable sequence
       $\overrightarrow{\AppearedVar_j}$ is linear w.r.t.\ variable
       occurrences, the term
       $\Usym^\rho_j(s_j,\overrightarrow{\AppearedVar_j})$ is linear.
       This contradicts the non-linearity of
       $\Usym^\rho_j(s_j,\overrightarrow{\AppearedVar_j})$. 
       Therefore, the {\it if}\/ part holds.
 \item Suppose that $\rho$\/ is $\U$-NE.
       Then, by definition, the rule
       $\Usym^\rho_k(t_k,\overrightarrow{\AppearedVar_k}) \to r$\/ is NE,
       and hence $\Var(l,t_1,\ldots,t_k)$ $=$ $\Var(t_k) \cup \AppearedVar_k$
       $\subseteq$ $\Var(r)$. 
       Therefore, the {\it only-if}\/ part holds.
       
       Suppose that $\Var(l,t_1,\ldots,t_n)$ $\subseteq$ $\Var(r)$.
       Then, since $\AppearedVar_k$ $=$ $\Var(l,t_1,\ldots,t_{k-1})$ by
       definition,  $\Usym^\rho_k(t_k,\overrightarrow{\AppearedVar_k})
       \to r$\/ is NE. 
       It follows from $\Var(l)$ $=$ $\AppearedVar_1$ that 
       $l \to \Usym^\rho_1(s_1,\overrightarrow{\AppearedVar_1})$ is NE. 
       Since $\Var(t_i) \cup \AppearedVar_i$ $=$ $\AppearedVar_{i+1}$ for
       all $1$ $\leq$ $i$ $<$ $k$\/ by definition, 
       $\Usym^\rho_i(t_i,\overrightarrow{\AppearedVar_i}) \to
       \Usym^\rho_{i+1}(s_{i+1},\overrightarrow{\AppearedVar_{i+1}})$ is
       NE for all $1$ $\leq$ $i$ $<$ $k$. 
       Thus, $\U(\rho)$ is NE, and hence $\rho$\/ is $\U$-NE. 
       Therefore, the {\it if}\/ part holds.
\qed
\end{enumerate}

\subsection{Proof of Lemma~\ref{lem:LL-soundness}}
\label{subsec:proof:lem:LL-soundness}

We first prepare a technical lemma to help us to prove
Lemma~\ref{lem:LL-soundness}. 
Let $X$\/ be a finite set of variables, $\sigma$\/ and $\theta$\/ be
substitutions, and $\to$ be a binary relation on terms.
Then, we write $X\sigma$ $\to$ $X\theta$ if $x\sigma$ $\to$ $x\theta$
for any $x$ $\in$ $X$.
\begin{lem}
\label{lem:Uopt-LL}
Let $R$ be an eDCTRS, $\rho: l \to r \Leftarrow s_1 \tto t_1; \ldots;
 s_k \tto t_k$ be a $\Uopt$-LL conditional rewrite rule in $R$,
 and $\sigma_1,\ldots,\sigma_{k+1}$ be substitutions. 
 If $s_i\sigma_i$ $\to^*_R$ $t_i\sigma_{i+1}$ and
 $\AppearedLaterUsedVar_i\sigma_i$ $\to^*_R$
 $\AppearedLaterUsedVar_i\sigma_{i+1}$ for all $1$ $\leq$ $i$ $\leq$
 $k$, then $l\sigma_1$ $\to^+_R$ $r\sigma_{k+1}$.
\end{lem}
\proof
Let $\sigma$\/ be the substitution $\sigma_1|_{\Var(l)} \cup
 \sigma_2|_{\AppearedLaterUsedVar_1\setminus \Var(l)} \cup \cdots \cup \sigma_k|_{\AppearedLaterUsedVar_k
 \setminus \AppearedLaterUsedVar_{k-1}} \cup \sigma_{k+1}|_{\Var(t_k,r) \setminus \AppearedLaterUsedVar_k}$.
Then, $l\sigma$ $=$ $l\sigma_1$.
It follows from  $\AppearedLaterUsedVar_i\sigma_i$ $\to^*_R$ $\AppearedLaterUsedVar_i\sigma_{i+1}$ that
 $\AppearedLaterUsedVar_i\sigma$ $\to^*_R$ $\AppearedLaterUsedVar_i\sigma_{i+1}$ for all $1$ $\leq$ $i$ $\leq$ $k$.
Moreover, it follows from the $\Uopt$-LL property and
 Theorem~\ref{th:Uopt-P} that $\Var(t_i) \cap \left(
 \Dom(\sigma_1|_{\Var(l)}) \cup   
\cdots \cup \Dom(\sigma_{i-1}|_{\AppearedLaterUsedVar_{i-1} \setminus
 \AppearedLaterUsedVar_{i-2}}) \right)$ 
 $=$ $\emptyset$ for all $1$ $\leq$ $i$ $\leq$ $k$, and hence
 $t_i\sigma_i$ $=$ $t_i\sigma$ for all $1$ $\leq$ $i$ $\leq$ $k$.

Now we show that $s_i\sigma$ $\to^*_R$ $s_i\sigma_i$ for all $1$ $\leq$
$i$ $\leq$ $k$, \ie, $x\sigma$ $\to^*_R$ $x\sigma_i$ for all variables
$x$ $\in$ $\Var(s_i)$.
The case that $i$ $=$ $1$ is trivial, so let $i$ $>$ $1$.
We make a case distinction depending on where $x$\/ appears.
\begin{enumerate}[$\bullet$]
 \item Consider the case that $x$ $\in$ $\Var(l)$.
       By definition, $x$ $\in$ $\AppearedLaterUsedVar_j$ for all $1$ $\leq$
       $j$ $<$ $i$, and hence we have the derivation
       $x\sigma$ $=$ $x\sigma_1$ $\to^*_R$ $x\sigma_2$ $\to^*_R
       \cdots \to^*_R$ $x\sigma_i$.
 \item Consider the remaining case that $x$ $\in$ $\Var(t_j)$ for some
       $j$\/ with $1$ $\leq$ $j$ $<$ $i$.
       It follows from the $\Uopt$-LL property of $\rho$\/ that $x$ $\in$
       $\AppearedLaterUsedVar_j\setminus \AppearedLaterUsedVar_{j-1}$.  
       By definition, $x$ $\in$ $\AppearedLaterUsedVar_{j'}$ for all $j$
       $\leq$ $j'$ $<$ $i$, and hence we have the derivation $x\sigma$
       $=$ $x\sigma_j$ $\to^*_R$ $x\sigma_{j+1}$ $\to^*_R \cdots
       \to^*_R$ $x\sigma_i$. 
\end{enumerate}
Thus, $x\sigma$ $\to^*_R$ $x\sigma_i$ for all variables
$x$ $\in$ $\Var(s_i)$, and hence $s_i\sigma$ $\to^*_R$ $s_i\sigma_i$.
It follows from the assumption that $s_i\sigma$ $\to^*_R$ $s_i\sigma_i$
$\to^*_R$ $t_i\sigma_{i+1}$ $=$ $t_i\sigma$ for all $1$ $\leq$ $i$
$\leq$ $k$.
Similarly, we have the derivation $r\sigma$ $\to^*_R$ $r\sigma_{k+1}$.
Therefore, we have the derivation $l\sigma_1$ $=$ $l\sigma$ $\to_R$ $r\sigma$
 $\to^*_R$ $r\sigma_{k+1}$.
\qed

Next, we show the proof of Lemma~\ref{lem:LL-soundness}.

\begin{againtheorem}[Lemma~\ref{lem:LL-soundness}]
Let $R$ be a\/ $\Uopt$-LL 3-eDCTRS over a signature $\cF$, $s$ be a term in
 $\cT(\cF,\cV)$, $t$ be a linear term in $\cT(\cF,\cV)$, and
 $\sigma$ be a substitution in $\Subst(\cF_{\Uopt(R)},\cV)$. 
Suppose that $R$ is non-LV or non-RV.
 If $s$ $\pto^n_{\Uopt(R)}$ $t\sigma$ for some $n$ $\geq$
 $0$, then there exists a substitution $\theta$ in $\Subst(\cF,\cV)$
 such that 
 \begin{enumerate}[$\bullet$]
  \item $s$ $\to^*_R$ $t\theta$ $\pto^{n'}_{\geq \Pos_\cV(t),\Uopt(R)}$
	$t\sigma$ for some $n'$ $\leq$ $n$, and  
  \item if $t\sigma$ $\in$ $\cT(\cF,\cV)$, then $t\theta$ $=$ $t\sigma$.
 \end{enumerate}
\end{againtheorem}

\proof
We prove this lemma by induction on the lexicographic product $(n,s)$
 of the length $n$\/ and the structure of $s$.
The case that $n$ $=$ $0$ is trivial, so let $n$ $>$ $0$.

We first consider the case that $s$ $\pto^n_{\Uopt(R)}$
 $t\sigma$\/ does not contain any reduction step at the root position.
 In this case, $s$\/ is not a variable.
 Let $s$\/ be of the form $f(s_1,\ldots,s_m)$ with $f$ $\in$ $\cF$. 
We make a case distinction depending on whether $t$\/ is a variable or
 not.
 \begin{enumerate}[$\bullet$]
  \item Consider the case that $t$\/ is not a variable.
	In this case, $s$ $=$ $f(s_1,\ldots,s_m)$
	$\pto^n_{\Uopt(R)}$ $f(t_1,\ldots,t_m)\sigma$
	$=$ $t\sigma$, and thus, $s_i$
	$\pto^{n_i}_{\Uopt(R)}$ $t_i\sigma$, where $n_i$
	$\leq$ $n$, for all $1$ $\leq$ $i$ $\leq$ $m$. 
	By the induction hypothesis, for all $1$ $\leq$ $i$ $\leq$ $m$,
	there exists a substitution $\theta_i$ $\in$ $\Subst(\cF,\cV)$
	such that
	\begin{enumerate}[--]
	 \item $s_i$ $\to^*_R$ $t_i\theta_i$ $\pto^{n'_i}_{\geq
	       \Pos\cV(t_i),\Uopt(R)}$ $t_i\sigma$ for some $n'_i$ $\leq$ $n_i$, and
	 \item if $t_i\sigma$ $\in$ $\cT(\cF,\cV)$, then $t_i\theta_i$
	       $=$ $t_i\sigma$.
	\end{enumerate}
	Let $\theta$ $=$ $\theta_1|_{\Var(t_1)} \cup \cdots \cup
	\theta_m|_{\Var(t_m)}$. 
	Then, it follows from the linearity of $t$\/ that $\theta$\/ is a
	substitution in $\Subst(\cF,\cV)$. 
	Thus, we have the derivation $s$ $=$ $f(s_1,\ldots,s_m)$
	$\to^*_R$ $f(t_1,\ldots,t_m)\theta$ $=$ $t\theta$
	$\pto^{n'}_{\geq \Pos_\cV(t),\Uopt(R)}$ $t\sigma$\/ where 
	\begin{enumerate}[--]
	 \item $n'$ is the maximum of $n'_1,\ldots,n'_m$, and 
	 \item if $t\sigma$ $\in$ $\cT(\cF,\cV)$, then $t\theta$ $=$
	       $t\sigma$. 
	\end{enumerate}
	Moreover, it follows from $n'_i$ $\leq$ $n$\/ that $n'$ $\leq$ $n$.
  \item Consider the remaining case that $t$\/ is a variable $x$.
	In this case, we can let $\sigma$ $=$ $\{ x \mapsto
	f(t_1,\ldots,t_m)\}$. 
	Now, let $t'$ be a linear term $f(x_1,\ldots,x_m)$ with
	$x_1,\ldots,x_m$ $\in$ $\cV$, and $\sigma'$ $=$ $\{ x_i \mapsto
	t_i \mid 1 \leq i \leq m \}$. 
	Then, $s$ $=$ $f(s_1,\ldots,s_m)$
	$\pto^n_{>\varepsilon,\Uopt(R)}$ $f(x_1,\ldots,x_m)\sigma'$.
	Similarly to the previous case, we have a substitution $\theta'$
	such that
	\begin{enumerate}[--]
	 \item $f(s_1,\ldots,s_m)$ $\to^*_R$ $t'\theta'$
	       $\pto^{n'}_{\geq \Pos_\cV(f(x_1,\ldots,x_m)),\Uopt(R)}$
	       $t'\sigma'$, and
	 \item if $t'\sigma'$ $\in$ $\cT(\cF,\cV)$, then $t'\theta'$ $=$
	       $t'\sigma'$,
	\end{enumerate}
	for some $n'$ $\leq$ $n$.
	Let $\theta$ $=$ $\{ x \mapsto t'\theta' \}$.
	Then, we have the derivation $s$ $=$ $f(s_1,\ldots,s_m)$ $\to^*_R$
	$t'\theta'$ $=$ $t\theta$
	$\pto^{n'}_{\geq \{\varepsilon\},\Uopt(R)}$ $t\sigma$\/ with $n'$ $\leq$
	$n$, and $t\theta$ $=$ $t'\theta'$ $=$ $t\sigma$\/ whenever
	$t\sigma$ $\in$ $\cT(\cF,\cV)$.
 \end{enumerate}

 Next we consider the remaining case that at least one rule is applied
 at the root position. 
 In the following, we make a case distinction depending on whether $R$\/
 is non-LV or non-RV.
 In the case that $R$\/ is non-LV, we focus on the first rule applied at
 the root position, and otherwise (\ie, $R$\/ is non-RV), we focus on the last
 rule applied at the root position. 
 The case that the focused rule does not contain a U symbol is simpler than the
 other case that the focused rule contains a U symbol since the rule is
 contained not only in $\Uopt(R)$ but also in $R$.
 For this reason, we only consider the case that the focused rule
 contains a U symbol.
 Now we assume that the focused rule is of the following form:
  \begin{enumerate}[$\bullet$]
   \item $l \to
	 \Usym^\rho_1(s_1,\overrightarrow{\AppearedLaterUsedVar_1})$ if
	 $R$\/ is non-LV, and
   \item $\Usym^\rho_i(t_i,\overrightarrow{\AppearedLaterUsedVar_i}) \to
	 \Usym^\rho_{i+1}(s_{i+1},\overrightarrow{\AppearedLaterUsedVar_{i+1}})$ or 
	$\Usym^\rho_k(t_k,\overrightarrow{\AppearedLaterUsedVar_k}) \to
	 r$\/ if $R$\/ is non-RV. 
  \end{enumerate}
 For the sake of readability, we assume w.l.o.g.\ that $k$ $=$ $2$.

 Let us start the case distinction mentioned above. 
  \begin{enumerate}[(1)]
   \item Consider the case that $R$\/ is non-LV.
	 In this case, we have the following subcases depending on where
	 $t\sigma$\/ appears.
   \begin{enumerate}[a.]
    \item Consider the case that
	  \[
	  \begin{array}{@{}l@{\>}l@{}}
	   s 
	    \pto^{n_0}_{>\varepsilon,\Uopt(R)}
	    l\sigma_1
	    &
	    \to_{\varepsilon,\Uopt(R)}
	    \Usym^\rho_1(s_1,\overrightarrow{\AppearedLaterUsedVar_1})\sigma_1
	    \pto^{n_1}_{>\varepsilon,\Uopt(R)}
	    t\sigma
	    \\
	  \end{array}
	  \]
	  where $n_0 + n'' + 1$ $=$ $n$.
	 By the induction hypothesis, there exists a substitution
	 $\theta_1$ $\in$ $\Subst(\cF,\cV)$ such that $s$ $\to^*_R$
	 $l\theta_1$ $\pto^{n'_1}_{\geq \Pos_\cV(l),\Uopt(R)}$ $l\sigma_1$
	 for some $n'_1$ $\leq$ $n_0$.
 	 Thus, $l\theta_1$ $\pto^{n'_1}_{\Uopt(R)}$ 
	 $\l\sigma_1$ $\to_{\varepsilon,\Uopt(R)}$
	 $\Usym^\rho_1(s_1,\overrightarrow{\AppearedLaterUsedVar_1})\sigma_1$ 
	 $\pto^{n''}_{>\varepsilon,\Uopt(R)}$ $t\sigma$\/ 
	 with $n'_1+1+n''$ $\leq$ $n$.
 	 Since $\Usym^\rho_1(s_1,\overrightarrow{\AppearedLaterUsedVar_1})\sigma_1$
	 $\pto^{n''}_{>\varepsilon,\Uopt(R)}$ $t\sigma$\/ 
	 does not contain a rewrite step at the root position,
	  $\Root(t\sigma)$ $=$ $\Usym^\rho_1$. 
	 It follows from the assumption $t$ $\in$ $\cT(\cF,\cV)$ that
	  $t$\/ is a variable $x$.
	 Now let $\theta$ $=$ $\{ x \mapsto l\theta_1 \}$.
	 Then, $\theta$\/ is a substitution in $\Subst(\cF,\cV)$ such that
	 $s$ $\to^*_R$ $l\theta_1$ $=$ $t\theta$
	 $\pto^{n'_1+1+n''}_{>\{\varepsilon\},\Uopt(R)}$ $t\sigma$\/
	 with $n'_1+1+n''$ $\leq$ $n$\/ and $t\sigma$ $\not\in$
	 $\cT(\cF,\cV)$. 
   \item Consider the case that
	 \[
	 \begin{array}{@{}l@{\>}l@{}}
	  s 
	   \pto^{n_0}_{>\varepsilon,\Uopt(R)}
	   l\sigma_1
	   &
	   \to_{\varepsilon,\Uopt(R)}
	   \Usym^\rho_1(s_1,\overrightarrow{\AppearedLaterUsedVar_1})\sigma_1
	   \pto^{n_1}_{>\varepsilon,\Uopt(R)}
	   \Usym^\rho_1(t_1,\overrightarrow{\AppearedLaterUsedVar_1})\sigma_2
	   \\
	  &
	   \to_{\varepsilon,\Uopt(R)}
	   \Usym^\rho_2(s_2,\overrightarrow{\AppearedLaterUsedVar_2})\sigma_2
	   \pto^{n''}_{>\varepsilon,\Uopt(R)}
	   t\sigma
	   \\
	 \end{array}
	 \]
	 where $n_0 + n_1 + n'' + 2$ $=$ $n$.
	 This case is proved similarly to the previous case.
    \item Consider the remaining case that
	  \[
	  \begin{array}{@{}l@{\>}l@{}}
	   s 
	    \pto^{n_0}_{>\varepsilon,\Uopt(R)}
	    l\sigma_1
	    &
	    \to_{\varepsilon,\Uopt(R)}
	    \Usym^\rho_1(s_1,\overrightarrow{\AppearedLaterUsedVar_1})\sigma_1
	    \pto^{n_1}_{>\varepsilon,\Uopt(R)}
	    \Usym^\rho_1(t_1,\overrightarrow{\AppearedLaterUsedVar_1})\sigma_2
	    \\
	   &
	    \to_{\varepsilon,\Uopt(R)}
	    \Usym^\rho_2(s_2,\overrightarrow{\AppearedLaterUsedVar_2})\sigma_2
	    \pto^{n_2}_{>\varepsilon,\Uopt(R)}
	    \Usym^\rho_2(t_2,\overrightarrow{\AppearedLaterUsedVar_2})\sigma_3
	    \\
	   &
	    \to_{\varepsilon,\Uopt(R)}
	    r\sigma_3
	    \pto^{n''}_{\Uopt(R)}
	    t\sigma
	    \\
	  \end{array}
	  \]
	 where $n_0 + n_1 + n_2 + n'' + 3$ $=$ $n$.
	 By the induction hypothesis, there exists a substitution
	 $\theta_1$ $\in$ $\Subst(\cF,\cV)$ such that $s$ $\to^*_R$
	 $l\theta_1$ $\pto^{n'_1}_{\geq \Pos_\cV(l),\Uopt(R)}$ $l\sigma_1$
	 for some $n'_1$ $\leq$ $n_0$.
	 Since $l\theta_1$ $\pto^{n'_1}_{\geq \Pos_\cV(l),\Uopt(R)}$
	 $l\sigma_1$, it follows from the well-known standard property of the
	 parallel reduction \cite[Lemma~6.4.2]{BN98} that
	 $\Usym^\rho_1(s_1,\overrightarrow{\AppearedLaterUsedVar_1})\theta_1$
	 $\pto^{n'_1}_{\geq
	 \Pos_\cV(\Usym^\rho_1(s_1,\overrightarrow{\AppearedLaterUsedVar_1})),\Uopt(R)}$
	 $\Usym^\rho_1(s_1,\overrightarrow{\AppearedLaterUsedVar_1})\sigma_1$.
	 Thus, 
	 $\Usym^\rho_1(s_1,\overrightarrow{\AppearedLaterUsedVar_1})\theta_1$
	 $\pto^{n'_1}_{>\varepsilon,\Uopt(R)}$
	 $\Usym^\rho_1(s_1,\overrightarrow{\AppearedLaterUsedVar_1})\sigma_1$
	 $\pto^{n_1}_{>\varepsilon,\Uopt(R)}$ 
	 $\Usym^\rho_1(t_1,\overrightarrow{\AppearedLaterUsedVar_1})\sigma_2$ with
	 $n'_1+n_1$ $<$ $n$, and hence $s_1\theta_1$
	 $\pto^{n'_1+n_1}_{\Uopt(R)}$ 
	 $t_1\sigma_2$ and $\AppearedLaterUsedVar_1\theta_1$
	 $\pto^{n'_1+n_1}_{\Uopt(R)}$ $\AppearedLaterUsedVar_1\sigma_2$.
	 Since the $\Uopt$-LL property provides 
	 the linearity of $t_1$, by the induction hypothesis, there
	 exists a substitution $\theta'_2$ $\in$ $\Subst(\cF,\cV)$ such
	 that $s_1\theta_1$ $\to^*_R$ $t_1\theta'_2$
	 $\pto^{n''_2}_{\geq \Pos_\cV(t_1),\Uopt(R)}$ $t_1\sigma_2$ for some
	 $n''_2$ $\leq$ $n'_1+n_1$.
	 Also, by the induction hypothesis, for any variable $y$ $\in$
	  $\AppearedLaterUsedVar_1$, 
	 there exists a substitution $\delta_y$ $\in$ $\Subst(\cF,\cV)$
	 such that $y\theta_1$ $\to^*_R$ $y\delta_y$
	 $\pto^{j_y}_{\geq \{\varepsilon\},\Uopt(R)}$ $y\sigma_2$ for some $j_y$
	 $\leq$ $n'_1+n_1$.
	 Let $\theta_2$ $=$ $\theta'_2|_{\Var(t_1)} \cup \{ y \mapsto
	 y\delta_y \mid y \in \AppearedLaterUsedVar_1 \}$.
	 Then, since the $\Uopt$-LL property provides $\Var(t_1) \cap
	 \AppearedLaterUsedVar_1$ $=$ $\emptyset$, 
	 we have the derivations $s_1\theta_1$ $\to^*_R$
	 $t_1\theta_2$ $\pto^{n'_2}_{\Uopt(R)}$ $t_1\sigma_2$ and
	 $\AppearedLaterUsedVar_1\theta_1$ $\to^*_R$ $\AppearedLaterUsedVar_1\theta_2$
	 $\pto^{n'_2}_{\Uopt(R)}$ $\AppearedLaterUsedVar_1\sigma_2$ for some
	 $n'_2$ $\leq$ $n'_1+n_1$ that is the maximum of $n''_2$ and $j_y$ for
	 $y$ $\in$ $\AppearedLaterUsedVar_1$.
	  Thus, we have the derivation
	 $\Usym^\rho_2(s_2,\overrightarrow{\AppearedLaterUsedVar_2})\theta_2$
	 $\pto^{n'_2+n_2}_{>\varepsilon,\Uopt(R)}$
	 $\Usym^\rho_2(t_2,\overrightarrow{\AppearedLaterUsedVar_2})\sigma_3$
	 with $n'_2+n_2$ $<$ $n$. 
 
	 In the same way, we obtain a substitution $\theta_3$ in
	 $\Subst(\cF,\cV)$ such that $s_2\theta_2$ $\to^*_R$ $t_2\theta_3$,
	 $\AppearedLaterUsedVar_2\theta_2$ $\to^*_R$
	  $\AppearedLaterUsedVar_2\theta_3$,  
	 $\Usym^\rho_2(t_2,\overrightarrow{\AppearedLaterUsedVar_2})\theta_3$
	 $\pto^{n'_3}_{>\varepsilon,\Uopt(R)}$
	 $\Usym^\rho_2(t_2,\overrightarrow{\AppearedLaterUsedVar_2})\sigma_3$
	  for some $n'_3$ $\leq$ 
	 $n'_2+n_2$.
	 Moreover, in the same way, we obtain a substitution $\theta$ in
	 $\Subst(\cF,\cV)$ such that  
	 $r\theta_3$ $\to^*_R$ $t\theta$ $\pto^{n'}_{\geq \Pos_\cV(t),\Uopt(R)}$
	 $t\sigma$\/ and  $t\sigma$ $\in$ $\cT(\cF,\cV)$ implies $t\theta$
	  $=$ $t\sigma$, where $n'$ $\leq$ $n'_3+n''$ $<$ $n$.
	 It follows from Lemma~\ref{lem:Uopt-LL} that $l\theta_1$ $\to^+_R$
	 $r\theta_3$. 
	 Therefore, we have the derivation $s$ $\to^*_R$ $l\theta_1$ $\to^+_R$
	 $r\theta_3$ $\to^*_R$ $t\theta$
	 $\pto^{n'}_{\geq \Pos_\cV(t),\Uopt(R)}$ $t\sigma$\/ with $n'$
	  $\leq$ $n$, and $t\theta$ $=$ $t\sigma$\/ whenever $t\sigma$
	  $\in$ $\cT(\cF,\cV)$. 
   \end{enumerate}

   \item Consider the remaining case (\ie, $R$\/ is non-LV). 
	 Similarly to Case~(1), we have the following
	 subcases.
   \begin{enumerate}[a.]
    \item Consider the case that
	  \[
	  \begin{array}{@{}l@{\>}l@{}}
	   s 
	    \pto^{n_0}_{\Uopt(R)}
	    l\sigma_1
	    &
	    \to_{\Uopt(R)}
	    \Usym^\rho_1(s_1,\overrightarrow{\AppearedLaterUsedVar_1})\sigma_1
	    \pto^{n_1}_{>\varepsilon,\Uopt(R)}
	    t\sigma
	    \\
	  \end{array}
	  \]
	  where $n_0 + n'' + 1$ $=$ $n$.
	  The only difference from Case (1)-a is that $s$
	  $\pto^{n_0}_{\Uopt(R)}$ $l\sigma_1$ 
	  may contain a rewrite step at the root position.
	  This case can be proved similarly to Case (1)-a.
  \item Consider the case that
	\[
	\begin{array}{@{}l@{\>}l@{}}
	 s 
	  \pto^{n_0}_{\Uopt(R)}
	  l\sigma_1
	  &
	  \to_{\varepsilon,\Uopt(R)}
	  \Usym^\rho_1(s_1,\overrightarrow{\AppearedLaterUsedVar_1})\sigma_1
	  \pto^{n_1}_{>\varepsilon,\Uopt(R)}
	  \Usym^\rho_1(t_1,\overrightarrow{\AppearedLaterUsedVar_1})\sigma_2
	  \\
	 &
	  \to_{\varepsilon,\Uopt(R)}
	  \Usym^\rho_k(s_2,\overrightarrow{\AppearedLaterUsedVar_2})\sigma_2
	  \pto^{n''}_{>\varepsilon,\Uopt(R)}
	  t\sigma
	  \\
	\end{array}
	\]
	 where $n_0 + n_1 + n'' + 2$ $=$ $n$.
 	 Again, the only difference from Case~(1)-b is that $s$
	 $\pto^{n_0}_{\Uopt(R)}$ $l\sigma_1$ 
	 may contain a rewrite step at the root position.
	 This case can be proved similarly to Case~(1)-b.
    \item Consider the remaining case that
	  \[
	  \begin{array}{@{}l@{\>}l@{}}
	   s 
	    \pto^{n_0}_{\Uopt(R)}
	    l\sigma_1
	    &
	    \to_{\varepsilon,\Uopt(R)}
	    \Usym^\rho_1(s_1,\overrightarrow{\AppearedLaterUsedVar_1})\sigma_1
	    \pto^{n_1}_{>\varepsilon,\Uopt(R)}
	    \Usym^\rho_1(t_1,\overrightarrow{\AppearedLaterUsedVar_1})\sigma_2
	    \\
	   &
	    \to_{\varepsilon,\Uopt(R)}
	    \Usym^\rho_k(s_2,\overrightarrow{\AppearedLaterUsedVar_2})\sigma_2
	    \pto^{n_2}_{>\varepsilon,\Uopt(R)}
	    \Usym^\rho_k(t_2,\overrightarrow{\AppearedLaterUsedVar_2})\sigma_3
	    \\
	   &
	    \to_{\varepsilon,\Uopt(R)}
	    r\sigma_3
	    \pto^{n''}_{>\varepsilon,\Uopt(R)}
	    t\sigma
	    \\
	  \end{array}
	  \]
	  where $n_0 + n_1 + n_2 + n'' + 3$ $=$ $n$.
	  The only difference from Case (1)-c is that $s$
	  $\pto^{n_0}_{\Uopt(R)}$ $l\sigma_1$ may contain a
	  rewrite step at the root position, but $r\sigma_{3}$
	  $\pto^{n''}_{\Uopt(R)}$ $t\sigma$\/ does not contain
	   any rewrite step at the root position. 
	  This case can be proved similarly to Case~(1)-c.
	  \qed
   \end{enumerate}
  \end{enumerate}

\subsection{Proof of Theorem~\ref{th:RL-LL}}
\label{subsec:proof:th:RL-LL}

Theorem~\ref{th:RL-LL} is a direct consequence of
the following lemma, a rule-based variant of Theorem~\ref{th:RL-LL}.
\begin{lem}
\label{lem:inverse-syntax}
Let $\rho: l \to r \Leftarrow s_1 \tto t_1; \ldots; s_k \tto t_k$ be an
 (extended) deterministic rewrite rule. 
Then all of the following hold:
\begin{enumerate}[\em(1)]
 \item $\Var(t_i)$ $\subseteq$ $\Var(r,s_{i+1},\ldots,s_k)$ for all\/ $1$
       $\leq$ $i$ $\leq$ $k$ iff\/ $(\rho)^{-1}$ is deterministic,
 \item $\Var(l)$ $\subseteq$ $\Var(r,s_1,\ldots,s_k)$ iff\/ $(\rho)^{-1}$
       is of Type 3,
 \item\label{lem:inverse-syntax:U-equivalence}
      if\/ $\Var(t_i)$ $\subseteq$ $\Var(r,s_{i+1},\ldots,s_k)$
       for all\/ $1$ $\leq$ $i$ $\leq$ $k$, then
       \begin{enumerate}[\em a.]
	\item $\Uopt((\rho)^{-1})$ $=$ $(\Uopt(\rho))^{-1}$ up to the
	      renaming of U symbols, 
	\item $\rho$ is $\Uopt$-LL iff\/
	      $(\rho)^{-1}$ is $\Uopt$-RL, and
	\item $\rho$ is $\Uopt$-RL iff\/ $(\rho)^{-1}$ is $\Uopt$-LL, 
       \end{enumerate}
 \item $\rho$ is non-LV iff\/ $(\rho)^{-1}$ is non-RV, and
 \item $\rho$ is non-RV iff\/ $(\rho)^{-1}$ is non-LV.
\end{enumerate}
\end{lem}
\proof
Since the first, second, fourth, and fifth claims are trivial, we only
 prove the third claim. 
Recall that $\Uopt(\rho)$ $=$ $\{\> l \to
\Usym^\rho_1(s_1,\overrightarrow{\AppearedLaterUsedVar_1}), ~
 \Usym^\rho_1(t_1,\overrightarrow{\AppearedLaterUsedVar_1}) \to 
\Usym^\rho_2(s_2,\overrightarrow{\AppearedLaterUsedVar_2}), ~\ldots, ~
\Usym^\rho_k(t_k,\overrightarrow{\AppearedLaterUsedVar_k}) \to r \>\}$.
 We assume w.l.o.g.\ that $\Uopt((\rho)^{-1})$ $=$ $\{\>
 r \to \Usym^\rho_k(t_k,\overrightarrow{V_k}),~
 \ldots, ~\Usym^\rho_2(s_2,\overrightarrow{V_2}) \to
 \Usym^\rho_1(t_1,\overrightarrow{V_1}),$
 $\Usym^\rho_1(s_1,\overrightarrow{V_1}) \to l \>\}$ where $V_i$ $=$ 
 $\Var(r,s_k,\ldots,s_{i+1}) \cap
 \Var(l,s_i,t_{i-1},s_{i-1},\ldots,t_1,s_1)$.

 Since $\rho$\/ is deterministic, we have $V_i$
 $=$ $\Var(r,s_k,\ldots,s_{i+1}) \cap \AppearedVar_i$.
 Moreover, it follows from the assumption that
 $\LaterUsedVar_i$ $=$ $\Var(r,t_i,s_{i+1},t_{i+1},\ldots,s_k,t_k)$ $=$
 $\Var(r,s_{i+1},\ldots,s_k)$, and hence $V_i$ $=$ $\AppearedVar_i \cap
 \LaterUsedVar_i$ $=$ $\AppearedLaterUsedVar_i$ for all $1$ $\leq$ $i$
 $\leq$ $k$. 
 Therefore, the claim (\ref{lem:inverse-syntax:U-equivalence})-a holds.

 By the definition of $(\Blank)^{-1}$, $\Uopt(R)$ is LL (RL) iff
 $(\Uopt(R))^{-1}$ is RL (LL).  
 Therefore, the claims (\ref{lem:inverse-syntax:U-equivalence})-b,c
 follow from the claim (\ref{lem:inverse-syntax:U-equivalence})-a.
\qed

\subsection{Proof of Lemma~\ref{lem:EV-instantiated}}
\label{subsec:proof:lem:EV-instantiated}

We first prepare some technical lemmas to prove
Lemma~\ref{lem:EV-instantiated}. 
\begin{lem}
 \label{lem:subst-decomposition_for_EV-instantiated}
 Let $l$ be a linear term with U-symbol-free proper subterms, and
  $\theta, \sigma, \eta$ be substitutions such that $\theta$ $\in$
  $\Subst(\cF,\cV)$ and\/ $\Root(x\eta)$ is a U symbol for any $x$ $\in$
  $\Dom(\eta)$.
  If $s\theta\eta$ $=$ $l\sigma$, then there exists a substitution
  $\sigma'$ such that $s\theta$ $=$ $l\sigma'$ and $l\sigma$
  $=$ $l\sigma'\eta$. 
\end{lem}
\proof[Proof (Sketch).]
 This claim can be proved by induction on the term structure of $s$. 
\qed

Lemma~\ref{lem:EV-instantiated} is a direct consequence of the following
 lemma.
\begin{lem}
 Let $s$ be a term in $\cT(\cF_{\Uopt(R)},\cV)$, $t$ be a term in
  $\cT(\cF,\cV)$, and $\theta,\eta$ be substitutions such that $\theta$
  $\in$ $\Subst(\cF_{\Uopt(R)},\cV)$ and $\Root(x\eta)$ is a U symbol
  for any $x$ $\in$ $\Dom(\eta)$.
  If $\Pos_\cF(s\theta): s\theta\eta$ $\evbto^*_{\Uopt(R)}$ $t$, then
  there exists a substitution $\sigma$ $\in$ $\Subst(\cF,\cV)$ such that
  \begin{enumerate}[$\bullet$]
   \item $s\theta\sigma$ $\evbto^n_{\Uopt(R)}$ $t$, and 
   \item the derivation is EV-instantiated on $\cT(\cF,\cV)$.
  \end{enumerate}
\end{lem}
\proof
 We prove this lemma by induction on the length $n$\/ of the derivation
 $\Pos_\cF(s\theta): s\theta\eta$ $\evbto^*_{\Uopt(R)}$ $t$.
The case that $n$ $=$ $0$ is trivial, so let $n$ $>$ $0$.

From the EV-safe property of the derivation and
  Lemma~\ref{lem:subst-decomposition_for_EV-instantiated}, we can 
 assume w.l.o.g.\ 
 that 
\begin{enumerate}[$\bullet$]
 \item $s\theta$\/ is of the form $C[s']_p$ with $s'$ $=$ $l\delta$\/ and
       $p$ $\in$ $\Pos_\cF(s\theta)$, 
 \item $s\theta\eta$ $=$ $C\theta\eta[s'\eta]$ $=$
       $C\theta\eta[l\delta\eta]_p$ $\evbto_{p,l \to r }$
       $C\theta\eta[r\delta\eta]$ $\evbto^{n-1}_{\Uopt(R)}$ $t$, and
 \item $\delta$ $\in$ $\Subst(\cF,\cV)$, 
\end{enumerate}
where 
\begin{enumerate}[$\bullet$]
 \item $l \to r$ $\in$ $\Uopt(R)$ with $\Var(l,r) \cap \Var(s\theta)$
       $=$ $\emptyset$, 
 \item the set $B$\/ of EV-safe positions in $C\theta\eta[r\delta\eta]$ is
       $(\Pos_\cF(s\theta) \setminus \{ q \in \Pos_\cF(s\theta) \mid p
       \leq q \})$ $\cup$ $\{ pq \mid q \in \Pos_\cF(r) \}$ $\cup$ $\{
       pp'q \mid pp''q \in \Pos_\cF(s\theta),~ p'' \in 
       \Pos_\cV(l), ~ l|_{p''} = r|_{p'} \}$, and
 \item $B: C\theta\eta[r\delta\eta]$ $\evbto^{n-1}_{\Uopt(R)}$ $t$. 
\end{enumerate}
 Let $\delta'$ and $\delta''$ be substitutions such that $\delta'$ $\in$
 $\Subst(\cF,\cV)$, $\delta|_{\EVar(l \to r)}$ $=$ $\delta'\eta$,
 $\Dom(\delta'') \cap (\Var(l,r) \cup \Dom(\eta))$ $=$ $\emptyset$, and
 $\Root(x\delta'')$ is a U symbol for any $x$ $\in$ $\Dom(\delta'')$. 

Let $\theta'$ $=$ $\theta|_{\Var(C[~])} \cup \delta_{\Var(l)} \cup
 \delta'|_{\EVar(l \to r)}$ and $\eta'$ $=$ $\eta \cup \delta''$.
Then, $\theta'$ and $\eta'$ are substitutions such that
 $C\theta\eta[r\delta]$ $=$
 $(C[r])\theta'\eta'$.
 It follows from the definition of the EV-safe property that $B$ $=$
 $\Pos_\cF((C[r])\theta')$.  
 Thus, by the induction hypothesis, there exists a
 substitution $\sigma$\/ in $\Subst(\cF,\cV)$ such that
 $(C[r])\theta'\sigma$
 $\evbto^{n-1}_{\Uopt(R)}$ $t$\/ and the derivation is EV-instantiated on
 $\cT(\cF,\cV)$.
 Now, we have the derivation $s\theta\sigma$ $=$ $(C\theta[s'])\sigma$
 $=$ $(C\theta[l\delta])\sigma$ $=$ 
 $(C\theta'[l\theta'])\sigma$ $\evbto_{\Uopt(R)}$
 $(C\theta'[r\theta'])\sigma$ $=$ $(C[r])\theta'\sigma'$
 $\evbto^{n-1}_{\Uopt(R)}$ $t$. 
Since $\theta$\/ and $\sigma$\/ are in $\Subst(\cF,\cV)$, any extra
 variable in $r$\/ is instantiated by a term in $\cT(\cF,\cV)$.
Therefore, this derivation is EV-instantiated on $\cT(\cF,\cV)$.
\qed

\subsection{Proof of Lemma~\ref{lem:SRTd-simulates-U}}\ \\
\label{subsec:proof:lem:SRTd-simulates-U}

\begin{againtheorem}[Lemma~\ref{lem:SRTd-simulates-U}]
Let $R$ be a DCTRS over a signature $\cF$.
 Then, $\phi_{\SRTd(R)}(\to^*_{\U(R)})$ $\subseteq$
 $\to^*_{\SRTd^\to(R)}$ on terms in $\cT(\cF,\cV)$.
\end{againtheorem}

\proof
 We extend the operation $\overline{\Blank}$ by adding the following
 clause to the definition of $\overline{\phi}$ in Definition~\ref{def:SRTd}:
\[
 \begin{array}{@{}l@{}}
  \overline{\phi}(\Usym^{\rho_{f,j}}_i(x_i,\overrightarrow{\Var(f(w_1,\ldots,w_n),t_1,\ldots,t_{i-1})})
   \\ 
  \hspace{20ex}
 = 
 \overline{f}(\overline{w_1},\ldots,\overline{w_n},\underbrace{\bot,\ldots,\bot}_{j-1},[\{x_i\},\overline{t_{i-1}},\ldots,\overline{t_1},\bot,\ldots,\bot]_k,\underbrace{\bot,\ldots,\bot}_{n_f-j}) \\
 \end{array}
\]
where $\rho_{f,j}: f(w_1,\ldots,w_n) \to r \Leftarrow s_1 \tto t_1;
 \ldots; s_k \tto t_k$ $\in$ $R$, $x_i$ is a fresh variable, and
 $\overline{w_1},\ldots,\overline{w_n},\overline{t_1},\ldots,\overline{t_{i-1}}$
 are terms obtained by applying the original operation
 $\overline{\Blank}$ to $w_1,\ldots,w_n,t_1,\ldots,t_{i-1}$, respectively.
 We also extend $\phi_{\SRTd(R)}$ by introducing the extension of~$\overline{\Blank}$. 

 To prove this lemma, it suffices to show that for terms $s,t$ $\in$
 $\cT(\cF_{\U(R)},\cV)$, if $s$ $\to^*_{\U(R)}$ $t$, then
 $\phi_{\SRTd(R)}(s)$ $\to^*_{\SRTd^\to(R)}$ $\phi_{\SRTd(R)}(t)$. 
 We prove this claim by induction on the length $m$\/ of $s$
 $\to^*_{\U(R)}$ $t$.
The case $m$ $=$ $0$ is trivial, so let $m$ $>$ $0$.

Suppose that $s$ $=$ $C[u\sigma]$ $\to_{\U(R)}$ $C[v\sigma]$
$\to^{m-1}_{\U(R)}$ $t$\/ and $\overline{\sigma}$ $=$ $\{ x \mapsto
\overline{x\sigma} \mid x \in \Dom(\sigma)\}$.
In applying $\phi_{\SRTd(R)}$ to $C[~]$, by definition, the hole in $C[~]$ is
neither erased nor duplicated.
Thus, $\phi_{\SRTd(R)}(C[~])$ is a context of the form $\{\overline{C}\}[~]$.
Moreover, by definition, the position $p$\/ of the hole in
$\{\overline{C}\}[~]$ is structural.
Now, we make a case distinction depending on what $u \to v$ $\in$ $\U(R)$ is. 
\begin{enumerate}[$\bullet$]
 \item Consider the case that $u \to v$\/ is $f(w_1,\ldots,w_n) \to r$
       $\in$ $R$. 
       Since
       $\overline{f}(\overline{w_1},\ldots,\overline{w_n},z_1,\ldots,z_{n_f}) $
       $\to \{\overline{r}\}$ $\in$ $\SRTd^\to(R)$ by definition,
       we have the derivation $\overline{u\sigma}$ $=$
       $\overline{f}(\overline{w_1},\ldots,\overline{w_n},z_1,\ldots,z_{n_f})\overline{\sigma}$
       $\to_{\SRTd^\to(R)}$ 
       $\{\overline{r}\,\overline{\sigma}\}$ $=$
       $\{\overline{v\sigma}\}$.
 \item Consider the case that $u \to v$\/ is $f(w_1,\ldots,w_n) \to
       \Usym^{\rho_{f,j}}_1(s_1,\overrightarrow{\Var(l)})$ $\in$ $\U(R)$. 
       By definition, $\overline{u\sigma}$ $=$ 
       $\overline{f}(\overline{w_1},\ldots,\overline{w_n},\bot,\ldots,\bot)\overline{\sigma}$ 
       and
       \[
	\begin{array}{@{}l@{}}
	 \overline{f}(\overline{w_1},\ldots,\overline{w_n},z_1,\ldots,z_{j-1},\bot,z_{j+1},\ldots,z_{n_f})
       \to \\
	 \hspace{15ex}
       \overline{f}(\overline{w_1},\ldots,\overline{w_n},z_1,\ldots,z_{j-1},[\{\overline{s_1}\},\bot,\ldots,\bot],z_{j+1},\ldots,z_{n_f})
       \in \SRTd^\to(R).
	\end{array}
       \]
       Therefore, we have the derivation
       \[
       \begin{array}{@{}l@{\>}l@{}}
	\overline{u\sigma} & = \overline{f}(\overline{w_1},\ldots,\overline{w_n},\bot,\ldots,\bot)\overline{\sigma} \\
	& \to_{\SRTd^\to(R)}
	 \overline{f}(\overline{w_1},\ldots,\overline{w_n},\bot,\ldots,\bot,[\{\overline{s_1}\},\bot,\ldots,\bot],\bot,\ldots,\bot)\overline{\sigma} \\
	& = 
       \overline{\Usym^{\rho_{f,j}}_1(s_1,\overrightarrow{\Var(l)})}\overline{\sigma}
	=
       \overline{\Usym^{\rho_{f,j}}_1(s_1,\overrightarrow{\Var(l)})\sigma}
       = \overline{v\sigma}. \\
       \end{array}
       \]
 \item Consider the case that $u \to v$\/ is 
       the rule $\Usym^{\rho_{f,j}}_i(t_i,\overrightarrow{\Var(l,t_1,\ldots,t_{i-1})}) \to
       \Usym^{\rho_{f,j}}_{i+1}(s_{i+1},$ $\overrightarrow{\Var(l,t_1,\ldots,t_i)})$
       $\in$ $\U(R)$.
       By definition, 
       \[
       \overline{u\sigma} =
       \overline{f}(\overline{w_1},\ldots,\overline{w_n},\bot,\ldots,\bot,[\{\overline{t_i}\},\overline{t_{i-1}},\ldots,\overline{t_1},\bot,\ldots,\bot],\bot,\ldots,\bot)\overline{\sigma}
       \]
       and
       \[
       \begin{array}{@{}l@{}}
	\overline{f}(\overline{w_1},\ldots,\overline{w_n},z_1,\ldots,z_{j-1},[\{\overline{t_i}\},\overline{t_{i-1}},\ldots,\overline{t_1},\bot,\ldots,\bot],z_{j+1},\ldots,z_{n_f})
	 \to \\
       \hspace{4ex}
	\overline{f}(\overline{w_1},\ldots,\overline{w_n},z_1,\ldots,z_{j-1},[\{\overline{s_{i+1}}\},\overline{t_i},\ldots,\overline{t_1},\bot,\ldots,\bot],z_{j+1},\ldots,z_{n_f})
	\in \SRTd^\to(R).
       \end{array}
       \]
       Therefore, we have the derivation
       \[
       \begin{array}{@{}l@{\>}l@{}}
	\overline{u\sigma} 
	 & = 
	 \overline{f}(\overline{w_1},\ldots,\overline{w_n},\bot,\ldots,\bot,[\{\overline{t_i}\},\overline{t_{i-1}},\ldots,\overline{t_1},\bot,\ldots,\bot],\bot,\ldots,\bot)\overline{\sigma} \\
	& \to_{\SRTd^\to(R)}
       \overline{f}(\overline{w_1},\ldots,\overline{w_n},\bot,\ldots,\bot,[\{\overline{s_{i+1}}\},\overline{t_i},\ldots,\overline{t_1},\bot,\ldots,\bot],\bot,\ldots,\bot)\overline{\sigma} \\
	& =
       \overline{\Usym^{\rho_{f,j}}_{i+1}(s_{i+1},\overrightarrow{\Var(l,t_1,\ldots,t_i)})}\overline{\sigma} 
	=
       \overline{\Usym^{\rho_{f,j}}_{i+1}(s_{i+1},\overrightarrow{\Var(l,t_1,\ldots,t_i)})\sigma} 
       = \overline{v\sigma}. \\
       \end{array}
       \]
 \item Consider the remaining case that $u \to v$\/ is
       $\Usym^{\rho_{f,j}}_k(t_k,\overrightarrow{\Var(l,t_1,\ldots,t_{k-1})})
       \to r$ $\in$ $\U(R)$.
       By definition, 
       \[
	\overline{u\sigma} =
       \overline{f}(\overline{w_1},\ldots,\overline{w_n},\bot,\ldots,\bot,[\{\overline{t_1}\},\overline{t_{k-1}}\ldots,\overline{t_1}],\bot,\ldots,\bot)\overline{\sigma}
       \]
       and
       $\overline{f}(\overline{w_1},\ldots,\overline{w_n},z_1,\ldots,z_{j-1},[\{\overline{t_k}\},\overline{t_{k-1}},\ldots,\overline{t_1}],z_{j+1},\ldots,z_{n_f})
       \to \{\overline{r}\}$ $\in$ $\SRTd^\to(R)$.
       Therefore, we have the derivation
       $\overline{u\sigma}$ $=$ $\overline{f}(\overline{w_1},\ldots,\overline{w_n},\bot,\ldots,\bot,[\{\overline{t_1}\},\overline{t_{k-1}}\ldots,\overline{t_1}],\bot,\ldots,\bot)\overline{\sigma}$
       $\to_{\SRTd^\to(R)}$ $\{\overline{r}\}\overline{\sigma}$
       $=$ $\{\overline{r\sigma}\}$ $=$ $\{\overline{v\sigma}\}$.
\end{enumerate}
Now, we have either $\overline{u\sigma}$ $\to_{\SRTd^\to(R)}$
       $\{\overline{v\sigma}\}$ or $\overline{u\sigma}$ $\to_{\SRTd^\to(R)}$
       $\overline{v\sigma}$.
\begin{enumerate}[$\bullet$]
 \item Consider the case that $\overline{u\sigma}$ $\to_{\SRTd^\to(R)}$
       $\{\overline{v\sigma}\}$.
       Since $p$\/ is a structural position, it follows from
       Lemma~\ref{lem:structural-positions} that $\phi_{\SRTd(R)}(s)$
       $=$ $\{\overline{C[u\sigma]}\}$ $=$
       $\{\overline{C}\}[\overline{u\sigma}]$ $\to_{\SRTd^\to(R)}$
       $\{\overline{C}\}[\{\overline{v\sigma}\}]$
       $\to^*_{\SRTd^\to(R)}$
       $\{\{\overline{C}\}[\overline{v\sigma}]\}$ $=$
       $\{\{\overline{C[v\sigma]}\}\}$ $\to_{\SRTd^\to(R)}$
       $\{\overline{C[v\sigma]}\}$ $=$ $\phi_{\SRTd(R)}(C[v\sigma])$.
\item Consider the remaining case that $\overline{u\sigma}$ $\to_{\SRTd^\to(R)}$
       $\overline{v\sigma}$.
      Then, $\phi_{\SRTd(R)}(s)$ $=$
      $\{\overline{C[u\sigma]}\}$ $=$
      $\{\overline{C}\}[\overline{u\sigma}]$ $\to_{\SRTd^\to(R)}$
      $\{\overline{C}\}[\overline{v\sigma}]$ $=$
      $\{\overline{C[v\sigma]}\}$ $=$
      $\phi_{\SRTd(R)}(C[v\sigma])$.
\end{enumerate}
 By the induction hypothesis, $\phi_{\SRTd(R)}(C[v\sigma])$
 $\to^*_{\SRTd^\to(R)}$ $\phi_{\SRTd(R)}(t)$.
 Therefore, we have the derivation $\phi_{\SRTd(R)}(s)$
 $\to^*_{\SRTd^\to(R)}$ $\phi_{\SRTd^\to(R)}(C[v\sigma])$
 $\to^*_{\SRTd^\to(R)}$ $\phi_{\SRTd(R)}(t)$.  
\qed

\end{document}